\documentclass[aps,pra,showpacs,twocolumn,nofootinbib,floatfix]{revtex4}
\usepackage{graphicx}
\usepackage{bm}
\usepackage{amssymb}
\usepackage{amsmath}
\usepackage{times}

\newcommand{\md}{\,\mathrm{d}} 

\newcommand{\trace}{\mathop{\rm Tr}\nolimits}
\newcommand{\vek}{\mathop{\rm Vec}\nolimits}
\newcommand{\matt}{\mathop{\rm Mat}\nolimits}
\newcommand{\diag}{\mathop{\rm Diag}\nolimits}

\newcommand{\erf}{\mathop{\rm Erf}\nolimits}
\newcommand{\erfc}{\mathop{\rm Erfc}\nolimits}

\newcommand{\twomat}[4]{\left(\begin{array}{cc}#1&#2\\#3&#4\end{array}\right)}

\newcommand{\cS}{{\mathcal S}} 
\newcommand{\cH}{{\mathcal H}} 
\newcommand{\cP}{{\mathcal P}} 
\newcommand{\cL}{{\mathcal L}} 
\newcommand{\cM}{{\mathcal M}} 
\newcommand{\cN}{{\mathcal N}} 
\newcommand{\cU}{{\mathcal U}} 

\newcommand{\qed}{\hfill$\square$\par\vskip12pt}

\newcommand{\C}{{\mathbb{C}}}

\newcommand{\M}{{\mathbb{M}}}

\DeclareRobustCommand\openone{\leavevmode\hbox{\small1\normalsize\kern-.33em1}}
\newcommand{\id}{\mathrm{\openone}}

\newcommand{\be}{\begin{equation}}
\newcommand{\ee}{\end{equation}}
\newcommand{\bea}{\begin{eqnarray}}
\newcommand{\eea}{\end{eqnarray}}
\newcommand{\beas}{\begin{eqnarray*}}
\newcommand{\eeas}{\end{eqnarray*}}

\newtheorem{lemma}{Lemma}

\newcount\minute
\newcount\hour
\def\currenttime{%
    \minute\time
    \hour\minute
    \divide\hour60
    \the\hour:\multiply\hour60\advance\minute-\hour\the\minute}

\begin{document}
\title{Quantum Tomographic Reconstruction with Error Bars:
a Kalman Filter Approach}
\author{Koenraad M.R.~Audenaert}
\email{koenraad.audenaert@rhul.ac.uk}
\affiliation{Dept. of Mathematics, Royal Holloway, University of London,
Egham, Surrey TW20 0EX, UK}
\author{Stefan Scheel}
\email{s.scheel@imperial.ac.uk}
\affiliation{Quantum Optics and Laser Science, Blackett Laboratory,
Imperial College London, Prince Consort Road, London SW7 2AZ, UK}

\begin{abstract}
We present a novel quantum tomographic reconstruction method based on
Bayesian inference via the Kalman filter update equations. The method
not only yields the maximum likelihood/optimal Bayesian
reconstruction, but also a covariance matrix expressing the
measurement uncertainties in a complete way. From this covariance
matrix the error bars on any derived quantity can be easily
calculated. This is a first step towards the broader goal of devising
an omnibus reconstruction method that could be adapted to any
tomographic setup with little effort and that treats measurement
uncertainties in a statistically well-founded way.

In this first part we restrict ourselves to the important subclass of
tomography based on measurements with discrete outcomes (as opposed to
continuous ones), and we also ignore any measurement imperfections
(dark counts, less than unit detector efficiency, etc.), which will be
treated in a follow-up paper. We illustrate our general theory on real
tomography experiments of quantum optical information processing
elements.
\end{abstract}

\date{\today, \currenttime}

\maketitle

\section{Introduction}

Since the first proposal of optical quantum tomography by Vogel and
Risken \cite{vogel89}, and the first practical tomographic experiments
by Smithey \textit{et al} \cite{smithey93}, quantum tomography has
gone a long way, and is now being used in a variety of physical
setups, not restricted to optical systems, and many improvements have
been made to the original reconstruction methods
\cite{leonhardtBook,welschRev}.
While this is certainly a desirable evolution, it must also be said
that on the negative side this resulted in a proliferation of
tomography methods.
While it is unavoidable that every physical system has its own
peculiarities, and each particular setup calls for its own tomographic
measurement method, it is not
conceivable
that for every type of system and for every tomography method there
should also be a different tomographic reconstruction
method. Furthermore, when each time the reconstruction software is
written from scratch, that does not benefit reliability. After 20
years of tomographical experience it is not unreasonable to ask for a
unification of these methods, taking the best out of each and devising
a small set of ``omnibus'' reconstruction methods, that only need some
small adaptations to the particular tomographic setup.

An even more important remark concerning reconstruction methods is the
fact that error bars are seldom given. Measurements are worthless
without error bars.
When tomography is just a measurement, even though a complicated one,
why then treat tomography differently? Error bars are dearly needed
here as well, since the whole purpose of tomography is to come up with
a description of the quantum state that is sufficient to derive
further properties, and for these properties error bars would
certainly be needed. As \v{R}eha\v{c}ek, Mogilevtsev and Hradil
\cite{rehacek08} stated: ``The quantification of relevant statistical
errors is an indispensable but often neglected part of any tomographic
scheme used for quantum diagnostic purposes.''

Some theoretical papers mention error bars, but they are calculated
from simulated data, using Monte-Carlo methods and are only meant to
give an indication about how good the method performs. As measurement
errors depend on the actual system and its state, this is clearly not
enough. What we are after is error bars produced straight from the
experimental data and the underlying statistical model.

A widely used error criterion is the state fidelity (for quantum
states) or the process fidelity (for quantum processes), which
compares the reconstructed state to a predefined ``desired state''
(e.g. \cite{awg,blatt,martinis}). While it is easy to use, it clearly
gives no indication about the reconstruction alone but is the sum of
reconstruction and construction errors. As such, it cannot answer the
following two questions separately: ``Are we seeing the correct
state?'', and ``Are we seeing the state correctly?'' Furthermore, as
pointed out in \cite{rehacek08}, fidelity is just a single number and
one cannot expect it to describe the complete error structure of the
reconstruction.


One could argue, however, that error bars are not explicitly needed if one
just subsumes all measurement
noise into the estimated quantum state via statistical
mixing. We disagree with this point of view, and we claim that this throws away useful
information. There is a difference between, on one hand, preparing a pure state and
assuming it is mixed because the measurements do not allow to conclude
otherwise, and, on the other hand, not being able to prepare a pure state, obtaining a
mixed state instead, and knowing that state perfectly. In both cases the
outcome is the same, a mixed state, but in the former case the
real state is actually pure.

To make sense of the concept of error bars in the context of state
(process) estimation, we have to clearly distinguish between the roles
of the state preparer and the observer measuring the state, both of
which involve noise. Tomography is based on the assumption that every
time the observer measures the state, it is actually the same
state. Because of measurement noise, the observer cannot obtain
perfect knowledge of which state has been prepared in finite
time. However, nothing prevents him from doing so in
principle. Measuring for an infinite time, using a sufficient set of
measurements, will yield perfect knowledge. If the same state is being
prepared, it is ultimately knowable. On the other hand, the
preparation of the state will also involve noise. Every time the
preparer attempts to prepare the nominal state, noise will affect this
and some slightly other state will result. This kind of noise is
impossible to overcome, not even in principle. The only way one can
deal with it is by absorbing the preparation noise into the quantum
state that is being prepared, through statistical mixing. If the
preparation noise is ergodic, the observer will recover the average
state of the quantum state ensemble.

In short: the individual states
are not knowable, but their average is; and if the number of
measurements is finite, the observer will not obtain this average
preparation state perfectly. Then error bars, or more precisely a
density function over state space, are  needed to express this lack of
complete knowledge.

\bigskip

One of the first papers calculating error bars from the measurement data is
Ref.~\cite{leonhardt96}, for a specific reconstruction method of
optical homodyning tomography (OHT) using so-called pattern functions
\cite{dariano,leonhardt}. Using this method, the density matrix $\rho$ can be derived directly 
from the detection probabilities $\mbox{pr}(x,\theta)$ sampled over phase space, where $x$ and
$\theta$ are parameters representing the settings of the OHT apparatus.
To reconstruct $\rho$ from the measurement data, these probabilities are replaced by the relative
frequencies $f(x,\theta)/N$.
To obtain error bars, the fluctuations on the detection frequencies are modeled by a Poisson process,
by which the variance $\sigma^2$ equals the mean $\mbox{pr}$ divided
by the number of runs $N$.
The first drawback of this method is that only the marginal variances of the density matrix elements are
treated here, disregarding correlations between errors, and therefore
overestimating them.
A second drawback of this approach is that, due to $\mbox{pr}$ not being known, it is approximated
by the relative frequency, giving $\sigma^2=\mbox{pr}/N\approx f/N^2$, where $f$ is the absolute
detection frequency. This approximation actually underestimates the variance, especially for low-probability events. Indeed, for events with $f=0$ this approach assigns zero probability
to the event, with zero variance, which corresponds to an absolute certainty.
That certainty is not warranted given that only a finite number $N$ of experiments were done.
This problem is known more generally as the ``zero-eigenvalue problem'' and occurs in different
guises in many other reconstruction methods.

The present work is a first step towards overcoming the two
deficiencies described above: we propose to use Bayesian inference as
the unifying principle to calculate a probability density function
over state space, from the measurement data, and from that density
derive the first and second order moments: the mean state and the
state covariance matrix.  The
goal of the present paper is to outline a practical method for
calculating this mean state and state covariance directly from a set
of tomographic data, in a completely general and statistically
well-founded way.

During the course of this work, the paper \cite{rehacek08} appeared, in which the same goals were
aimed for and a method quite similar in spirit to ours was proposed. We refer the reader to Section
\ref{sec:discuss} for a discussion of the main differences between our method and the one in \cite{rehacek08}.

The input required by our method is a statistical model of the quantum
tomographic setup. Given a state and the measurement settings, how do
the statistical properties of the measurement outcomes depend on that?
For a measurement with a finite number $K$ of outcomes, a measurement
setting corresponds to a choice of operator elements $A^{(k)}$, one
element for each outcome. When the outcome is a continuous variable
(e.g. position $x$), the measurement operator is parameterised by the
continuous outcome value $x$. In either case, the laws of quantum
mechanics dictate that the outcome $k$ (or $x$) occurs in an
experiment with a probability (or probability density) given by Born's
rule $p_k = \trace[\rho A^{(k)}]$ (or $p(x) dx = \trace[\rho A(x)]dx$).
The measurements taken in a tomographic experiment relate to this
probability density, in one way or another. For any given setting a
number of runs $N$ would typically be performed, and in case of a
finite outcome experiment, the frequencies $f_k$ of the various
outcomes would be recorded, or the values of the measurement in case
of a continuous outcome. The relation between these frequencies and
the underlying probability distribution is governed by the laws of
statistics.

Alternatively, in some experiments the outcome could be the combined
effect of many small measurement events. For example, in tomography
of atomic/molecular clouds, the fluorescence response of the cloud to
an impinging probe beam could be observed, in which case the
experimental outcome is a fluorescence spectrum
\cite{dwm95,wwv97,zvww99}. The recorded spectrum is a random variate
with expected value proportional to the relevant probability density
(a marginal of the quasi-probability density describing the state) and
variance depending on the signal-to-noise ratio.
Another example is an optical homodyning tomography (OHT) experiment where the
probe beam intensity is so high that individual photons impinging on the detectors
can no longer be resolved and the measurement results are light intensities,
measured as voltages.

Once the statistical model of the tomographic setup has been supplied
in suitable form, a general-purpose Bayesian inference engine could in
principle take it from there, converting the measurement data into a
posterior probability density over quantum state space. However, the
actual calculations are typically too demanding to be at all practical.
The second ingredient of our proposed solution is to make the
calculations involved in Bayesian inference feasible by approximating
the statistical tomographic model by a so-called linear-Gaussian model
(explained below). For such models, the Bayesian inference simplifies
to a set of simple equations known as the \textit{Kalman filter update equations}.

Kalman filtering is a technique for dynamical state estimation that
allows to estimate a dynamical state from a sequence of noisy data
\cite{dsimon}. Kalman filtering has already been applied in the
context of continuous quantum measurement and quantum control
\cite{verstraete,geremia,belavkin}. For tomographic reconstruction,
applying Kalman filtering seems to be a novel idea.


The goals we have set out for this work are quite challenging. Rather
than trying to solve all the problems involved in one go, we focus
here on a particular, but important, class of tomography experiments,
namely those where the  measurements have (few) discrete outcomes, as
opposed to measurements of continuous variables. This class still
covers a wide variety of quantum systems, including single-photon
optical systems (e.g. optical quantum computing)
\cite{awg,banaszek07,zeilinger,alvaro}, spin systems based on ions
\cite{blatt}, atoms \cite{polzik}, or electrons  (spin echo
tomography) \cite{lukin}, superconducting \cite{martinis} and solid
state qubits \cite{julsgaard}, and tomography of atomic and molecular
states based on fluorescence spectra \cite{dwm95,wwv97,zvww99}. For
the purposes of exposition, we will restrict our attention here and
illustrate our reconstruction method for optical systems.
Important examples of tomography experiments not falling in this class
are optical homodyning and heterodyning
\cite{vogel89,smithey93,leonhardt}, where the outcome is the
continuous
variable $x$. We leave this for future work.

The second restriction we have imposed here is that we assume that the
apparatus performing the tomography is ideal. In reality any
measurement exhibits imperfections; e.g. photon detectors have less
than unit efficiency and may exhibit dark counts and input states for
process tomography may be imperfectly prepared. Although these
imperfections pose no deep fundamental problems, they would obscure
the presentation which is why we have chosen to treat them in a
follow-up paper \cite{kalman2}. In the present paper we will to
cover the main principles of our proposal and illustrate them using a
simple real application (based on actual data) as proof of principle.

A third (minor) restriction is that the number
of experimental runs per measurement setting is not too small, so that
the distributions governing the measurement outcomes can be
approximated by normal distributions without too great an error.

It goes without saying that our reconstruction method is suitable both
for state tomography and process tomography, because state tomography
lies at the basis of process tomography. Either one sends various
input states through the quantum process and measures the output
states (as in Ref.~\cite{awg}), or one half of a maximally entangled
state is sent through the process and the global output state
(including the other half) is measured (as in
Ref.~\cite{banaszek07}). Both methods are formally equivalent with
state tomography of the state representative of the process, under the
Choi-Jamiolkowski isomorphism. The presentation of our method will
therefore be mainly state based, for definiteness and simplicity.


As we bring together a number of concepts from statistics, engineering
mathematics and quantum mechanics, we begin our presentation with a
rather lengthy section (Sec.~\ref{sec:background}) on background
material, with an extensive list of notations, and brief overviews of
Bayesian inference and Kalman filtering. In Sec.~\ref{sec:theory} we
present the basic theory of our proposal, and show how the problem of
tomographic reconstruction can be made to fit the mould of Kalman
filtering, thereby proposing solutions to several problems that we
encounter along the way. The theory is then illustrated on two real
tomography applications in Secs.~\ref{sec:NathanState} and
\ref{sec:loopy}, based on actual experimental tomography data.
Finally, in Sec.~\ref{sec:discuss} we highlight the main benefits of our method,
its performance and the costs associated with it, and compare it to existing methods,
in particular to its sister-method proposed in \cite{rehacek08}.
\section{Theoretical Background\label{sec:background}}

\subsection{Notations}

Let us start with introducing the main notations and typographical
conventions that we will use in the paper, along with some fundamental
notions from applied probability theory.
We denote vectors and matrices by symbols in boldface,
$\bm{F},\bm{f}$, to distinguish them from scalar quantities which we
denote in roman, including vector and matrix components,
$f_{i}$. Exceptions to this convention are quantum-mechanical
quantities like states $\rho$, POVM elements $\Pi$ and maps
$\Phi$. The vector whose entries are all 1 is denoted by
$\bm{1}:=(1,\ldots,1)$, and the identity matrix by $\id$.

We adopt the usual convention from the statistics literature to denote
random variables with capital letters, $F$, and the values they can
take with lowercase, $f$. For example, the probability density
function (PDF) of a random variable $F$ is denoted $f_F(f)$. The first
$f$ is the general symbol for a PDF, the second $F$ is the random
variable, and the third $f$ is the argument of the PDF and represents
the values the random variable $F$ can take.
The mean and variance of a scalar random variable $X$ are denoted by
$\mu(X)$, $\sigma^2(X)$, and the mean and the covariance matrix of a
$d$-dimensional random variable $\bm{X}$ by $\mu(\bm{X})$ and
$\bm{\Sigma}(\bm{X})$.

In this paper, a number of distributions are prominent. Here we recall
definitions and notations about the multinomial and normal distributions.
Other distributions (chi-squared, Dirichlet and beta) will be described
in subsequent sections.

When a random $d$-dimensional variable $\bm{F}$ is distributed
according to a multinomial distribution with parameters $N$ (where
$\sum_{k=1}^d f_k=N$) and $\bm{p}$ this is denoted by
$\bm{F}\sim\text{Mtn}(N;\bm{p})$. The PDF of this distribution is
given by
\be
\label{eq:fFP}
f_{\bm{F}}(\bm{f}) = {N \choose \bm{f}} p_1^{f_1}\ldots p_d^{f_d}.
\ee
Here we denote the multinomial coefficient by
$$
{N \choose \bm{f}} = \frac{N!}{f_1!\ldots f_d!}.
$$
The mean of the multinomial distribution is given by
$$
\mu_k = Np_k,
$$
and the entries of its covariance matrix are
$$
\sigma_{k,l}^2 = \left\{
\begin{array}{ll}
N\,p_k(1-p_k),&k=l \\ [2mm]
-N\,p_k p_l,&k\neq l.
\end{array}
\right.
$$

When a random variable $X$ is distributed according to a univariate
normal distribution with mean $\mu$ and variance $\sigma^2$ we write
this as $X\sim\cN(\mu,\sigma^2)$. Similarly, for a multivariate normal
distribution with mean $\bm{\mu}$ and covariance matrix $\bm{\Sigma}$
we write $\bm{X}\sim\cN(\bm{\mu},\bm{\Sigma})$.

We will reserve the symbol $\phi$ for the PDF of a normal
distribution, while using $f$ for general PDFs. The PDF of the
univariate normal distribution will thus be denoted by
$$
\phi(x;\mu,\sigma^2) := \frac{1}{\sqrt{2\pi \sigma^2}}
\,\exp(-\frac{(x-\mu)^2}{2\sigma^2})
$$
Similarly, we will denote the PDF of an $n$-dimensional multivariate
normal distribution by
\be
\phi(\bm{x};\bm{\mu},\bm{\Sigma}) :=
\frac{\exp\left[-\frac{1}{2}\,(\bm{x}-\bm{\mu})^*\bm{\Sigma}^{-1}
(\bm{x}-\bm{\mu})\right]}
{\sqrt{(2\pi)^n |\det\bm{\Sigma}|}}. \label{eq:pdfnormal}
\ee
The quadratic form appearing in the exponent  is a proper
distance measure between the vectors $\bm{x}$ and $\bm{\mu}$ and is
called the \textit{squared Mahalanobis distance}, which we denote by $\cM^2$:
\be\label{eq:Maha}
\cM^2:=(\bm{x}-\bm{\mu})^*\bm{\Sigma}^{-1}(\bm{x}-\bm{\mu}).
\ee

One of the main statistical tools used in this paper is the
approximation of distributions by normal distributions, using the
technique of \textit{moment matching}, whereby a distribution is
approximated by a normal distribution with the same first and second
order moments as the original. While conceptually simple and easy to
use in practice, this method is also statistically well-founded
because it gives the approximation that minimises the Kullback-Leibler
distance $D_{KL}(p||q) := \int \md x\,\, p(x) \,\log(p(x)/q(x))$
between a given PDF $p$ and the approximating normal PDF $q$.


On the matrix analysis side, we will follow mathematical convention
(and not the physical one) of denoting Hermitian conjugates with an
asterisk $\bm{A}^*$ instead of a dagger, and reserve the dagger for
the Moore-Penrose (MP) inverse: $\bm{A}^\dagger$. Complex conjugation is
denoted by an overline: $\overline{\bm{A}}$.

We will have the occasion to apply the following formula for the
matrix  inverse of a rank-$k$ correction to a non-singular matrix:
\bea
\label{eq:milemma}
\lefteqn{\left(A+UCV^* \right)^{-1}} \nonumber\\
&=& A^{-1} - A^{-1}U \left(C^{-1}+V^*A^{-1}U \right)^{-1} V^*A^{-1}
\eea
Here, $A$ is $n\times n$ and non-singular, $C$ is $k\times k$ and
non-singular, and $U$ and $V$ are general $n\times k$ matrices. This
formula is alternately known as the \textit{Matrix inversion lemma}, or the
Woodbury matrix identity \cite{hager}.


While the main topic of this paper is the tomographic reconstruction
of quantum states, maps and POVMs, objects that are typically
represented by matrices (density matrices, Choi matrices, POVM
elements), the reconstruction technique we use is based on vector
representations of the state of the system. Therefore, more often than
not, we will need to convert the usual matrix representation of the
quantum objects to vector representation. For quantum states that
means we will be employing a Hilbert space representation. The space
$\M_d(\C)$ of $d\times d$ matrices will be considered as a Hilbert
space $\cH_d$ equipped with the Hilbert-Schmidt inner product
$\langle\bm{A},\bm{B}\rangle = \trace AB^*$. To distinguish between
the two representations, we write $\rho$ for a density matrix, and
$\bm{\rho}$ for its Hilbert space representative.

While many different bases could be used for $\cH$, by far the easiest
one for the purposes of this paper is the basis of matrix units
$\{e^{ij}\}_{i,j=1}^d$; in quantum physical notation
$e^{ij}=|i\rangle\langle j|$. Converting a density matrix $\rho$ to
its Hilbert space representative $\bm{\rho}$ amounts to the so-called $\vek$
operation, which works by just stacking the columns of the density matrix
into a single $d^2$-dimensional column vector:
$\bm{\rho} = \vek(\rho):=\sum_{i,j}\rho_{ij}|i\rangle|j\rangle$. The reverse
operation is the $\matt$ operation, which reshapes a $d^2$-dimensional
vector into a $d\times d$ matrix. The vector $\vek{\id}$ is denoted
$|\id\rangle$. That is, $|\id\rangle = \sum_{i=1}^d |i\rangle|i\rangle$.

In the same vein, the Hilbert representation of a linear map $\cL$ (be
it completely positive (CP) or not) acting on $d\times d$ density
matrices, expressed in the basis of matrix units, is a $d^2\times d^2$
matrix whose columns are the Hilbert space representations of the
matrices $\cL(e^{ij})$. More specifically, if the map is a CP map and
has the Kraus representation
$\rho\mapsto \cL(\rho) = \sum_{k=1}^K \bm{A}^{(k)}\rho \bm{A}^{(k),*}$,
then the Hilbert space representation of $\cL$ is the matrix $\bm{L}$
given by
$$
\bm{L} = \sum_k \bm{A}^{(k)}\otimes \overline{\bm{A}^{(k)}}.
$$

\subsection{The Dirichlet Distribution\label{sec:dirichlet}}

The Dirichlet distribution is the higher-dimensional generalisation of
the beta distribution. The importance of this distribution stems from
the fact that it is the conjugate distribution of the multinomial
distribution. That is, if $\bm{F}\sim \mbox{Mtn}(N,\bm{p})$ is the
distribution of $\bm{F}$ conditional on $\bm{P}=\bm{p}$, then Bayesian
inversion yields that $\bm{P}$ conditional on $\bm{F}=\bm{f}$ is
Dirichlet with parameter $\bm{f}$. Formally, the two distributions
only differ by their normalisation. The multinomial is normalised by
summing over all integer non-negative $\bm{f}$ summing up to $N$,
while the Dirichlet is normalised by integrating over the simplex of
non-negative $\bm{p}$ summing to 1.


The general form of the PDF of a $d$-dimensional Dirichlet
distribution with parameters $\alpha_i$ is (\cite{kbj}, Chapter 49)
$$
f_{\bm{P}}(\bm{p}) = \Gamma(\alpha_0) \prod_{i=1}^d
\frac{p_i^{\alpha_i-1}}{\Gamma(\alpha_i)},
$$
where $\alpha_0$ is defined as
\be
\alpha_0 := \sum_{i=1}^d \alpha_i.
\ee
The range of $\bm{P}$ is the simplex $p_i\ge0, \sum p_i=1$.
In our case $\alpha_i-1=f_i$, and as $\sum_i f_i=N$, the PDF is
$$
(N+d-1)! \prod_{i=1}^d \frac{p_i^{f_i}}{f_i!}.
$$
The mean values of the Dirichlet distribution are
\bea
\mu_i &=& \frac{\alpha_i}{\alpha_0} \nonumber \\
&=& \frac{f_i+1}{N+d},\label{eq:dirichletmean}
\eea
and the elements of its covariance matrix are
\bea
\sigma_{ij}^2 &=& \left\{
\begin{array}{ll}
\frac{\alpha_i(\alpha_0-\alpha_i)}{\alpha_0^2(\alpha_0+1)},&i=j \\[2mm]
\frac{-\alpha_i\alpha_j}{\alpha_0^2(\alpha_0+1)},&i\neq j
\end{array}
\right. \nonumber \\
&=& \left\{
\begin{array}{ll}
\frac{(f_i+1)(N+d-f_i-1)}{(N+d)^2(N+d+1)},&i=j \\[2mm]
\frac{-(f_i+1)(f_j+1)}{(N+d)^2(N+d+1)},&i\neq j
\end{array}
\right. \label{eq:dirichletcov}
\eea

For further reference, we note a few properties of the covariance
matrix $\bm{\Sigma}$ of the Dirichlet distribution. First of all,
$\bm{\Sigma}$ is singular; it has a zero eigenvalue pertaining to the
eigenvector $\bm{1} := (1,\ldots,1)$. As the inverse of $\bm{\Sigma}$
does not exist we will need its Moore-Penrose inverse
$\bm{\Sigma}^\dagger$. Because $\bm{\Sigma}$ can be written as a
diagonal matrix minus a symmetric rank-1 matrix,
$$
\bm{\Sigma} = \frac{(N+d)\diag(\bm{f}+1)-(\bm{f}+1)(\bm{f}+1)^*}%
{(N+d)^2(N+d+1)},
$$
its MP-inverse can be calculated analytically.

For non-singular differences of a matrix $\bm{D}$ and a rank 1 matrix
$\bm{x}\bm{x}^*$, the matrix inversion lemma (\ref{eq:milemma})
provides the formula
$$
(\bm{D}-\bm{x}\bm{x}^*)^{-1} =
\bm{D}^{-1} +(1-\bm{x}^*\bm{D}^{-1}\bm{x})^{-1}\,\,\bm{D}^{-1}
\bm{x}\bm{x}^*\bm{D}^{-1}.
$$
Even for invertible $\bm{D}$ the difference can still be singular
when $\bm{x}^*\bm{D}^{-1}\bm{x}=1$. In that case the MP-inverse is
given by
\beas
(\bm{D}-\bm{x}\bm{x}^*)^\dagger &=& \bm{G}\,\bm{D}^{-1}\,\bm{G} \\
\bm{G} &=& \id-(\bm{x}^*\bm{D}^{-2}\bm{x})^{-1}\,\,\bm{D}^{-1}
\bm{x}\bm{x}^*\bm{D}^{-1}.
\eeas
Here, $G$ is an orthogonal projector on the support of
$\bm{D}-\bm{x}\bm{x}^*$.

This gives in our case
\bea
\bm{\Sigma}^\dagger &=& (N+d)(N+d+1)\bm{G}\diag(\bm{f}+1)^{-1}\bm{G}
\nonumber\\
\bm{G} &=& \id-d^{-1}\bm{1}\,\,\bm{1}^*.\label{eq:dirichletsigmp}
\eea
Thus $\bm{G}$ is the projector on the subspace of vectors $\bm{x}$ for
which $\sum_i x_i=0$. In other words, on the subspace of differences
of probability vectors, $\bm{\Sigma}^\dagger$ reduces to the diagonal
matrix  $$(N+d)(N+d+1)\diag(\bm{f}+1)^{-1}.$$

\subsection{Bayesian Inference}

Our reconstruction procedure essentially amounts to performing Bayesian
inference, in conjunction with an approximation scheme for the statistical
properties of the measurement process. More precisely, the
measurement process is approximated by a so-called linear-Gaussian
model. Within the confines of this model, the actual calculations for
performing the Bayesian inference turn out to be very simple and are
given by the update equations of a Kalman filter; this will be
explained below. In this section we briefly describe the elements of
Bayesian inference. For an in-depth treatment we refer to the
excellent introductory work \cite{sivia}.

We describe the state of the system under investigation by a vector
and denote it by $\bm{x}$. For the time being, we ignore the fact that
the system is a quantum system. As our knowledge of $\bm{x}$ is
obtained from (quantum) measurements and is statistical in nature, we
describe it by a random variable, $\bm{X}$. Since in our setup
measurements are independent (each measurement operates on a private
copy of the quantum state under investigation), the reconstruction
procedure can be decomposed as an iterative scheme in which each
measurement is incorporated sequentially. We assume that in each
iteration any prior knowledge about $\bm{X}$, as well as any knowledge
obtained through the outcomes of the first $m-1$ measurement settings
has been incorporated into the probability density function
$f_{\bm{X}}$. In Bayesian terminology the PDF of $\bm{X}$ is called
the prior PDF. The goal of the inference procedure is to come up with
a posterior PDF that incorporates the knowledge obtained by the
measurement outcomes in the $m$-th measurement setting. We use the
random variable $\bm{X}'$ to describe the updated knowledge; its PDF
$f_{\bm{X}'}$ is called the posterior PDF.

We denote the ``knowledge obtained through a measurement'' by a vector
$\bm{z}$ describing the measurement outcome. This vector $\bm{z}$ is a
sample of the corresponding random variable $\bm{Z}$. It can give us
additional information about the system via the statistical relation
between $\bm{Z}$ and $\bm{X}$, which we express as the conditional PDF
$f_{\bm{Z}|\bm{X}}$. This conditional PDF is the statistical
description of the measurement model linking $\bm{X}$ to $\bm{Z}$; it
will be specified further in section~\ref{sec:charlik}. The posterior
$f_{\bm{X}'}$ is then nothing but the conditional PDF
$f_{\bm{X}|\bm{Z}}$.

At the heart of any Bayesian inference procedure lies Bayes' rule,
which expresses the relation between $f_{\bm{X}|\bm{Z}}$ and
$f_{\bm{Z}|\bm{X}}$:
$$
f_{\bm{X}|\bm{Z}}(\bm{x}|\bm{z}) = \frac{f_{\bm{Z}|\bm{X}}(\bm{z}|\bm{x})
f_{\bm{X}}(\bm{x})}{f_{\bm{Z}}(\bm{z})}.
$$
While $f_{\bm{Z}}$ is defined as the marginal of
$f_{\bm{X},\bm{Z}}=f_{\bm{Z}|\bm{X}} f_{\bm{X}}$, it is much more
convenient to interpret $f_{\bm{Z}}$ as a normalisation constant,
ensuring that $f_{\bm{X}|\bm{Z}}$ integrates to 1 over the probability
space of $\bm{X}$. Second, as the main random variable in Bayes' rule
is $\bm{X}$, while $\bm{z}$ plays the role as parameter and is given
by the observation, we have to consider $f_{\bm{Z}|\bm{X}}$ as a
function of $\bm{x}$ too. As a function of its second argument,
$f_{\bm{Z}|\bm{X}}$ is no longer a  conditional probability density
but a \textit{likelihood function}. We will denote this function by
$L_{\bm{X}|\bm{Z}}$. Because of the explicit  normalisation in Bayes'
rule, $L$ is defined up to proportionality. Note the reversal of the
arguments, which is customary in the Bayesian literature and reflects the
change of focus from $\bm{Z}$, being the measurement outcome causally related to
the underlying state $\bm{X}$, to $\bm{X}$, our guess of what the state should be,
given the measurement outcome $\bm{Z}=\bm{z}$ as evidence:
\be
L_{\bm{X}|\bm{Z}}(\bm{x}|\bm{z}) \propto f_{\bm{Z}|\bm{X}}(\bm{z}|\bm{x}).
\ee
We will henceforth rewrite Bayes' rule as
$$
f_{\bm{X}|\bm{Z}} = C\, L_{\bm{X}|\bm{Z}} f_{\bm{X}}.
$$
The Bayesian inference step, incorporating $\bm{z}$ as new knowledge,
is then expressed as
\be
f_{\bm{X}'}(\bm{x}) = C\, L_{\bm{X}|\bm{Z}}(\bm{x}|\bm{z})
f_{\bm{X}}(\bm{x}). \label{eq:bayes}
\ee

For a sequence of measurement settings $(\bm{Z}_i)_{i=1}^n$, this step
has to be iterated $n$ times, leading through a sequence of $n+1$ PDFs
that describe the state $\bm{X}$ in a way that is consistent with the
additional knowledge obtained through the measurements. If we
describe the prior information by the PDF of $\bm{X}_0$, the $i$-th
measurement by the likelihood function with parameter $\bm{z}_i$, and
the updated information after $i$ iterations by the PDF of $\bm{X}_i$,
we get
\bea
f_{\bm{X}_n}(\bm{x}) &=& C\,
   L_{\bm{X}_{n-1}|\bm{Z}_n}(\bm{x}|\bm{z}_n) \,
   L_{\bm{X}_{n-2}|\bm{Z}_{n-1}}(\bm{x}|\bm{z}_{n-2}) \ldots \nonumber
   \\
&& L_{\bm{X}_0|\bm{Z}_1}(\bm{x}|\bm{z}_1)
f_{\bm{X}_0}(\bm{x}). \label{eq:bayesfull}
\eea
As one can see, this is merely a product of the $n$ likelihood
functions and the prior PDF, and can therefore be calculated in any
order. This will turn out to be important later on.

Despite the apparent simplicity of the Bayesian update formula, actual
calculations based on it are in general very complicated because the
variable $\bm{X}$ appears both as main variable of a continuous PDF
(the prior) and as continuous parameter of the (discrete) likelihood
functions. The resulting product is in general an extremely
complicated continuous function of $\bm{x}$. In the context of
reconstruction of tomographic data, for example, the likelihood
functions are polynomials of very high degree.

The calculations do become feasible in the specific case of so-called
\textit{Linear-Gaussian Models}. In such models the dynamics of the system can be described by
a time-discrete Markov chain with a linear evolution operator
perturbed by zero-mean Gaussian noise. Similarly, the measurement
also depends linearly on the system state and any perturbation must
also be zero-mean Gaussian noise.
In other words, all variables ($\bm{X}_0$,
and all $\bm{Z}_i$) are normally distributed, and the parameter
$\bm{x}$ enters only in the value of the mean of
$f_{\bm{Z}_i|\bm{X}_{i-1}}$; moreover, it does so in a linear way
only. A typical example is a classical measurement whose output
depends linearly on the state and is perturbed by additive white
Gaussian noise.

For these models, and for more complicated dynamical
models where the state $\bm{X}$ varies over time, the Bayesian update
formula reduces to a set of simple equations called the \textit{Kalman
filter equations}.
A Kalman filter is the optimal state estimator when the
system and the measurement can be modelled by linear-Gaussian
models. The Kalman filter equations consist
of two sets of equation; one set (the predictor equations) predicts
the evolution of the Markov chain, while the other set (the update
equations) updates the state of the system based on the measurements
taken. For the purposes of the present paper only the update equations
will be used because the basic assumption of tomography is that the
system is static.
A very good account of Kalman filtering is given in
Ref.~\cite{dsimon}.

The reason for the feasibility of the linear-Gaussian model is that
all distributions occurring in the calculations are normal, including
the intermediate products of the factors of (\ref{eq:bayesfull}). This
will be explained in the next section, where we describe the Kalman
update equations in detail.

\subsection{Kalman Filtering for Static Systems}
Let us return again to the Bayesian update formula (\ref{eq:bayes})
$$
f_{\bm{X}'}(\bm{x}) = C\, L_{\bm{X}|\bm{Z}}(\bm{x}|\bm{z})
f_{\bm{X}}(\bm{x}).
$$
A linear-Gaussian model can be represented pictorially as follows:
$$
\bm{X} \stackrel{\bm{H}}{\longrightarrow} \bm{Y}
\stackrel{+\cN(0,\bm{\Theta})}{\longrightarrow} \bm{Z}.
$$
For the purposes of this paper it turns out to be beneficial to split
the model into two parts: a linear mapping, represented by a matrix
$\bm{H}$, followed by the noise process, which consists simply of
adding zero-mean Gaussian white noise with given covariance matrix
$\bm{\Theta}$. A further model assumption is that the prior PDF
$f_{\bm{X}}$ is Gaussian, so that the posterior PDF will be Gaussian,
too.

Suppose now that an observation of $\bm{Z}$ is made, giving the value
$\bm{z}$. Bayesian inference of the noise process then yields that the
distribution of $\bm{Y}$ conditional on this observation will be
$\cN(\bm{z},\bm{\Theta})$. In other words, the moments of $\bm{Y}$ are
given by
\begin{subequations}
\label{eq:momentsY}
\bea
\mu(\bm{Y}) &=& \bm{z} \\
\Sigma(\bm{Y}) &=& \bm{\Theta}.
\eea
\end{subequations}
We are now left with performing Bayesian inference on the linear part,
with the variables distributed as
\bea
\bm{X}' &\sim& \cN(\bm{\mu}',\bm{\Sigma}') \\
\bm{X} &\sim& \cN(\bm{\mu},\bm{\Sigma}) \\
\bm{Y} &\sim& \cN\left(\mu(\bm{Y}),\Sigma(\bm{Y})\right).
\eea
Here, $\bm{\mu}$ and $\bm{\Sigma}$ are the already known first and
second order moments of $\bm{X}$ (mean and covariance matrix) and
$\bm{\mu}'$ and $\bm{\Sigma}'$ are the unknown first and second order
moments of $\bm{X}'$.

Taking into account the explicit formula (\ref{eq:pdfnormal}) for the PDF of a Gaussian distribution,
the logarithm of the likelihood function is given by
$$
-\frac{1}{2}[\bm{y}-\mu(\bm{Y})]^* \Sigma(\bm{Y})^{-1} [\bm{y}-\mu(\bm{Y})]
$$
plus some constant.
We get similar expressions for the logarithm of $f_{\bm{X'}}$ and the logarithm of $f_{\bm{X}}$.
Combining everything, dropping the factors $-1/2$, and using the relation $\bm{y}=\bm{H}\,\bm{x}$,
the Bayesian update formula (\ref{eq:bayes}) for the linear mapping becomes:
\bea
\lefteqn{(\bm{x}-\bm{\mu}')^* {\bm{\Sigma}'}^{-1} (\bm{x}-\bm{\mu}')}
\nonumber\\
&=& c + [\bm{H}\,\bm{x}-\mu(\bm{Y})]^* \Sigma(\bm{Y})^{-1}
[\bm{H}\,\bm{x}-\mu(\bm{Y})] \nonumber\\
&& + (\bm{x}-\bm{\mu})^*
\bm{\Sigma}^{-1}(\bm{x}-\bm{\mu}). \label{eq:KalmanSource}
\eea
Here, all additive constants have been absorbed in the constant $c$. This constant
is actually irrelevant because it is ultimately absorbed in the Bayesian
normalisation factor $C$ of (\ref{eq:bayes}).
Both sides of the equation are therefore degree-2 polynomials in $\bm{x}$,
which confirms our earlier statement that all distributions
occurring in the calculations for linear-Gaussian models are Gaussian.
Eliminating $\bm{x}$ from this equation gives us
the two update equations we need, one for the mean, and one for the
covariance.

\bigskip

\textbf{Remark.} Although the probability space of $\bm{X}$ is a real vector
space, the vector entries of $\bm{x}$ need not be real. This will give
us more freedom in choosing a basis when $\bm{X}$ is a Hilbert space
representation of density matrices; to get real vector entries one is
forced to choose Pauli matrices (or generalisations thereof) as basis
vectors. This is the reason why we have applied the Hermitian
conjugate $*$ rather than the transpose.

\bigskip

Combining the terms that are \textit{quadratic} in $\bm{x}$ yields the
equation
$$
{\bm{\Sigma}'}^{-1} = \bm{H}^* \Sigma(\bm{Y})^{-1}\bm{H}
+\bm{\Sigma}^{-1}.
$$
Inverting gives the solution for $\bm{\Sigma}'$ [using the matrix
inversion lemma, Eq.~(\ref{eq:milemma})]
\bea
\bm{\Sigma}' &=& [\bm{H}^* \Sigma(\bm{Y})^{-1}\bm{H}
+\bm{\Sigma}^{-1}]^{-1}
\label{eq:KalmanSigmaAlt}\\
&=& \bm{\Sigma} - \bm{\Sigma} \bm{H}^*[\bm{H}\bm{\Sigma} \bm{H}^*
+\Sigma(\bm{Y})]^{-1}\bm{H}\bm{\Sigma}. \label{eq:Pprime}
\eea
The factor
$\bm{\Sigma} \bm{H}^*(\bm{H}\bm{\Sigma}\bm{H}^*+\Sigma(\bm{Y}))^{-1}$
is customarily called the \textit{Kalman gain factor} and is denoted
by $\bm{K}$.

Combining the terms \textit{linear} in $\bm{x}$ yields the equation
$$
(\bm{\Sigma}')^{-1}\bm{\mu}' = \bm{H}^* \Sigma(\bm{Y})^{-1}\mu(\bm{Y})
+\bm{\Sigma}^{-1}\bm{\mu},
$$
with the solution for $\bm{\mu}'$:
\bea
\bm{\mu}' &=& \bm{\Sigma}'[\bm{H}^* \Sigma(\bm{Y})^{-1}\mu(\bm{Y})
+\bm{\Sigma}^{-1}\bm{\mu}] \nonumber \\
&=& \bm{\mu} + \bm{K}[\mu(\bm{Y})-\bm{H}\bm{\mu}]. \label{eq:muprime}
\eea
In the last line we have used equation (\ref{eq:Pprime}) and the easily
verified relation
$(\bm{\Sigma}-\bm{K}\bm{H}\bm{\Sigma})\bm{H}^*\Sigma(\bm{Y})^{-1}=\bm{K}$.

The three relevant equations are commonly known as the
\textit{Kalman update equations}, and they form the backbone for the
method of this paper. Combined with the equations (\ref{eq:momentsY})
for the moments of $\bm{Y}$ they read
\begin{subequations}
\label{eq:Kalman}
\bea
\bm{K} &=& \bm{\Sigma} \bm{H}^*[\bm{H}\bm{\Sigma} \bm{H}^*
+\bm{\Theta}]^{-1}
\label{eq:KalmanK} \\
\bm{\mu}' &=& \bm{\mu} + \bm{K}(\bm{z}-\bm{H}\bm{\mu})
\label{eq:KalmanMu}\\
\bm{\Sigma}' &=& \bm{\Sigma} -
\bm{K}\bm{H}\bm{\Sigma}. \label{eq:KalmanSigma}
\eea
\end{subequations}

It is instructive to consider the special case where the measurement
mean $\bm{y}$ is exactly the state $\bm{x}$, i.e.\ $\bm{H}=\id$.
This gives the simplified equations
\begin{subequations}
\label{eq:KalmanSimple}
\bea
\bm{K} &=& \bm{\Sigma} (\bm{\Sigma}+\bm{\Theta})^{-1} \\
\bm{\mu}' &=& \bm{\mu} + \bm{K}(\bm{z}-\bm{\mu})\\
\bm{\Sigma}' &=& \bm{\Sigma} - \bm{K}\bm{\Sigma}.
\eea
\end{subequations}
After some algebra one finds that $\bm{\Sigma}'$ is given by the
parallel sum of $\bm{\Sigma}$ and $\bm{\Theta}$:
$$
\bm{\Sigma}' = (\bm{\Sigma}^{-1}+\bm{\Theta}^{-1})^{-1}.
$$
In other words, in this special case, the inverses of the covariance
matrices add up.

\textbf{Remarks.}
\begin{enumerate}
\item
In the
expression for the Kalman gain we use a matrix inversion, and that
requires invertibility of the argument. In our setting (quantum
mechanics), it turns out that the argument is actually never
invertible. The reason for that is a certain set of exact constraints
that the state has to obey. For example, when the state is a quantum
state, its trace must be equal to 1. In addition, the measurement
vector also has to satisfy certain exact constraints. For example, in
an $N$-run experiment, the number of clicks must add up to $N$. This
eventually has an impact on $\bm{\Sigma}$, $\bm{H}$ and $\bm{\Theta}$,
causing $\bm{H}\bm{\Sigma}\bm{H}^*+\bm{\Theta}$ to be
non-invertible. How to deal with this will be described later on, in
Section \ref{sec:exact}.
\item
From an analytic viewpoint it is not readily clear that the expression
(\ref{eq:KalmanSigma}) for $\bm{\Sigma}'$ always yields a positive
semi-definite matrix. More
importantly, the expression is not the best one from a numerical
viewpoint; since a subtraction is made, numerical roundoff may produce
a non-positive semi-definite matrix. This is more likely to happen for
precise measurements, yielding small variances. For that reason, the
alternative formula (\ref{eq:KalmanSigmaAlt}) involving addition
rather than subtraction is preferable. In that form it is also obvious
that the obtained $\bm{\Sigma}'$ is always positive semi-definite.
\end{enumerate}
\section{Kalman Filter Reconstruction of Quantum Tomographic Data}
\label{sec:theory}

In this Section we present the basic theory of our reconstruction
method, based on the concepts of Bayesian inference and Kalman
filtering, which were described in the previous Section.
This Section contains the bulk of the material, and includes the mathematical
underpinnings of our method. To assist the readers who are more interested in
the method itself and how to apply it, we have included Sec.~\ref{sec:digest},
containing a self-contained explanation of the method. Those readers are advised to
read that Section only and then fast-forward to the applications and discussions Sections
\ref{sec:NathanState}, \ref{sec:loopy} and \ref{sec:discuss}.

We start in Sec.~\ref{sec:charlik} with a characterisation of
the likelihood function for quantum measurements (characterised
themselves by a POVM $\{\Pi^{(k)}\}$) and obtain a normal
approximation that allows to approximately fit a quantum measurement
in the mould of a linear-Gaussian model. We find that incorporating the information
obtained from the quantum measurements into the likelihood function
amounts to applying the Kalman filter update equations (\ref{eq:Kalman})
where the entries of the measurement mean $\bm{z}$ are given by
formula (\ref{eq:dirichletmean}), those of the measurement covariance
$\bm{\Theta}$ by (\ref{eq:dirichletcov}), and the measurement matrix
$\bm{H}$ by the matrix representing the linear mapping
$\rho\mapsto \bm{p} = (\trace[\rho\Pi^{(k)}])_{k=1}^K$.

Apart from calculating the likelihood, Bayesian inference requires a
choice of a prior, and this is partially treated in
Sec.~\ref{sec:prior}. In a full treatment one would have to
incorporate the structure of the physical set into the prior, i.e.
the prior should be zero outside of the set of physical states. This,
however, causes the prior to be non-Gaussian and it therefore does not
fit the requirements of Kalman filtering. For that reason, the
restriction to the physical set will be done in a post-processing
step, as described in Sections \ref{sec:phys} and \ref{sec:phys2}, and instead a Gaussian
dummy prior is chosen. Directly after the Kalman updates, a simple
correction step removes the effects of this dummy prior again
[Sec.~\ref{sec:prior}, Eqs.~(\ref{eq:corrSigma}) and
(\ref{eq:corrMu})]. At this point, the eigenvalues of the corrected
$\Sigma$, before restricting to the physical set, already give useful
information as they are variances of certain linear combinations of
state components and give an indication about how many more
measurements are needed to reduce the measurement error. For example,
to bound the maximal error on any state component from above, the
largest eigenvalue of the covariance matrix should be bounded.

Once the posterior PDF, restricted to the physical set is calculated,
via its first and second order moments, one can calculate the
confidence region for the reconstruction, i.e. the set within which
the actual state should lie with high probability (say, 95\%).
The basic quantity expressing this confidence region is the
Mahalanobis distance. This is explained in Sec.~\ref{sec:cr}.

Next, the problem of restricting the posterior
PDF to the physical set is treated. This is actually a very difficult
numerical problem, especially in high dimensions, and turns out to be
a challenge even for current state-of-the-art Bayesian integration
methods. In Sections \ref{sec:phys} and \ref{sec:phys2} we give two
simple algorithms that perform the task
reasonably well, if one is willing to give up exactness of the
solution.

For most quantum tomography problems, the physical set is partially
defined by exact constraints. For example, quantum states have trace
equal to 1. In Sec.~\ref{sec:exact} we show how such exact
constraints are best dealt with, in order to avoid numerical
problems. The Kalman update equations have to be slightly modified.

Finally, Sec.~\ref{sec:represent} deals with the issue of
graphically representing the calculated results in a meaningful way,
based on mean values and error bars.
The first moment of the posterior PDF roughly corresponds to the
maximum likelihood solution, and as frequently happens with this kind
of solutions, exhibits reconstruction artifacts,
which are not features of the actual state, but are not ruled out by the
measurements either. A number of methods have been developed to remove
these artifacts, all based on maximisation of entropy or related
functionals, and we discuss these in our context.

\subsection{Overview}\label{sec:digest}
In this Section we present a self-contained overview of our method for the purpose
of implementation. Mathematical details, as well as the underlying rationale are
explained in subsequent sections, which could be skipped on first reading.
For the sake of clarity we restrict ourselves to the setting of state reconstruction.

The first step in implementing the method is to gather all relevant information about
the tomography process and cast it in an appropriate mathematical form.
The following are needed:
\begin{itemize}
\item dimension $D$ of the state (or of its representation);
\item the various sets of POVM elements $\{\Pi^{(j)}\}_j$, one set per measurement setting;
\item the number of runs $N$ per measurement setting, if applicable; in continuous wave (CW)
experiments, there is no such $N$;
\item the measurement data: the frequencies $f_j$ (the ``number of clicks'') of each of the outcomes;
\item any exact linear constraints on the state; by default, $\trace\rho=1$ is the only such constraint;
\item any exact linear constraints on the measurement outcomes; for example, $\sum_j f_j=N$;
\item a statistical model of the measurement process in the form of a PDF of the frequencies $f_j$
in terms of the POVM elements, $N$, or any other parameter; typically, this PDF is multinomial or Poissonian;
\item a reference state $\rho_0$ satisfying the linear constraints on the state;
typically, this would be the maximally mixed state, $\rho_0=\id_D/D$.
\end{itemize}

The reconstruction algorithm consists of a number of phases, and we describe each in the following.
It is important to note that we represent states by vectors. The most convenient basis is the standard
basis, in which $\rho$ is represented by the $D^2$-dimensional vector $\vek(\rho)$.
\subsubsection{Setup phase}
The exact linear constraints are enforced through the use of two projectors $\bm{T}_X$ and $\bm{T}_Z$.
These have to be calculated first.
The projector $\bm{T}_X$ is a projector on a subspace $\cS_X$ in state space (that is, the vector space representation thereof),
while $\bm{T}_Z$ is a projector on a subspace $\cS_Z$ in measurement space
(the space of measurement vectors $\bm{z}$).
In principle, the latter could differ per measurement setting, but we will assume here
that it does not.
The interpretation of the subspace $\cS_X$ is that a state $\rho$ satisfies the linear state constraints
if and only if $\vek(\rho-\rho_0)$ is in the subspace $\cS_X$. For example, if the only linear constraint
is $\trace\rho=1$, then $\cS_X$ is the space of vector representations of Hermitian matrices of trace zero;
that is, the space of $D^2$-dimensional vectors orthogonal to $\bm{x}_0=\vek(\rho_0)= \vek(\id_D/D)=|\id_D\rangle/D$.
We will henceforth represent a state $\rho$ by the vector $\vek(\rho-\rho_0)$ in $\cS_X$.
The reference state $\rho_0$ is thus represented by the null vector.

The interpretation of $\cS_Z$ is similar. If, for example,
the only linear constraint on the measurements is $\sum_j f_j=N$, resulting
in the constraint $\sum_j z_j=1$, then $\cS_Z$ is the space of
real vectors (with dimension equal to the number of measurement outcomes) whose components sum up to zero.

The corresponding projectors can be calculated by constructing orthonormal bases (onb) supporting
each of these subspaces.
If $\{\bm{x}_i\}$ is an onb for $\cS_X$, we construct the matrix $X_1$ whose columns are the vectors
$\bm{x}_i$. This matrix is a so-called partial isometry. The projector $\bm{T}_X$ is then given as
$\bm{T}_X = X_1X_1^*$.
Similarly we have $\bm{T}_Z=Z_1Z_1^*$, where the columns of $Z_1$ form an onb for the subspace $\cS_Z$.
In the actual algorithm we don't need the projectors but only these partial isometries $X_1$ and $Z_1$.

Conversely, if the projectors are given,
we can calculate $X_1$ and $Z_1$ from them via the singular value decomposition (SVD).
Let $\bm{T}_X = USU^*$ be the singular value decomposition (SVD) of $\bm{T}_X$,
where $U$ is unitary and $S$ is diagonal, the diagonal elements of which are the
singular values of $\bm{T}_X$.
For a projector $\bm{T}_X$ of rank $k$, the first $k$ singular values are 1, while the others are zero.
Then $X_1$ consists of the first $k$ columns of $U$.
The partial isometry $Z_1$ is calculated in the same way from the SVD of $\bm{T}_Z$.

For the example where the linear state constraint is $\trace\rho=1$, the projector $\bm{T}_X$ is
easily constructed as
$\bm{T}_X = \id_{D^2}-|\id_D\rangle\langle\id_D|/D$. Using an SVD, this gives $X_1$.
Similarly, when the measurement constraint is $\sum_{j=1}^d z_j=1$, with $d$ the number of outcomes,
$\bm{T}_Z = \id_d-|1\rangle\langle 1|/d$, and $Z_1$ is also calculated using an SVD.

The exact constraints can be dealt with by expressing the relevant quantities in terms of
the bases $X_1$ and $Z_1$. A tilde will be used to indicate this.
For example, a state $\rho$ satisfying the constraint can be represented by the tilde vector
$\tilde{\bm{x}}=X_1^*(\bm{x}-\bm{x}_0) = X_1^*\vek(\rho-\rho_0)$.

\bigskip

\noindent\textbf{Initial mean value and covariance:}
Let $k$ be the rank of $\bm{T}_X$; thus $k$ is $D^2$ minus the number of independent constraints
on $\rho$.
The initial mean value is set to the $k$-dimensional null vector $\tilde{\bm{\mu}}_0=\bm{0}$,
representing the reference state $\rho_0$.
The initial covariance matrix is $\tilde{\bm{\Sigma}}_0 = b\id_k$, where $b$ is a scalar
with a ``large'' chosen value.
\subsubsection{Kalman Filter phase}
The following is applied iteratively, for each measurement setting:

The measurement matrix $\bm{H}$ represents the POVM elements of the current measurement setting.
The $j$-th column of $\bm{H}$ is given by $\vek(\Pi^{(j)})$.
From $\bm{H}$ we calculate $\tilde{\bm{H}} = Z_1^*\bm{H}\,X_1$, to incorporate the exact state
and measurement constraints.

The reference measurement vector is then $\bm{z}_0 = \bm{H}\,\bm{x}_0$.
The measurement vectors will be represented by the tilde vectors
$\tilde{\bm{z}}= Z_1^*(\bm{z}-\bm{z}_0)$.
The original measurement vector $\bm{z}$ is derived from the actual measurements $f_j$,
along with the measurement covariance matrix $\bm{\Theta}$, as follows.

For a measurement with multinomial statistics (as for $N$ runs of single-photon tomographic measurements)
$\bm{z}$ is given by $z_j = (f_j+1)/(N+d)$ (formula (\ref{eq:dirichletmean})),
and $\bm{\Theta}$ by the formula (\ref{eq:dirichletcov}).
The tilde quantities then follow using
$\tilde{\bm{z}} = Z_1^*(\bm{z}-\bm{z}_0)$ and $\tilde{\bm{\Theta}}=Z_1^*\bm{\Theta} Z_1$.

For measurements with Poissonian statistics, in the case that the POVM elements add up to either
the identity matrix or a scalar multiple thereof, the same formulas hold, but with $N$ given by
$N=\sum_j f_j$.
When the POVM elements do not add up to a scalar matrix (a situation that better be avoided),
the modified formulas (\ref{eq:nonpovm}) and (\ref{eq:nonpovm2}) have to be used for $\bm{z}$
and $\bm{\Theta}$.

We now apply the Kalman filter update equations (\ref{eq:KalmanModif2}),
written entirely in terms of tilde quantities:
\beas
\tilde{\bm{K}} &=& \tilde{\bm{\Sigma}}
\tilde{\bm{H}}^*(\tilde{\bm{H}}\tilde{\bm{\Sigma}} \tilde{\bm{H}}^*
+\tilde{\Theta})^{-1} \\
\tilde{\bm{\mu}}' &=& \tilde{\bm{\mu}} +
\tilde{\bm{K}}(\tilde{\bm{z}}-\tilde{\bm{H}}\tilde{\bm{\mu}}) \\
\tilde{\bm{\Sigma}}' &=& \tilde{\bm{\Sigma}} -
\tilde{\bm{K}}\tilde{\bm{H}}\tilde{\bm{\Sigma}}.
\eeas
For the first iteration, we set $\tilde{\bm{\mu}}=\tilde{\bm{\mu}}_0$ and $\tilde{\bm{\Sigma}}=\tilde{\bm{\Sigma}}_0$.
The primed quantities are the updated values and have to be used as $\tilde{\bm{\mu}}$
and $\tilde{\bm{\Sigma}}$ in the next iteration.

After the final iteration, the ``untilded'' quantities can be calculated, if one so wishes,
using the formulas $\bm{\mu}'=\bm{\mu}_0+X_1\tilde{\bm{\mu}}'$
and $\bm{\Sigma}'=X_1\tilde{\bm{\Sigma}}'X_1^*$.

\subsubsection{First Interpretation}
The Kalman filter reconstruction procedure yields a Gaussian distribution over the linear space
containing the physical state space as a subset. In most cases, this distribution will have a
non-negligible probability outside this physical set. In fact, more often than not the mean
of the distribution will be non-physical.
At this point, the reconstructed mean and covariance only summarise what the measurements
tell us, regardless of the physical significance of the outcome.
In the next phase, the physicality constraint has to be combined with the reconstruction,
as a kind of prior knowledge.

However, the unphysical covariance matrix already gives us interesting diagnostic information about the
tomography itself, namely about the inherent accuracy of the measurements. To that purpose one
can investigate the eigenvalues of the covariance matrix $\bm{\Sigma}$. If some (or all) of these
are large, that means the tomography is not able to give accurate information about the state
in the direction of the corresponding eigenvectors. Application 2 (Section \ref{sec:loopy})
gives a particularly nice example of this.
\subsubsection{Restriction phase}
Restricting the reconstruction to physical state space can be done in a number of ways, but is always
more complicated than in the MaxLik case, because the covariance matrix also has to be treated.
The easiest way is to first calculate the maximum likelihood physical state. This is the state $\rho_{ML}$
satisfying $\trace\rho_{ML}=1$ and $\rho_{ML}\ge0$ and minimising the squared Mahalanobis distance
$\cM^2=(\rho-\bm{\mu})^*\bm{\Sigma}^{-1}(\rho-\bm{\mu})$
to the reconstructed unphysical mean state.
In the presence of exact constraints, these calculations are best done using tilde quantities.
This is a constrained minimisation problem that can be reformulated as a semi-definite problem (SDP)
(see Section \ref{sec:phys}) and can be solved efficiently using SDP solvers.

The obtained minimal Mahalanobis distance $\cM^2_{ML}$ has diagnostic value.
If nothing has gone wrong, the MaxLik
solution should be well within the confidence region of the reconstructed likelihood distribution.
Taking a 95\% confidence value, the confidence region is given by the
inequality $\cM^2\le\gamma_\nu:=(\sqrt{\nu-1/2}+1.5)^2$, where $\nu$ is the dimension
of the subspace $\cS_X$ supporting the reconstructed state $\rho$ (the number of degrees-of-freedom).
If the value of $\cM^2_{ML}$ is much larger than that, this indicates that something has gone wrong
either with the tomographic measurements or with the reconstruction, in the sense that the underlying
assumptions are violated, for example if certain additional noise sources haven't been accounted for.
Application 1 illustrates this aspect very clearly.

If $\cM^2_{ML}$ falls within the confidence interval related to the
unphysical reconstruction, we can calculate the confidence region for the physical restriction.
A good approximation for that region is given by the intersection of the ellipsoid
$(\rho-\rho_{ML})^*\bm{\Sigma}^{-1}(\rho-\rho_{ML})\le (\sqrt{\gamma_\nu}+\cM_{ML})^2=:\gamma'_\nu$
with the physical set. A drawback of this approach is that to get error bars
the intersection has to be calculated explicitly, which is a non-trivial task.

In Sec.~\ref{sec:phys2} we present an alternative algorithm for performing the restriction
to the physical set. This algorithm does not rely on an SDP solver and, furthermore, yields
an explicit confidence region,  allowing to calculate error bars in a straightforward way.

Finally, it is possible to calculate a regularised solution. This is a physical state within
the physical confidence region that optimises a certain regularisation functional.
Possible choices are the entropy, and then we get the so-called maximum entropy (MaxEnt) solution, or a
functional expressing the smoothness of a solution. An example of the latter is given in Application 2.
The calculations for this again require solving a semi-definite program (see Sec.~\ref{sec:regul}).
\subsection{Approximation of the Measurement Process in
Quantum Tomography by a Linear-Gaussian Model}
\label{sec:charlik}

\subsubsection{The Likelihood Function in Quantum Tomography}

Any quantum measurement, be it in state tomography or process
tomography, can be characterised by the application of a POVM
$\{\Pi^{(k)}\}_{k=1}^K$ to a certain state $\rho$; this state could be
the state under investigation, or the output of a quantum process
given a certain applied input state. From the point of view of the
experimenter, the state $\rho$ is initially unknown, even though the
experimenter may have certain preconceptions about it. Because the
tomographic experiments reveal information about the state of a
statistical nature, the state has to be treated as a
random variable. Henceforth, $\rho$ will denote an observed quantum
state corresponding to a random variable denoted by $R$.

Quantum mechanics predicts the probabilities of each of the $K$ POVM
outcomes on a state $\rho$ to be $p_k=\trace[\rho\Pi^{(k)}]$. We
define the vector of probabilities as $\bm{p}=(p_k)_{k=1}^K$. When the
state is described by a random variable $R$, the vector
$\bm{p}$ is an observation of an underlying random variable
$\bm{P}$, with $P_k=\trace[R\Pi^{(k)}]$.

In reality, $P$ is never observed directly. We will consider two types
of optical tomography experiments in this paper. In \textit{pulsed mode}
tomography experiments, $N$ individual light pulses are sent into the
system, each pulse prepared in a state $\rho$. The POVM measurement is
repeated $N$ times, presumably on a sequence of $N$ independent
identically prepared states $\rho$. For every pulse, a detector either
clicks or does not click. The results of these $N$ runs can then be
combined into a vector of frequencies $\bm{f}=(f_k)_{k=1}^K$ of the
respective outcomes. This vector is an observation of a random
variable, $\bm{F}$. As is well-known, for fixed $N$ and $\bm{p}$,
$\bm{F}$ has a multinomial distribution with parameters $N$ and
$\bm{p}$: $\bm{F}\sim \text{Mtn}(N;\bm{p})$.

In \textit{continuous wave (CW) mode} optical experiments, the incoming
laser beam is turned on for a relatively long but fixed time, and the
number of times the detectors click in that time span are taken as the
frequencies $f_k$. The elements $F_k$ are Poisson distributed with
mean value $\mu_k=A p_k$, where $A$ is a proportionality factor
called the \textit{brightness factor}.
This incorporates the intensity of the laser beam, the duration of the
experiment, detector losses, etc. Obviously, the sum of frequencies
$N=\sum_k f_k$ is not fixed but is a Poissonian random variable as
well.

Combining all this with the relation $P_k=\trace[R\Pi^{(k)}]$ we
obtain the PDF $f_{\bm{F}|R}(\bm{f}|\rho)$ of $\bm{F}$ conditional on
$R$, or the likelihood function when considered as a function
of $\rho$. Pictorially, we have the following (for pulsed mode
experiments):
$$
R \stackrel{\Pi^{(k)}}{\longrightarrow} \bm{P}
\stackrel{\mbox{Mtn}(N,p)}{\longrightarrow} \bm{F}.
$$
The first step is a linear mapping, and the second step is the quantum
noise model. In comparison, recall that Kalman filtering is based on
the linear-Gaussian model:
$$
\bm{X} \stackrel{\bm{H}}{\longrightarrow} \bm{Y}
\stackrel{+\cN(0,\Sigma)}{\longrightarrow} \bm{Z}.
$$
The first step is again a linear mapping, but the second step is an
additive Gaussian white noise (AGWN) model.

As mentioned, the basic idea explored in this paper is to approximate
the quantum model by a linear-Gaussian model in order to open the door
for Kalman filtering. To do that, the following incompatibilities have to be
overcome: first, in linear-Gaussian models there are typically no
restrictions on the state vector $\bm{X}$, while in quantum mechanics
$R$ is confined to quantum state space (positive semi-definite and
trace equal to 1). We postpone the solution to this problem until
Section~\ref{sec:phys} and just pretend for the time being that the
variable $R$ is unrestricted.

The second difference is of course the different noise model. While in
both measurement models the first step is a linear mapping, the
quantum noise model is non-additive and non-Gaussian. In spite of this
apparently rather large difference, the simplicity of the Kalman
filter equations is so appealing that one is enticed to try and
approximate $f_{\bm{F}|R}$ by a linear-Gaussian model anyhow. Indeed,
many distributions, including the multinomial and Poisson
distributions, can be approximated by a normal distribution, and
according to the law of large numbers the approximation gets better
when the number of observations increases. Incidentally, this is why
we have imposed the requirement that the number of experimental runs
per measurement setting should not be too small.

So the main problem we are faced with is to reconcile the two models
in a statistically sound way, but without losing sight of the
practical issues. In the next subsection we first present a deceptively simple
``solution'', one that comes to mind almost automatically, but which
suffers from a number of serious drawbacks. An observation that is
more than 200 years old will then provide a way out of this conundrum,
paving the way to a more satisfactory solution.

\subsubsection{A na{\"\i}ve approach}

If we make the straightforward identifications $\bm{Z}=\bm{F}$,
$\bm{Y}=\bm{P}$ and $\bm{X}=\bm{R}$, then
$L_{\bm{R}|\bm{F}}(\rho|\bm{f})$ provides the likelihood function
$L_{\bm{X}|\bm{Z}}$ required for the Bayesian update formula
(\ref{eq:bayes}). There are two problems with this, however,
preventing a direct mapping to a linear-Gaussian model:
\begin{enumerate}
\item $\bm{P}$ enters in the moments of $\bm{F}$ of all order, and not
just the mean.
\item $\bm{P}$ enters in these moments in a highly non-linear way.
\end{enumerate}
A first na{\"\i}ve approach could be to simply approximate the
multinomial distribution by a multivariate normal with mean
$N p_k = N\trace[\rho\Pi^{(k)}]$ (which is linear in $\rho$, as
required) and with covariance matrix the one obtained by taking the
covariance matrix of the multinomial distribution and replacing every
occurrence of $p_k$ by $f_k/N$ (which is independent of $\rho$, as
required).

Although this superficially seems to solve the above problems, a
serious drawback of this approach is that the assignment of the
covariance matrix is very ad-hoc; for example, $p_k$ is replaced by
its estimator $f_k/N$ in the covariance but not in the mean. Even more
importantly, this approach is statistically ill-founded and, in fact,
underestimates the actual variance of $\bm{F}$.

This is most apparent when some of the components of $\bm{f}$ are
zero. Indeed, consider an $N$-trial 2-outcome measurement, where
$\bm{f}=(f_0,f_1)=(f_0,N-f_0)$, and suppose $f_0=0$. In the na{\"\i}ve
approach, the variance assigned to $f_{\bm{F}|R}$ would be $Np(1-p)$,
with $p$ replaced by $f_0/N$, hence giving zero. This is clearly a
mistake because a variance of zero amounts to perfect knowledge, and a
confidence interval of zero width. However, never having seen outcome
`0' is no guarantee that `0' will not occur in the future, no matter how
high the value of $N$ may be.
This has also been noted in Ref.~\cite{rehacek08}.

The first documented encounter of this phenomenon appeared in a 1774
paper by Laplace, as the so-called ``sunrise problem'': calculate the
probability of a sunrise, solely based on the information that it has
risen $N$ days before \cite{laplace}. The answer is not 1. Instead, the
correct value of this conditional probability is given by a formula
that  is now known as \textit{Laplace's rule of succession} (see,
e.g.\ Ref.~\cite{thatcher}). In a more wider context we can consider
the ``visible sunrise problem'' and calculate the probability $p$ that
we can see the sun rise (unhindered by clouds), given that we have
done so in $f$  of the $N$ days before. In the modern interpretation
of Laplace's rule, $p$ is a random variable with a uniform prior, and
conditional on the $N$ observations has a posterior PDF that according
to Bayes' rule is a beta-distribution with parameters $f$ and $N-f$,
whose mean is $(f+1)/(N+2)$. While useful for predicting sunrises,
beta distributions will also offer the solution to
our reconstruction problem.

\subsubsection{Bayesian Solution}

Essentially, Kalman filtering can be seen as Bayesian inference,
simplified to the case of linear-Gaussian models. When the noise is no
longer Gaussian, as in our case, but we still want to reap the
benefits from the simplicity of the Kalman filter equations, we really
should be looking at the Bayesian inference equations and suitably
approximate these, rather than approximate the model and apply Kalman
filtering to that. In this way we can avoid the problems
of the na{\"\i}ve approach.

More precisely, what we will do is match the two models after Bayesian inversion of
their noise processes. Recall, for the linear-Gaussian model this gave
$$
\bm{X} \stackrel{\bm{H}}{\longrightarrow} \bm{Y}, \quad
\bm{Y}\sim\cN\left(\mu(\bm{Y}),\Sigma(\bm{Y})\right),
$$
with the moments of $\bm{Y}$ determined by the observation $\bm{z}$,
Eqs.~(\ref{eq:momentsY}). For the quantum measurement model we have
$$
R \stackrel{\Pi^{(k)}}{\longrightarrow} \bm{P}
\stackrel{\mbox{Mtn}(N,p)}{\longrightarrow} \bm{F},
$$
Bayesian inversion yields the PDF of $\bm{P}$ conditional on the
observation $\bm{f}$ of $\bm{F}$. As explained in
Section~\ref{sec:dirichlet} this is a Dirichlet distribution with
parameter $\bm{f}$, $\bm{P}\sim\mbox{Dirichlet}(\bm{f})$, and moments
given by (\ref{eq:dirichletmean}) and (\ref{eq:dirichletcov}).

The solution to the matching problem has now become very simple. We
match the partially inverted quantum measurement model to the
partially inverted linear-Gaussian model, and to do so we approximate
the Dirichlet distribution by a Gaussian distribution with same first
and second order moments (moment matching). The upshot of all this is
is the following rule:\\
\textit{In the Kalman filter update equations (\ref{eq:Kalman}) replace
$\bm{z}$ by formula (\ref{eq:dirichletmean}), and $\bm{\Theta}$ by
(\ref{eq:dirichletcov}).}

\bigskip

\noindent\textbf{Remarks.}
\begin{enumerate}
\item
In our context, the formula for the mean of a Dirichlet random variate
(Laplace's rule of succession) could be paraphrased as ``each outcome
gets one click for free''. In statistics this extra count is called a
\textit{pseudocount} \cite{fienberg}. In comparison, the mode (the
position of the maximum of the PDF) is given by $p_i=f_i/N$.
\item
We would like to point out that in maximum likelihood reconstruction
one takes the mode as the basic quantity (the relative frequencies of
the outcomes), as that is the point of maximum likelihood, while in
our approach we use the confidence region, which is approximately
centered around the mean (the \textit{modified} relative frequencies,
with pseudocounts included). To counter potential objections against
this approach we already mention here that, for any non-unreasonably
small confidence value, the mode is well within this confidence
region. This will be shown in the appendix, Section~\ref{sec:modeCR}.
\item
In fact, even if one is not going to calculate the confidence region,
Laplace's rule tells us that one should really use the modified
relative frequencies, because ``the mean of a posterior can be thought
of as being more representative [than its mode] as it takes into
account the skewness of the PDF.'' (\cite{sivia}, p.~25).
\item
In this whole discussion we have quietly disregarded the fact that in
the quantum setting the probabilities $\bm{P}$ do not necessarily
range over the whole probability space. The range is essentially
determined by the relations $p_j = \trace[\rho \Pi^{(j)}]$. Thus, to
be completely correct, all Bayesian integrations should be carried out
over this range. However, in many cases, integrating over the exact
range complicates the calculations too much. On the other hand, in the
quantum tomography setting, performing exact integrations does not
guarantee that the final solution, in terms of the state, belongs to
the physical set anyway. Therefore, we integrate over the full
probability space (the $d$-dimensional probability simplex) and
restrict the solution to the exact physical set afterwards (see
Sections \ref{sec:phys} and \ref{sec:phys2}).
More about this is discussed in Section~\ref{sec:nopovm}.
\end{enumerate}


We have illustrated our approach here for the case of pulsed
experiments, where the distribution of clicks is governed by the
Multinomial distribution. More generally, the approach can be
described as follows. Let $f$ be the PDF of the distribution of the
outcome frequencies, conditional on the probabilities $\bm{p}$. From
this, derive the conjugate PDF, i.e. the PDF of $\bm{P}$ conditional
on the observed frequencies. Then approximate this conjugate PDF by a
Gaussian using moment matching. This amounts to replacing $\bm{z}$ and
$\bm{\Theta}$ in the Kalman update equations by the first and second
order moments (which are functions of the observed outcomes) of the
conjugate PDF, respectively.

\subsubsection{Poissonian counts}
Consider, as a second example, CW experiments, where the outcome
frequencies are governed by a Poisson distribution. More precisely,
the frequencies $f_j$ of the outcomes are independent Poisson variates
with parameters $\mu_j = Ap_j$, where $A$ is the brightness factor
of the experimental setup.

The conjugate distribution of the Poissonian
$f(k|p) = e^{-\mu}\mu^k/k!$ with $\mu=Ap$ is the PDF
\be
f(p|k)=\frac{A e^{-Ap}(Ap)^k}{k![1-Q(k+1,A)]}
\ee
where $Q(k+1,A)$ is the regularised incomplete Gamma function
\cite{abramowitz}.

We can apply this to find the PDF of $\bm{P}$ conditional on the
outcome frequencies $f_j$. While the latter are independently
Poissonian distributed, the $p_j$ have to add up to 1 and are
therefore correlated. Thus, the PDF of $\bm{P}$ equals the product
$\prod_j f(p_j|f_j)$ but renormalised to 1 over the probability
simplex of $p$. A short calculation yields the surprising result that,
again, the conjugate PDF is given by the Dirichlet distribution, with
$N=\sum_{j=1}^d f_j$. Maybe even more surprising is the fact that the
brightness factor $A$ cancels out completely. This is rather
convenient, since, in general, $A$ is not known, or at least not with
great precision. We can therefore carry over the formulas for the
pulsed mode case to the CW case wholesale, with the one addition that
$N$ has to be explicitly defined as $N=\sum_{j=1}^d f_j$.

\subsubsection{Non-POVM Measurements\label{sec:nopovm}}
As is well-known, the most general measurement one can perform is a
POVM measurement, described by POVM-elements, positive-semidefinite
operators that add up to the identity operator. In practical
experiments, however, one is not bound to implement the full POVM. For
example, one could just implement one element of the POVM at a time,
make $N$ measurements with it, and leave the other elements for
subsequent runs. Under the assumption that the measurements are always
made on identically prepared state, this makes no difference in the
end result (the vector of outcome frequencies). Because of this, one
can simulate measurements that cannot be performed in a single shot,
namely POVM measurements where the elements do not add up to identity,
as long as the elements $\Pi^{(j)}$ themselves obey the condition
$0\le \Pi^{(j)}\le\id$ (so that they form part of some POVM-proper).

However, for the purposes of tomographic reconstruction, in particular
for the kind we consider here, this potentially poses a problem in the
CW case. When the POVM elements add up to a multiple of identity, the
unknown brightness factor $A$ drops out of the calculations, just as in the
case of proper POVMs. When the elements do not add up to a multiple of
identity, this is no longer the case, and the calculations become more
complicated. The brightness factor $A$ is now a so-called nuisance parameter and has
to be removed from the likelihood function by integrating it out, as
shown below.

This situation has already been considered in \cite{mogilevtsev06,mogilevtsev07}
for Maximum Likelihood reconstruction. It was noted there that the sum of the POVM elements,
the matrix $\Pi^{(0)}:=\sum_j \Pi^{(j)}$, determines the field-of-view of the tomography experiment.
That is, if $\Pi^{(0)}$ is supported only on a restricted subspace, the reconstructed state will also
be supported on that subspace only. The tomography will be ``blind'' to state components outside
that subspace. Moreover, the eigenvalues of $\Pi^{(0)}$ determine the sensitivity of the tomography
along the corresponding eigenvectors. The larger the eigenvalue, the more accurate the tomography
in that direction will be.

When the POVM elements $\Pi^{(j)}$ no longer add up to identity, the
corresponding ``probabilities'' $p_j = \trace[\rho \Pi^{(j)}]$ do not
add up to 1 either. Let us then define
$p_0=\sum_{j=1}^d p_j$. Similarly, define $f_0 = \sum_{j=1}^d f_j$.
The maximum likelihood reconstruction method can be extended to cover this situation
by renormalising $p_j$: one just replaces $p_j$ by $p_j/p_0$ in the original MaxLik formulas.
This so-called \textit{extended maximum likelihood principle}
was first suggested by Fermi (see \cite{barlow}, p.\ 90).
For our reconstruction method, however, we need the mean and variance
of the likelihood function, and not just the mode,
and these depend on $\Pi^{(0)}$ in a more complicated way.

If the POVM elements add up to a multiple of identity, $\Pi^{(0)}=M\id$, then $p_0=M$, a constant.
Otherwise, $p_0$ is not a constant, but depends on the state $\rho$.
The PDF of the corresponding measurement outcomes (CW case) is the product
of Poissonians
\beas
f_{\bm{F}|\bm{P}}(\bm{f}|\bm{p})
&=& \prod_{j=1}^d \exp(-A p_j) \frac{(A p_j)^{f_j}}{f_j!} \\
&=& \exp(-A p_0) A^{f_0} \prod_{j=1}^d \frac{p_j^{f_j}}{f_j!}.
\eeas
The corresponding likelihood function is proportional to this, and can
be converted to a PDF by normalising over the set of allowed values of
$\bm{P}$. Now this is exactly the problem: how should the probability
space $\cP$ of $\bm{P}$ look like when the $p_j$ no longer add up to 1?
When $p_0$ is a constant, $p_0=M$, it is reasonable to
take the set of non-negative vectors adding up to $M$ as probability
space. Then the normalisation constant becomes
$$
\exp(-A M) A^{f_0} \int_{\cP}\md p_1\ldots \md p_d \prod_{j=1}^d
\frac{p_j^{f_j}}{f_j!},
$$
and all references to $A$ indeed cancel. The end result is then a
Dirichlet distribution in terms of the probability vector $p/M$. The
mean and covariance matrix of $\bm{P}$ are thus given by the formulas
for the Dirichlet moments, multiplied by $M$ and $M^2$, respectively.

\bigskip

\textit{The remainder of this subsection can be skipped on first reading,
and can be skipped altogether if one always makes
sure that the POVM elements used add up to a multiple of identity;
this design rule is recommended.}

\bigskip

When $p_0$ is not a constant, this magical cancellation of $A$ no
longer takes place. The standard way to deal with this in Bayesian
inference is to consider $A$ as a random variable, too, and calculate
the joint distribution of $A$ and $\bm{P}$:
$$
f_{A,\bm{P}|\bm{F}}(a,\bm{p}|\bm{f}) \propto \exp(-a p_0) a^{f_0}
\prod_{j=1}^d \frac{p_j^{f_j}}{f_j!}.
$$
Since we are not really interested in $A$ (it is a nuisance
parameter) we then take the marginal distribution of $\bm{P}$ by
integrating out $a$. Using the integral
$\int_0^\infty da \exp(-p_0 a) a^{f_0} = f_0!/p_0^{f_0+1}$, this gives
$$
f_{\bm{P}|\bm{F}}(\bm{p}|\bm{f}) \propto
\frac{f_0!}{p_0^{f_0+1}}\prod_{j=1}^d \frac{p_j^{f_j}}{f_j!}.
$$

A second problem that occurs when $p_0$ is not a constant is the geometrical shape
of the probability space $\cP$. In principle, this shape can be derived from
the relations $p_j=\trace[\rho \Pi^{(j)}]$, but this nearly always
yields a complicated set and integrating over it is extremely
difficult (as has been remarked before). For example, let $\rho$ be a
qubit state $\rho=\twomat{x}{z}{\overline{z}}{1-x}$, and take the POVM
elements
$$
\Pi^{(1)} = \twomat{1}{0}{0}{0},
\Pi^{(2)} = \twomat{0}{0}{0}{1},
$$
$$
\Pi^{(3)} = \frac{1}{2}\twomat{1}{1}{1}{1},
\Pi^{(4)} = \frac{1}{2}\twomat{1}{i}{-i}{1}.
$$
Then $\cP$ is defined by $p_1+p_2=1$ and
$(p_1-1/2)^2+(p_3-1/2)^2+(p_4-1/2)^2\le 1/4$ (essentially a Bloch
sphere). The reader is invited to try and integrate the function
$p_1^{f_1} p_2^{f_2} p_3^{f_3} p_4^{f_4}/p_0^{f_0+1}$ over this set.

As before, we propose to integrate over the smallest set containing
$\cP$ and giving easy integrals, and restrict to the physical set in a
later phase. The easiest way to do this is to first fix $p_0$ and
include all points in the simplex
$\cP(p_0):=\{\bm{p}: p_j\ge 0, \sum_{j=1}^d p_j=p_0\}$, perform all the
calculations conditional on this assumption, and then average over the
range of $p_0$. This range can be easily determined from the extremal
eigenvalues of $\Pi^{(0)}:=\sum_j \Pi^{(j)}$. Let $m$ and $M$ be the
smallest and largest eigenvalue of $\Pi^{(0)}$, respectively, then
$m\le p_0\le M$.

To average over $p_0$ we need a measure; to get
$\cP=\bigcup_{m\le p_0\le M}\cP(p_0)$, with all points in the set
equally weighted, this measure has to be proportional to the volume of
the simplex $\cP(p_0)$. As this volume is proportional to $p_0^{d-1}$,
the measure is $p_0^{d-1} \md p_0/K$, with
$$
K=\int_m^M p_0^{d-1} dp_0 = (M^d-m^d)/d.
$$

For fixed $p_0$, the calculations show that $\bm{P}/p_0$ follows a
Dirichlet distribution, with $N=\sum_j f_j$. Thus the moments of
$\bm{P}$ are:
\beas
\mu(P_i|P_0=p_0) &=& p_0\,\frac{f_j+1}{N+d} \\
\mu(P_i P_j|P_0=p_0) &=& p_0^2 \,\frac{(f_i+1)(f_j+1)}{(N+d)(N+d+1)} \\
\mu(P_i^2|P_0=p_0) &=& p_0^2 \,\frac{(f_i+1)(f_i+2)}{(N+d)(N+d+1)}.
\eeas

Now we average over $P_0$. The average of $P_0^k$ is
\beas
\mu(P_0) &=& \frac{\int_m^M \md p_0 \,p_0^{d-1}\,p_0^k}{(M^d-m^d)/d} \\
&=& \frac{d}{d+k}\,\frac{M^{d+k}-m^{d+k}}{M^d-m^d} =: \phi_k M^k,
\eeas
where we have defined
$$
\phi_k := \frac{d}{d+k}\,\frac{1-(m/M)^{d+k}}{1-(m/M)^d}.
$$
The range of $\phi_k$ is between $d/(d+k)$ and $1$, obtained when $m=0$
and $m=M$, respectively.

Hence we get
\begin{subequations}
\label{eq:nonpovm}
\bea
\mu(P_i)     &=& M\phi_1\,\frac{f_j+1}{N+d} \\
\mu(P_i P_j) &=& M^2\phi_2 \,\frac{(f_i+1)(f_j+1)}{(N+d)(N+d+1)} \\
\mu(P_i^2)   &=& M^2\phi_2 \,\frac{(f_i+1)(f_i+2)}{(N+d)(N+d+1)}.
\eea
\end{subequations}
This yields the covariance matrix elements via the relations
\begin{subequations}
\label{eq:nonpovm2}
\bea
\sigma^2_{i,j}(\bm{P}) &=& \mu(P_i P_j) - \mu(P_i)\mu(P_j) \\
\sigma^2_{i,i}(\bm{P}) &=& \mu(P_i^2)   - \mu(P_i)^2.
\eea
\end{subequations}
For the extreme case $m=0$, this gives
\beas
\sigma^2_{i,j}(\bm{P}) &=& M^2
\frac{da(f_i+1)(f_j+1)}{(d+1)^2(d+2)(N+d)^2(N+d+1)} \\
\sigma^2_{i,i}(\bm{P}) &=& M^2
\frac{d(f_i+1)(af_i+b)}{(d+1)^2(d+2)(N+d)^2(N+d+1)} \\
a &:=& N-d^2-d \\
b &:=& (d^2+2d+2)N+d^3+d^2.
\eeas
As a small check, for $m=M$ ($\Pi^{(0)}=M\id$), we get the mean and covariance
matrix  of the Dirichlet distribution, multiplied by $M$ and $M^2$,
respectively.

\subsection{Choosing a Prior}
\label{sec:prior}

In this Section, we tackle the problem of choosing an appropriate
prior distribution, as required for starting the Kalman filter
process. We also have to solve problem of restricting the solution of
the Kalman filter to physical space. Both problems could have been
solved in one go by choosing the prior to be a uniform distribution
over state space, and setting it equal to 0 outside of
it. Unfortunately, Kalman filtering requires a Gaussian prior, and we
leave the solution of the restriction problem to the next section. In
this section, therefore, we ignore the restriction to physical state
space.

When we have no prior information about the quantum state apart from
the tomography data, we have to construct a prior that reflects this
total lack of knowledge. Moreover, to allow for the application of
Kalman filtering, this prior has to be Gaussian. One such prior could
be a Gaussian with an infinite covariance (the mean would then be
irrelevant): $\bm{\Sigma}=\infty\id$. In numerical computations, this
infinity of course has to be replaced by a finite, but still big
number $b$, giving $\bm{\Sigma} = b\id$. On the other hand, to avoid
numerical instability, $b$ should be not too big.

But what does big enough and not too big mean? Fortunately, as we are
dealing with quantum state estimation, we know that the state belongs
to a bounded set: its eigenvalues are positive numbers summing up to
1. To illustrate this, let us restrict to diagonal $d$-dimensional
states, i.e. distributions. With the choice $\bm{\Sigma}=b\id$, the
squared 2-norm distance between two such distributions $p$ and
$q$ is given by $||p-q||_2^2/b$. (Why we take the 2-norm distance will
become clear in the next Section). The maximum value is therefore
$2/b$. To minimise the influence of the prior, this distance should be
small enough, and certainly much smaller than $d$. Hence, we need
$b\gg 2/d$. In our numerical experiments, we have chosen the value
$b=1$.

For the mean of the prior, the best choice is to take a state
``in the middle'' of state space. For distributions this would be the
uniform distribution $(1,\ldots,1)/d$, for quantum states the
maximally mixed state $\rho_0=\id/d$. More generally, one could take
the state that has the largest entropy within the physical set.


An alternative solution to the problem of choosing a prior
is based on the observation that the Bayesian
update equation (\ref{eq:bayes}) is basically a multiplication and
therefore all Bayesian updates commute. We can therefore start with
any suitable prior and divide it out again after all Kalman filter
updates have been performed. This amounts to the same thing as
starting off with the infinite width prior. This division is easy when
the covariance matrix of the chosen prior is a scalar matrix,
$\bm{\Sigma}_0=b\id$, with some finite choice of $b$.

Let $\bm{\mu}_0$ represent some fixed state and let us consider measurement
parameters of a very specific form $\bm{z}=\bm{H}\bm{\mu}_0$ and
$\bm{\Theta}=b\bm{H}\bm{H}^*$. In that case the Kalman update equations simplify to
\bea
\bm{\mu}'    &=& b(b+\bm{\Sigma})^{-1}\bm{\mu}
+ \bm{\Sigma}(b+\bm{\Sigma})^{-1}\bm{\mu}_0 \label{eq:K1} \\
\bm{\Sigma}' &=& b(b+\bm{\Sigma})^{-1}\bm{\Sigma}.\label{eq:K2}
\eea
Using this it is easy to calculate that starting off the Kalman filtering
sequence with the ``infinite width'' Gaussian prior ($\bm{\mu}=0$ and
$\bm{\Sigma}=\infty\id$) and applying the Kalman update step
($\bm{z}=\bm{H}\bm{\mu}_0$ and $\bm{\Theta}=b\bm{H}\bm{H}^*$) yields as
updated state a Gaussian with $\bm{\mu}'=\bm{\mu}_0$ and
$\bm{\Sigma}'=b\id$. Starting off with this Gaussian as prior
($\bm{\mu}=\bm{\mu}_0$ and $\bm{\Sigma}=b\id$) is therefore equivalent
to starting off with an infinitely wide prior and applying this particular Kalman
update step once, anywhere during the sequence (anywhere, because of
commutativity). In particular, this update can be done at the end of
the sequence.

Undoing the narrow prior can therefore be done after the final Kalman
update by applying the inverses of equations (\ref{eq:K1}) and
(\ref{eq:K2}). Denoting by $\bm{\mu}$ and $\bm{\Sigma}$ the quantities
obtained at the end of the Kalman filter sequence, and by
$\bm{\mu}_{\text{corr}}$ and $\bm{\Sigma}_{\text{corr}}$ the corrected
ones, with infinite prior, we have the \textit{correction equations}
\bea
\bm{\Sigma}_{\text{corr}} &=& (\bm{\Sigma}^{-1}-1/b)^{-1}
\label{eq:corrSigma} \\
\bm{\mu}_{\text{corr}} &=&
\bm{\mu}+(\bm{\Sigma}_{\text{corr}}/b)(\bm{\mu}-\bm{\mu}_0).
\label{eq:corrMu}
\eea
In practice, one could for example choose $\bm{\mu}_0$ to be a
representation of the maximal entropy (maximally mixed) quantum state
($\id/d$).

The problem with these correction equations is the extreme sensitivity
of $\bm{\mu}_{\text{corr}}$ to even the tiniest variations in
$\bm{\mu}$ when $\bm{\Sigma}_{\text{corr}}$ has very large
components. While this is not necessarily a numerical artifact---large
uncertainties on certain components of the covariance should go hand
in hand with equally large uncertainties on the mean---it may cause
numerical problems further down the line. For that reason we try to
avoid this situation by choosing a slightly larger value for $b$ in
the correction equations than was used in the construction of the
prior. To obtain a corrected covariance matrix
$\bm{\Sigma}_{\text{corr}}$ with an upper bound $\sigma^2_{\max}$ on
the  variances we can choose a value $b'$ satisfying
$1/b-1/b'=1/\sigma^2_{\max}$.

\subsection{Calculation of the Confidence Region}
\label{sec:cr}
The mean value and covariance matrix that we have been able to calculate using the Kalman filter
method are not ends in themselves.
One possible use of these is to calculate mean and variance of certain operators
when applied to the state.
In Ref.~\cite{rehacek08} it is shown that the
mean value and variance of an operator $Z$ depends on the mean state
$\mu$ and state covariance $\Sigma$ (the inverse of the Fischer
information matrix $F$) via the relations
\beas
\langle z\rangle &=& \trace \mu Z \\
\langle(\Delta z)^2\rangle &=& \bm{z}^*\Sigma\bm{z},
\eeas
where $\bm{z}$ is a vector representation of the operator $Z$.
The error bars on $Z$ can then be derived by setting appropriate condifence levels.

In this subsection, we derive more generally an expression for the confidence
regions for the complete state, corresponding to the reconstructed mean value and covariance.
The confidence region is the region around the mean value
obtained from the Kalman filter procedure within which the actual
state can be found ``with high probability''. The value of this
probability is called the  confidence value and is typically chosen to
be 95\%.
Stated otherwise, the probability that the actual state is outside the
confidence region should be ``low'', e.g. 5\%.

For the multivariate normal distribution, with mean $\bm{\mu}$ and
covariance matrix $\bm{\Sigma}$, the confidence region is an ellipsoid
centered around the mean. This is quite clear as the surfaces where
the PDF assumes a constant value are governed by the quadratic
equation $(\bm{x}-\bm{\mu})^*\bm{\Sigma}^\dagger(\bm{x}-\bm{\mu})=c$
(note the Moore-Penrose inverse, as required when there are exact linear
constraints; see Section \ref{sec:exact}).
The quantity of the left-hand side is the
\textit{squared Mahalanobis distance} $\cM^2$ between points $\bm{x}$ and
$\bm{\mu}$, as defined by (\ref{eq:Maha}).
The surface of the confidence region is thus the set of
points at a certain Mahalanobis distance from the mean. To find which
value $\cM$ should take for which confidence value, we have to
consider the  distribution of $\cM$.

It is well-known that the squared Mahalanobis distance has a
chi-square distribution with $\nu$ degrees of freedom (DoF):
$\cM^2\sim\chi_\nu^2$, where $\nu$ is the rank of $\Sigma$, equalling
the dimension of $\bm{x}$ minus the number of independent exact
constraints (zero variance components). A proof of this basic fact
goes as follows:

\bigskip

\noindent\textit{Proof.}
Suppose $\bm{\Sigma}$ has rank $\nu$ and is bounded (all eigenvalues
are finite), then its MP inverse has rank $\nu$ as well and can
therefore be written as $\bm{\Sigma}^\dagger=\bm{Q}^*\bm{Q}$, where
$\bm{Q}$ is a $\nu\times d$ matrix. Introduce the $\nu$-dimensional
random vector $\bm{U} = \bm{Q}(\bm{X}-\bm{\mu})$. Then the entries of
this vector are independent and distributed as $U_i\sim_{\text{ind}}
\cN(0,1)$. The sum-of-squares of these entries is then, by definition,
$\chi_\nu^2$ distributed \cite{jkb94}. Since the squared Mahalanobis
distance $\cM^2$ is equal to
\beas
\cM^2 &=& (\bm{x}-\bm{\mu})^*\bm{\Sigma}^\dagger(\bm{x}-\bm{\mu}) \\
&=& (\bm{x}-\bm{\mu})^*\bm{Q}^*\bm{Q}(\bm{x}-\bm{\mu})\\
&=& u^*u=\sum_i u_i^2,
\eeas
it follows that, indeed, $\cM^2\sim\chi_\nu^2$.
\qed

We summarise the main properties of the chi-squared distribution
\cite{jkb94}. The PDF as a function of $x$, with $x\ge0$, is given by
$$
\frac{1}{2^{\nu/2}\Gamma(\nu/2)}x^{\nu/2-1}\exp(-x/2),
$$
and the cumulative distribution function (CDF) by
$$
1-\frac{\Gamma(\nu/2,x/2)}{\Gamma(\nu/2)},
$$
where $\Gamma(a,y)$ is the incomplete gamma function
\footnote{In Matlab, the CDF can be calculated using the built-in
function {\tt gammainc(x/2,nu/2)}.}.

The mean of a variable $X^2\sim\chi_\nu^2$ is $\nu$ and its variance
is $2\nu$. The variable $X$ itself is distributed as $X\sim\chi_\nu$.
For not too small values of $\nu$, $X$ is approximately normal with
mean $\sqrt{\nu-1/2}$ and variance $1/2$. The 95\%
confidence interval of $X$ is therefore approximately given by $0\le
x\le \sqrt{\nu-1/2}+1.16309$, where $1.16309=\text{InvErf}(0.9)$, the
root of $[1+\erf(x)]/2=0.95$. Even for the smallest $\nu$ that we will
encounter, this approximation turns out to be very good; for $\nu=3$
(a single qubit state, for example) the value $x\le
\sqrt{3-1/2}+1.16309$ yields the only slightly smaller confidence
value of 94.3\%.

One immediately obtains that 95\%
of the probability mass of a multivariate normal is contained in the
ellipsoid consisting of points whose $\cM^2$ lies in the 95\%
$\chi_\nu^2$ confidence interval. In other words, the 95\%
confidence region is the ellipsoid
\be
\cM^2:=(\bm{x}-\bm{\mu})^*\bm{\Sigma}^\dagger(\bm{x}-\bm{\mu})
\le (\sqrt{\nu-1/2}+1.16309)^2=:\gamma_\nu. \label{eq:MahBound}
\ee
This formula lies at the heart of many statistical procedures, for
example outlier detection.

Now, since we are approximating the actual posterior PDF by a
Gaussian, the true confidence region will be different from the one
just obtained. However, we show in Appendix \ref{sec:waldproof} that
the difference will not be dramatic. Even in the worst case, the
actual distribution of $\cM^2$ is very close to chi-squared, but has
a variance that is larger by a factor of about 30\%
(unless $N$, the number of measurements per run, is pathologically
small). This means that the confidence region will be slightly larger
than the value given by (\ref{eq:MahBound}). A  conservative estimate
is to take
\be
\gamma_\nu = (\sqrt{\nu-1/2}+1.5)^2.
\ee
\subsection{Restricting to Physical State Space}
\label{sec:phys}

In this Section and the next, we treat the problem of restricting the solution of
the Kalman filter to the physical region $\cS$; when the state is a
quantum state, this means restricting solutions to state space, the
set of positive semidefinite matrices of trace 1. We will perform this
restriction as a post-processing step after the Kalman filtering
calculations.

At the level of the PDF, the restriction involves setting the values
of the obtained PDF equal to zero outside $\cS$, and then
renormalising the  PDF (as it should now integrate to 1 over $\cS$
instead of the whole space). The whole art is to determine the value
of the new normalisation factor. This requires the integration of the
posterior PDF over $\cS$, which is a very complicated problem,
especially for high dimensions; this problem is known as the Bayesian
integration problem. Likewise, similar integrals are necessary to
obtain the moments of the restricted PDF.

Various numerical methods have been proposed to approximate such
integrals; for an overview see \cite{EvansSwartz,mckaybook}. The
particular problem faced here is that the intersection of the
unconstrained confidence region with the physical set has extremely
low volume both within the confidence region and within the physical
set, in part due to the high dimensionality of the problem. This turns
out to be a very challenging situation for all existing integration methods.

In the following, we present two approximative methods. The first
method is very simple but rather crude and actually circumvents the
Bayesian integration problem. It is not a very powerful method,
because it only allows to check whether a state is in the restricted
confidence region. For some situations this might be already
enough. If one desires to know the shape of the restricted confidence
region, e.g.\ via its moments, then this method does not
suffice. Nevertheless, the method is extremely simple to apply.

A second method, discussed in the next Section, is more
powerful and yields an approximation of the restricted confidence
region in explicit form. Perhaps not surprisingly, it is based again on Kalman
filtering and yields an approximative confidence region expressed by
a mean value and a covariance matrix.
This method, while still in its experimental stages, seems
to work amazingly well in practice.

\bigskip

The simple method consists of keeping the first and second moments of
the unphysical posterior PDF but modifying the $\cM^2_{CR}$ value to
take the renormalisation over the physical set into
account. Furthermore, rather than calculate the exact value of the new
$\cM^2_{CR}$, a conservative upper bound is taken that is easy to
calculate.

The method consists of two parts, of which the first one is optional.
First, the maximum likelihood (MaxLik) solution is calculated. That
is, the physical state for which the unphysical posterior PDF is
maximal is calculated. This corresponds to the following semi-definite
program: find the minimal value of $t$ for which a state $\rho$ exists
satisfying the mixed semi-definite/quadratic constraints
\beas
\rho &\ge& 0 \\
\trace\rho&=&1 \\
(\bm{\rho}-\bm{\mu})^*\bm{\Sigma}^\dagger(\bm{\rho}-\bm{\mu}) &\le & t.
\eeas
For this minimal $t$, the state $\rho$ in question is the MaxLik
solution.

Even though very efficient semi-definite program solvers exist
\cite{sedumi}, this part can still be very time consuming when the
dimension of the state is high. Nevertheless, finding the MaxLik
solution is interesting enough in its own right to warrant inclusion
of this part. After all, this solution is what most reconstruction
algorithms try to find. In our context, the MaxLik solution also
allows to check the validity of the tomography data. Indeed, given
that the MaxLik solution corresponds to the best physical ``guess'' of
the actual state, the former should lie within the ``raw''
(i.e. unconstrained) confidence region allowed by the
measurements. Thus, the Mahalanobis distance between the MaxLik
solution $\rho_{ML}$ and the mean of the unconstrained posterior PDF
should be below $\cM_{CR}$. If not, this could be an indication that
something is wrong, either with the data, or with the underlying
assumptions (e.g. the noise model).

For the second step, we consider the Mahalanobis distance just
calculated,
\be
\cM^2_{ML} := (\rho_{ML}-\mu)^*\Sigma^\dagger(\rho_{ML}-\mu).
\label{eq:cmml}
\ee
The confidence region for the constrained PDF is the intersection of
the  physical set with the ellipsoid
$(\rho-\mu)^*\Sigma^\dagger(\rho-\mu)\le \cM_{CR,phys}^2$, where
$\cM_{CR,phys}^2$ is the confidence value for the constrained
posterior PDF. Note that $\mu$ and $\Sigma$ are still the moments of
the unconstrained PDF. Because the constrained posterior has to be
normalised over the physical set only, $\cM_{CR,phys}$ will be larger
than $\cM_{CR}$. Calculating the exact value of this $\cM_{CR,phys}$
is an extremely difficult problem, but we can prove the validity of a
very simple upper bound:
\be
\cM_{CR,phys} \le \cM_{CR,unphys} + \cM_{ML},
\label{eq:crbound}
\ee
with $\cM_{ML}$ given by Eq.~(\ref{eq:cmml}). The proof of this bound
is given in Appendix \ref{sec:boundproof}.

In case one does not even want to calculate the MaxLik solution, and
one is willing to believe this solution is in the unconstrained
confidence region, $\cM_{ML}$ can be replaced by its (presumed) upper
bound $\cM_{CR,unphys}$, giving the simple result that
\textit{the constrained confidence limit is at most twice the unconstrained one:}
\be
\cM_{CR,phys} \le 2\cM_{CR,unphys}.
\label{eq:crbound2}
\ee

In conclusion, the simple method consists of doing the following:
\begin{enumerate}
\item Depending on resources and taste, choose between steps 2 or 3,
then proceed to step 4.
\item Calculate the physical MaxLik solution, i.e.\ the solution in
$\cP$ that minimises the Mahalanobis distance to the (unphysical)
$\mu$. Record $\cM_{ML}$, the minimal Mahalanobis distance just
found.
\item Or, just take $\cM_{ML} = \cM_{CR,unphys}$.
\item A physical state $\hat{\rho}$ is in the physical confidence
region if its Mahalanobis distance to $\mu$ is not (much) above
$\cM_{CR,unphys} + \cM_{ML}$.
\end{enumerate}

\subsection{Restricting to Physical State Space; Kalman Filter Method}
\label{sec:phys2}

Let us now move on to our second method for restricting the posterior
PDF to physical state space. It is a more involved method but gives more
information.
To simplify the discussion, consider an example where $\bm{X}$ is a
$d$-dimensional real variable, and the physical region consists of the
positive orthant $x_i\ge0, i=1,\ldots,d$. We assume that, after
incorporating the measurement data, the unrestricted PDF of $\bm{X}$
is (approximately) given by its mean $\bm{\mu}$ and its covariance
matrix $\bm{\Sigma}$. We assume that the corresponding confidence
region is not completely contained in the physical region; otherwise,
there would be nothing to do here.

Consider now the marginal distributions of each of the components
$X_i$ of $\bm{X}$. The marginal distribution of $X_i$ is of course
normal, with mean $\mu_i$ and variance $\Sigma_{ii}$. We can then
easily calculate the confidence interval of each $X_i$ for given
confidence levels. There will at least be one such $X_i$ whose
confidence interval will not be completely contained in the physical
interval $x_i\ge0$. Broadly speaking, this is the one-dimensional
marginal version of our restriction problem.

The first key idea of our proposal is to consider the marginal
distribution of this $X_i$. If more than a fixed amount $\alpha$ of
probability mass of this marginal is outside the physical interval
$X_i\ge0$, we truncate the marginal to that interval. That means, we
set the density equal to 0 outside the physical interval, and
renormalise to 1. We then approximate this truncated normal
distribution by an ordinary normal distribution, with appropriate mean
$\tilde{\mu}$ and variance $\tilde{\sigma}^2$.
How we will do this is described in the next subsection.

\subsubsection{The marginal restriction problem}
\label{sec:1d}

The obvious idea of using moment matching to approximate the truncated
normal is of no use here because we need a procedure that gives a
stable result when applied twice or more. Approximating a truncated
normal with moment matching does not yield a distribution with
controlled tails (the tail being that part of the distribution outside
the physical interval). Truncating it again and approximating that
truncated normal with moment matching for a second time will typically
yield yet another set of values for mean and variance. There is no
guarantee at all that this process would converge. For that reason we
seek an approximation procedure that controls the tail probability
explicitly.

\begin{figure}
\includegraphics[width=8.6cm]{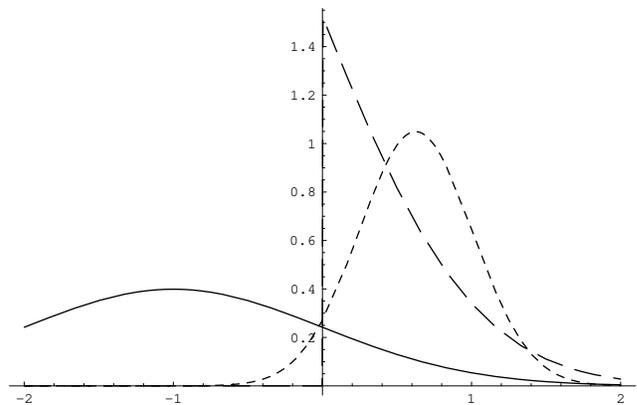}
\caption{Illustration of the approximation process of
Sec.~\ref{sec:1d}. Normal distribution with moments $\mu=-1$ and
$\sigma=1$ (full line). The corresponding truncated normal
distribution, truncated to $x\ge0$ (dashed line). The best
approximating normal distribution with 5\%
of probability mass outside the interval $x\ge0$ (dotted line).
\label{fig:approx}
}
\end{figure}

We will require that the approximating normal distribution has exactly
an amount $\alpha$ of probability mass outside the physical interval
$x\ge0$, where $\alpha$ is small, say 5\%. For an illustration of this
approximation process, see Fig.~\ref{fig:approx}.

As a quality measure of the approximation, we will use the
Kullback-Leibler distance again $D_{KL}(p||q)$ with the truncated
distribution as first argument (thus, with integration interval
restricted to $x\ge0$) and the approximating distribution as second
argument. Without the restriction on the approximating distribution
this would result in the moment matched approximation.

\begin{figure}[ht]
\includegraphics[width=8cm]{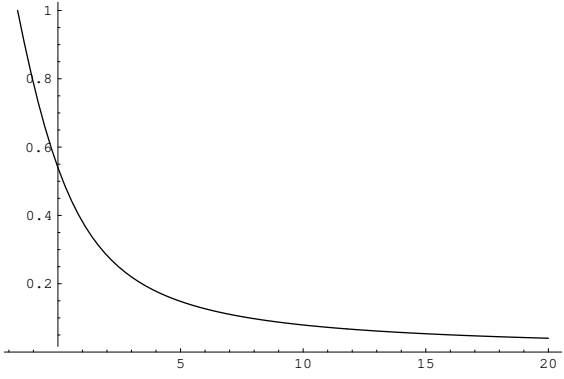}
\caption{
\label{fig:sigtilde}
The ratio $\tilde{\sigma}/\sigma$ plotted as a function of $t$
[Eq.~(\ref{eq:sigtilde})] for $\alpha_0=1.64485$ and
$t\ge-\alpha_0$. }
\end{figure}

Minimising this distance over all choices of approximating
distributions amounts to maximising the following function over the
parameters $\tilde{\mu}$ and $\tilde{\sigma}$:
$$
\int_0^\infty \md x \,\, \phi(x;\mu,\sigma) \,\log
\phi(x;\tilde{\mu},\tilde{\sigma}).
$$
The requirement we imposed on the probability mass in the left tail of
the approximating normal translates to the equality
\be
\tilde{\mu}=\alpha_0\,\,\tilde{\sigma}, \label{eq:mutilde}
\ee
where $\alpha_0=\sqrt{2}\,\,\text{InvErf}((\alpha+1)/2)$. For
$\alpha=5\%$ we find $\alpha_0\approx 1.64485$. After some algebraic
manipulations one finds the following complicated set of formulas for
the optimal $\tilde{\sigma}$:
\begin{subequations}
\label{eq:sigtilde}
\bea
\tilde{\sigma} &=& \sigma\left(-\alpha_0 g+\sqrt{\alpha_0^2
g^2-2c}\right)\\
c&=&t\tau-(1+t^2)/2\\
g&=&\tau-t/2 \\
\tau&=& \exp(-t^2/2)/[\sqrt{2\pi}\,\erfc(t/\sqrt{2})] \\
t&=& -\mu/\sigma.
\eea
\end{subequations}
In Fig.~\ref{fig:sigtilde}, we plot the ratio
$\tilde{\sigma}/\sigma$ as a function of $t$.

This one-dimensional solution has now to be translated back to the
original setting of restricting the state vector $\bm{X}$ to the
physical set. As this translation is basically an inverse problem
involving normal distributions only, we will again use a Kalman filter
to solve it, as explained in the next subsection.

\subsubsection{Backprojecting the marginal restriction using Kalman filtering}
\label{sec:back}

Our second key idea is to enforce the original unrestricted
distribution to have a marginal distribution (for component $X_i$)
given by the one-dimensional approximating normal, with parameters
$\tilde{\mu}$ and $\tilde{\sigma}$. Formally, this is equivalent to
performing a linear one-dimensional measurement, and we can
incorporate its effects on the state $\bm{X}$ by a Kalman filter
update step. The parameters of the measurement (measurement matrix
$\bm{H}$, mean $\bm{z}$ and covariance $\bm{\Theta}$) will have to be
such that the effect of the measurement is the required enforcement of
the marginal mentioned above.

As we only want to enforce the marginal of the $X_i$ component, we
will set $\bm{H}$ equal to the row vector $\langle i|$. Thus,
$\bm{H}\bm{x}=x_i$. This is a one-dimensional measurement, so that its
mean $z$ and covariance  $\Theta$ will be one-dimensional
too. Correspondingly, the Kalman gain is a column vector. Inserting
this into the Kalman filter update equations (\ref{eq:Kalman})
gives
\beas
\bm{K} &=& \bm{\Sigma} \bm{H}^*(\bm{H}\bm{\Sigma}\bm{H}^*+\Theta)^{-1}
= \bm{\Sigma}|i\rangle\,\, (\Sigma_{ii}+\Theta)^{-1} \\
\bm{\mu}' &=& \bm{\mu}+\bm{K}(z-\bm{H}\bm{\mu})
= \bm{\mu}+(z-\mu_i)\bm{K} \\
\bm{\Sigma}' &=& (\id-\bm{K}\bm{H})\bm{\Sigma}
= \bm{\Sigma} - \bm{K}\langle i|\bm{\Sigma}.
\eeas
The $X_i$ marginal of the updated distribution will then have moments
$\mu'_i$ and $\Sigma'_{ii}$ given by
\beas
\mu'_i &=& \mu_i+(z-\mu_i)\langle i|\bm{K}
= \mu_i + (z-\mu_i)\Sigma_{ii}(\Sigma_{ii}+\Theta)^{-1} \\
\Sigma'_{ii} &=& \Sigma_{ii} -\langle i|\bm{K}\,\Sigma_{ii}
= [1-\Sigma_{ii}(\Sigma_{ii}+\Theta)^{-1}]\Sigma_{ii}.
\eeas
We find the required values for the measurement parameters $z$ and
$\Theta$ by solving the equations $\mu'_i = \tilde{\mu}$ and
$\Sigma'_{ii} = \tilde{\sigma}^2$.

The solution is
\bea
\Theta &=& (1/\kappa-1)\Sigma_{ii} \label{eq:theta}\\
z &=& [\tilde{\mu}-(1-\kappa)\mu_i]/\kappa \label{eq:z} \\
\kappa &=& 1-\tilde{\sigma}^2/\Sigma_{ii}. \label{eq:kappa}
\eea
Here, $\tilde{\mu}$ and $\tilde{\sigma}$ are given by the equations
(\ref{eq:mutilde}) and (\ref{eq:sigtilde}), with $\mu=\mu_i$ and
$\sigma=\sqrt{\Sigma_{ii}}$, the marginal moments of $X_i$ in the
original distribution.

\subsubsection{The Restriction Procedure}

In the previous subsections, we have shown how a single variable $X_i$
can be restricted to its physical interval $X_i\ge0$. In general, if
the physical set is convex, the set is defined by a number of such
inequalities, possibly an infinite number. For simplicity, we first treat
the case that the physical interval is defined by a finite set of
inequalities $X_i\ge0,\forall i$, and treat the more general case
below. This case corresponds for example to diagonal quantum states,
and also to the optical POVM of Section~\ref{sec:loopy}.

To restrict the complete state vector $\bm{X}$ to the physical set, we
repeat the above-mentioned procedure for every component of $X$, or at
least for those components for which $\mu_i/\sigma_i<\alpha_0$. In
general, however, because of correlations, multiple components of
$\bm{X}$ will be affected by a single step of the procedure, and it
could very well be that the work of previous steps is partially undone
by the current step. For example, forcing $X_2$ to be positive could
bring $X_1$ back into the non-physical region.

Therefore, a number of runs of the algorithm will be necessary,
stopping when all marginals have their confidence intervals
approximately within the physical interval. As the quantity
$\min_i \mu_i/\sigma_i$ will converge to $\alpha_0$ from below, a good
stopping criterion is $\min_i \mu_i/\sigma_i>(1-\epsilon)\alpha_0$,
where $\epsilon$ is a small positive number. In practice, $\epsilon$
should not be chosen too small, so that the algorithm terminates in
reasonable time; in our applications  we chose $\epsilon=0.003$.

The order of the iterations, namely which marginal $X_i$ to treat
first, does not seem to influence the end result very much. In one set
of experiments we treated the marginals in fixed order, and in another
we always chose the marginal with smallest $\mu_i/\sigma_i$
first. It is not clear that the latter order should converge faster
because of the correlations between the $X_i$; in our experiments it
only did marginally so. While this and other convergence issues are
still under investigation, they appear not to be of major importance.

\begin{figure*}[ht]
\begin{tabular}{ccc}
\includegraphics[width=7.5cm]{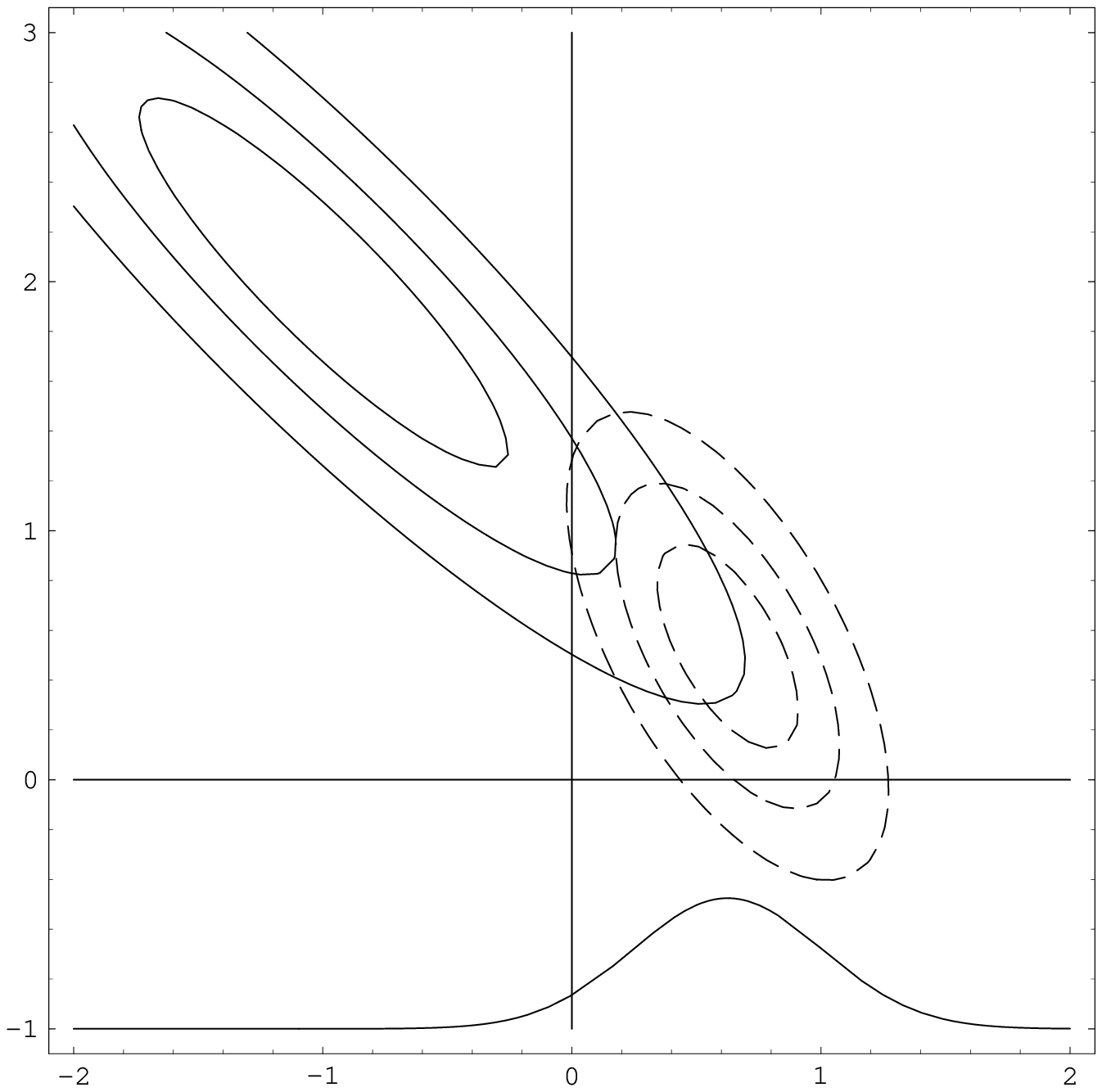}
&$\qquad$& \includegraphics[width=7.5cm]{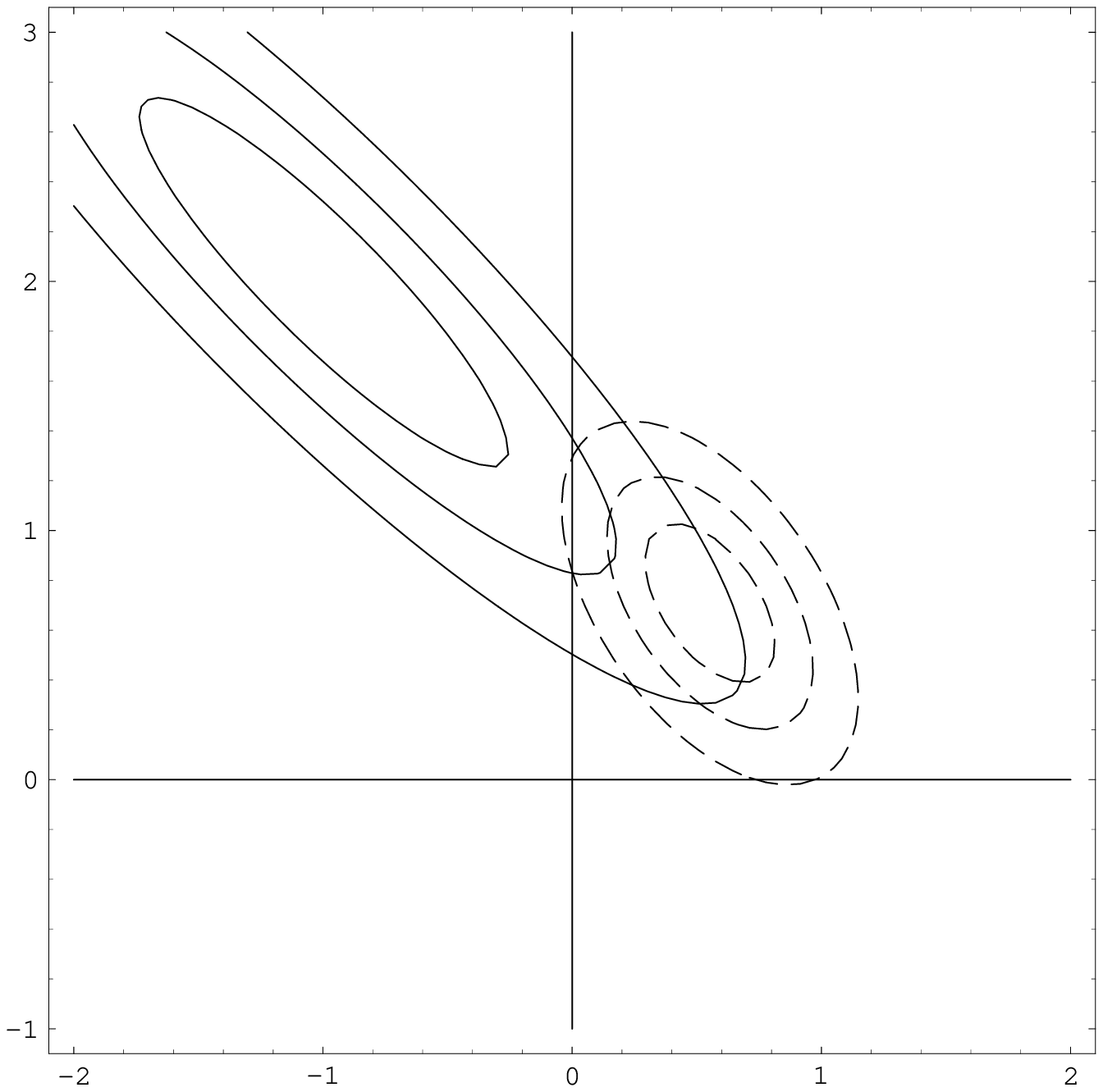}
\end{tabular}
\caption{
\label{fig:trunc2d}
Illustration of the Kalman filter update step of section
\ref{sec:back}. The physical set consists of the positive orthant. We
start with an unrestricted normal distribution with mean $(-1,2)$ and
covariance $\bm{\Sigma}=\protect\twomat{1}{-0.9}{-0.9}{1}$. The
contour lines of the  PDF are plotted in full lines in the upper left
of both graphs. The $X_1$ marginal of this distribution is the same
one as plotted in Fig.~\ref{fig:approx}. The normal distribution
approximating the truncated marginal is reproduced here as the full
line curve at the bottom of the left diagram. After applying one step
of the Kalman filter, with parameters given by
Eqs~(\ref{eq:theta}-\ref{eq:kappa}), we obtain a normal distribution
whose PDF is plotted in the left diagram in dashed lines. In the graph
on the right the result of a second Kalman filter update step is
shown, now for component $X_2$.}
\end{figure*}
An illustration on a small example is shown in Fig.~\ref{fig:trunc2d}.
The left graph shows the result of the restriction of $X_1$ to the
physical interval $X_1, X_2\ge0$. One sees that the new distribution again crosses
the border of the physical set, but now component $X_2$ is involved,
although it did not before the update. A second Kalman filter update
step will therefore be necessary, on $X_2$. The result of that second
step is shown in the right graph of Fig.~\ref{fig:trunc2d}.


In the case that the number of inequalities defining the physical set
is infinite, for example for the set of quantum states, where the
inequalities are $\trace\rho X\ge0, \quad\forall X\ge0$, the fixed
order rule is obviously infeasible. The smallest-first rule, on the
other hand, requires the complicated minimisation of
$\bm{X}^*\bm{\rho}/\sqrt{\bm{X}^* P\bm{X}}$ over all $X\ge0$. A third,
and much simpler possibility is to choose $X$ at random. However, that
method exhibits slow convergence, especially in the final stages when
the number of unsatisfied constraints becomes small.

A better option is to consider a combination of two rules in the
following double iteration: the inner iteration consists of, given a
unitary $U$, performing the restriction on the diagonal of
$U\rho U^*$, that is with  $H= \vek(U^*e^{ii}U)^*$, and varying
$i$. This inner iteration can be performed either using the fixed
order rule $i=1,\ldots,d$ or the smallest-first rule. The outer
iteration consists of choosing a new random unitary $U$ each time,
until a suitable stopping criterion is satisfied, e.g.\ until the inner
iterations achieve no further reduction of $\mu/\sigma$.
Although the smallest $\mu/\sigma$ component does not necessarily
occur for $U$ diagonalising $\rho$, it is a good idea to choose the
unitary $U$ diagonalising the current $\rho$ every now and then.

\subsection{Incorporating Exact Linear Constraints}
\label{sec:exact}

In many cases, the description of the physical state is subject to one
or more exact constraints. For quantum states the trace of the density
matrix is 1, for trace preserving quantum processes the partial trace
of the state representative over the output Hilbert space is the
maximally mixed state $\id/d$, and for POVMs the sum of all the
elements must be the identity matrix. Depending on the physical
system, there may be further constraints like this. These exact linear
constraints can be incorporated into the reconstruction process in a
number of ways.

A first approach is to incorporate exact constraints via dummy
measurements with zero measurement covariance, and replace inverses by
Moore-Penrose (MP) inverses. The benefit of this method is that
virtually no changes to the Kalman filter implementation  are needed.
A serious drawback is
that the state covariance matrix becomes ill-conditioned, since exact
constraints correspond to zero variance components. In reality,
numerical round-off causes these components to have non-zero variance,
which makes it hard to discriminate between variances that
are nominally zero and those that are not. This is a notorious problem
for actual implementations of Kalman filters and may cause serious
numerical instabilities. Later on in the calculations, the Mahalanobis
distance $(\bm{x}-\bm{\mu})^*\bm{\Sigma}^{-1}(\bm{x}-\bm{\mu})$ has to
be calculated (see Sec.~\ref{sec:cr}) and even the smallest
deviation in $\bm{x}-\bm{\mu}$ from the exact constraints is blown up
by the inverse of $\bm{\Sigma}$.


A second approach is to store exact constraints in two additional
matrices, along with state mean and state covariance. In general,
exact constraints may have an impact on the state but also on the
measurement. For example, when the state is a quantum state, we have
the exact constraint on the state $\trace\rho=1$, and a corresponding
exact constraint on the measurement probabilities $\sum_i p_i=1$. This
implies that the difference between any two states, e.g.\ the mean
$\bm{X}$ and its update $\bm{X}'$, should lie in a subspace, namely
the one for which the trace is zero. Similarly, the difference between
two measurements, e.g.\ the actual measurement outcome $\bm{z}$ and the
expected outcome $\bm{H\mu}$, should also lie in a subspace, namely the
one for which the sum of all components is zero.

Both subspaces can be represented in calculations by two projectors,
$\bm{T}_X$ and $\bm{T}_Z$. The projector $\bm{T}_X$ projects on the
subspace in state space, and $\bm{T}_Z$ on the subspace in measurement
space. The Kalman filter update equations can be made more resistant
to numerical inaccuracies using these projectors, ensuring that the
exact constraints are obeyed in any iteration of the update process,
as follows:
\begin{subequations}
\label{eq:KalmanModif}
\bea
\bm{y}&=&\bm{T}_Z(\bm{z}-\bm{H\mu}) \\
\bm{S}&=&\bm{T}_Z(\bm{H}\bm{\Sigma} \bm{H}^*+\bm{\Theta})\bm{T}_Z\\
\bm{K}&=&\bm{\Sigma} \bm{H}^* \bm{S}^\dagger\\
\bm{\mu}'&=&\bm{T}_X(\bm{\mu}-\bm{\mu}_0+ \bm{K} \bm{y})+ \bm{\mu}_0\\
\bm{\Sigma}'&=&\bm{T}_X(\bm{\Sigma}- \bm{K} \bm{H} \bm{\Sigma})\bm{T}_X
\eea
\end{subequations}
Here, $\bm{\mu}_0$ is a reference state, e.g.\ the maximally mixed
state $\id/d$.

Note that the ordinary inverse in the formula for the Kalman gain
$\bm{K}$ has been replaced by the MP inverse. Likewise, the inverse of
$\bm{\Sigma}$ appearing in the formula for the posterior PDF
corresponding to the Kalman filter solution has to be replaced by an
MP inverse too.


A third approach is to parameterise the state such that the exact
constraints are inherently satisfied. The obvious benefit is that the
exact constraints do not have to be explicitly imposed. A second benefit is
higher numerical stability, and straightforward invertibility of all
matrices that have to be inverted.

We start again from the projectors $\bm{T}_X$ and $\bm{T}_Z$. From
these projectors we can derive two partial isometries, $X_1$ and
$Z_1$, such that the following holds: the number of columns of $X_1$
and $Z_1$ must be equal to the ranks of $\bm{T}_X$ and $\bm{T}_Z$,
respectively; $X_1 X_1^* = \bm{T}_X$, $Z_1 Z_1^* = \bm{T}_Z$; and
$X_1^* X_1=\id$, $Z_1^* Z_1=\id$. Numerically, these partial
isometries can be calculated from a singular value decomposition (SVD) of
the projectors. For example, let $\bm{T}_X = USU^*$; the partial
isometry $X_1$ is then obtained from the unitary $U$ by dropping those
columns that correspond to the zero-valued singular values.

Roughly speaking, using these partial isometries,  the matrices
$\bm{\Theta}$, $\bm{\Sigma}$ and $\bm{H}$ can be ``cut down'' to their
invertible parts, which we will denote by a tilde. Define
\be
\tilde{\bm{\Theta}}:=Z_1^*\bm{\Theta} Z_1.
\ee
Since the support of $\bm{\Theta}$ is exactly the support of
$\bm{T}_Z$, we also have the reversed equality
$\bm{\Theta} = Z_1 \tilde{\bm{\Theta}} Z_1^*$. Furthermore,
$\tilde{\bm{\Theta}}$ is full rank and therefore invertible.
In a similar way we define
\be
\tilde{\bm{\Sigma}} := X_1^* \bm{\Sigma} X_1,
\ee
which is also full rank and invertible and satisfies
$\bm{\Sigma} = X_1 \tilde{\bm{\Sigma}} X_1^*$.

Furthermore,
$\bm{\mu}$ and $\bm{z}$ live in certain affine subspaces. If
$\bm{\mu}_0$ and $\bm{z}_0$ are fixed reference vectors in these
affine subspaces, we find that $\bm{\mu}-\bm{\mu}_0$ is a vector in
the support of $\bm{T}_X$, and $\bm{z}-\bm{z}_0$ a vector in the
support of $\bm{T}_Z$. Then we can define
\bea
\tilde{\bm{\mu}} &:=& X_1^* (\bm{\mu}-\bm{\mu}_0) \\
\tilde{\bm{z}}   &:=& Z_1^* (\bm{z}-\bm{z}_0),
\eea
and these again obey $\bm{\mu} = \bm{\mu}_0+X_1\tilde{\bm{\mu}}$ and
$\bm{z} = \bm{z}_0+Z_1\tilde{\bm{z}}$. In addition, it is possible,
and best, to choose $\bm{z}_0$ such that
$\bm{z}_0=\bm{H}\bm{x}_0$. Finally, we define
\be
\tilde{\bm{H}} := Z_1^* \bm{H} X_1.
\ee

Using these definitions (and a little work), the Kalman filter update
equations can be rewritten as follows:
\begin{subequations}
\label{eq:KalmanModif2}
\bea
\bm{K} &=& X_1 \tilde{\bm{K}} Z_1^* \\
\tilde{\bm{K}} &:=& \tilde{\bm{\Sigma}}
\tilde{\bm{H}}^*(\tilde{\bm{H}}\tilde{\bm{\Sigma}} \tilde{\bm{H}}^*
+\tilde{\Theta})^{-1} \label{eq:KalmanKModif2} \\
\bm{\mu}' &=& \bm{\mu}_0+X_1[\tilde{\bm{\mu}} +
\tilde{\bm{K}}(\tilde{\bm{z}}-\tilde{\bm{H}}\tilde{\bm{\mu}})]
\label{eq:KalmanMuModif2}\\
\bm{\Sigma}' &=& X_1 (\tilde{\bm{\Sigma}} -
\tilde{\bm{K}}\tilde{\bm{H}}\tilde{\bm{\Sigma}}) X_1^*
\label{eq:KalmanSigmaModif2} \\
&=& X_1 (\tilde{\bm{\Sigma}}^{-1}+\tilde{\bm{H}}\tilde{\bm{\Theta}}^{-1}
\tilde{\bm{H}}^*)^{-1}
X_1^*
\label{eq:KalmanSigmaAltModif2}
\eea
\end{subequations}
It has to be stressed again that all inversions here are ordinary ones, not
MP inverses. One sees that the equations reduce to the
original Kalman filter update equations provided one always works with
the ``tilde'' quantities. For the sake of reference, we combine all
definitions here again:
\begin{subequations}
\label{eq:tildedefs}
\bea
\tilde{\bm{\Theta}}&:=&Z_1^*\bm{\Theta} Z_1,    \quad
\bm{\Theta} = Z_1 \tilde{\bm{\Theta}} Z_1^* \\
\tilde{\bm{\Sigma}}&:=& X_1^* \bm{\Sigma} X_1,  \quad
\bm{\Sigma} = X_1 \tilde{\bm{\Sigma}} X_1^*\\
\tilde{\bm{\mu}} &:=& X_1^* (\bm{\mu}-\bm{\mu}_0) \\
\tilde{\bm{z}}   &:=& Z_1^* (\bm{z}-\bm{z}_0) \\
\tilde{\bm{H}}   &:=& Z_1^* \bm{H} X_1.
\eea
\end{subequations}

All required calculations can be expressed directly in terms
of tilde quantities. For the initial (prior) $\bm{\Sigma}_0$, we
choose $b\bm{T}_X$ (rather than $b\id$), which amounts to setting
$\tilde{\bm{\Sigma}}_0 = b\id$. Concerning the Mahalanobis distance,
if we define $\tilde{\bm{x}}:=X_1^* \bm{x}$, we have for any vector
$\bm{x}$ (as long as it is in the support of $\bm{T}_X$; if not, the
Mahalanobis distance will be infinite)
$$
(\bm{x}-\bm{\mu})^*\bm{\Sigma}^{\dagger}(\bm{x}-\bm{\mu})
= (\tilde{\bm{x}}-\tilde{\bm{\mu}})^*\tilde{\bm{\Sigma}}^{-1}
(\tilde{\bm{x}}-\tilde{\bm{\mu}}).
$$

As an example, we consider the case of $N$-run pulsed mode state
tomography. Then the constraints on the state are
$\trace\rho=1$. Denoting the dimension of the underlying Hilbert space
by $D$, this translates to $\bm{x}_0 = |\id_D\rangle/D$ and
$\bm{T}_X = \id_{D^2}-\frac{1}{D}|\id_D\rangle\langle\id_D|$.
The measurement vector $\bm{z}$ must in turn satisfy the constraint
$\sum_{i=1}^d z_i=1$, with $d$ the number of POVM elements of the
measurement POVM, i.e.\ $\langle 1|\bm{z}\rangle=1$. Hence
$\bm{T}_Z = \id_{d}-\frac{1}{d}|1\rangle\langle 1|$.

The corresponding partial isometries can be found numerically using an
SVD, as indicated, but for this particular case analytical formulas
can be found. Let $\cU$ be the $d$-dimensional discrete Fourier
transform-kernel
$$
\cU_{j,k} = \frac{1}{\sqrt{d}}\exp[2\pi i(j-1)(k-1)/d], \quad 1\le
j,k\le d.
$$
Let $\cU'$ be the matrix obtained from this by dropping the first
column (which has constant entries). This $\cU'$ is a good choice for
$Z_1$, as can be readily checked. Similarly, for $X_1$ we can choose
the matrix obtained from
$$
X_{j,k} =
\left\{
\begin{array}{ll}
\delta_{j,k},&\quad d+1\not|\,j-1 \\
\cU_{j',k'} ,&\quad j-1=(d+1)(j'-1),\\
             &\quad k-1=(d+1)(k'-1).
\end{array}
\right.
$$
by dropping the first column.

\subsection{Graphical Representations of the Reconstruction}
\label{sec:represent}

In the previous sections we have presented a methodology for state
reconstruction from tomographic data by which a Kalman filter is used
to obtain a normal approximation to the likelihood function
$f_{\bm{X}|\bm{F}}$, where $\bm{X}$ is the state and $\bm{F}$ is the
measurement data. The normal approximation is defined by its two
moments: the mean state vector $\bm{\mu}$, and the covariance matrix
$\bm{\Sigma}$. These two moments should in principle suffice as a
complete statistical description of the reconstructed state (within
the limits of the normal approximation).

When it comes to presenting the reconstruction, however, there are a
number of problems with the use of mean and covariance matrix alone.
Consider, for example, the reconstruction of an optical POVM using our
method, as discussed in Section \ref{sec:loopy} below. The
reconstruction of the diagonal elements of the first element is shown
in Fig.~\ref{fig:loopy1}. The reconstructed mean is plotted as the
centerline in the figure. On top of that, we would like a depiction of
the covariance matrix, because this matrix essentially describes the
reconstruction uncertainties.

\subsubsection{Depicting the covariance matrix}

The first problem one is faced with is that the covariance matrix
$\bm{\Sigma}$, being a matrix, cannot really be depicted in a very
meaningful way. Nevertheless, as the whole purpose of calculating it
is to provide some kind of error bars on the reconstruction, it is
desirable to have some means of representation. One can do this by
plotting its diagonal elements $\sqrt{\Sigma_{ii}}$ as error bars on
the mean value. This is meaningful because the diagonal element
$\Sigma_{ii}$ is exactly the variation of the marginal distribution of
$X_i$. Of course, such a plot has to be accompanied by the proviso
that the plot can only be indicative, because the variations on the
elements are in general correlated.

\subsubsection{Avoiding reconstruction artifacts}

The second, and more important problem we want to address in this
Section is the appearance of reconstruction artifacts in the
reconstructed mean. Closer inspection of Fig.~\ref{fig:loopy1}
reveals the presence of a wave-like pattern in the centerline, while
from theoretical considerations of the underlying POVM model there
really is no reason why that pattern should be there. Such artifacts
are typical for maximum-likelihood reconstruction methods and are
well-known in image restoration \cite{sivia}.
%
Even though the wave pattern in the POVM reconstruction stays well
within the error bars, which is already a clear counter-indication to
its statistical significance, it would be better to have a
reconstruction not showing such artifacts at all. Two methods for
obtaining artifact-free solutions (or at least for suppressing the
artifacts) are described below.

\subsubsection{MaxEnt reconstruction}

A widely used method for suppressing reconstruction artifacts is the
MaxEnt method, first proposed by Skilling in the context of image
reconstruction \cite{sivia}. Originally, the method was formulated as
choosing a special prior PDF based on the entropy $S(\bm{x})$ of the
states (provided such an entropy exists). In many cases a state can be
formally identified with a probability distribution, after suitable
normalisation. This is possible whenever the state consists of a set
of positive numbers. For digital images, the PDF is the list of
intensities of each pixel. For quantum states, it could be the list of
eigenvalues of the density matrix. In those cases one can assign a
meaningful entropy functional to the state space. For quantum states,
the von Neumann entropy is the obvious choice.

The MaxEnt method then
consists of choosing the function $\exp(\alpha S(\bm{x}))$ (properly
normalised), where $\alpha$ is a fixed parameter, as prior
PDF. Inference then proceeds in the normal way, by calculating the
posterior PDF and finding the maximum likelihood solution. The upshot
of this choice of prior is that in the absence of other information,
preference is given to states with higher entropy. The parameter
$\alpha$ characterises the amount of preference. Jaynes' principle of
maximum entropy \cite{jaynes} could be seen as a legitimisation of
this approach.


In the context of quantum tomography, Hradil and \v{R}eha\v{c}ek
\cite{rehacek} advocated a combination of the maximum entropy method
with the maximum likelihood (MaxLik) reconstruction method, which they
called \textit{MaxEnt assisted MaxLik (MEML) tomography}.
This method can be seen as a special case of Skilling's
MaxEnt method. In their
paper, they considered the situation of incomplete measurements. This
corresponds to a likelihood function whose covariance matrix has a
certain number of eigenvalues that are (almost) zero, while the others
are infinitely large. The MaxLik reconstruction is thus known with
certainty to lie in a certain subspace, but its position within that
subspace is completely unknown. In other words, there exists not a
single state maximising the likelihood function, but a whole plateau
of states. The proposal of \cite{rehacek} consists of finding the
point on that plateau (i.e.\ in the MaxLik subspace) for which the
entropy $S(\rho) = -\trace\rho\log\rho$ is maximised and take that
point as the reconstruction.

From an experimental viewpoint, the
situation considered in Ref.~\cite{rehacek}, of variances that are
either zero or infinite, is an idealised one. In practical
experiments, the number of measurements is finite, so that even the
most precisely known state components have a non-negligible
variance. Second, there may be practical and/or technical limitations
on the kind of measurements that can be performed, so that some
variances may be very large, but still finite. In Sec.~\ref{sec:loopy}
we will see a clear example of this. In that section the
reconstruction of an optical POVM is described. While the elements of
this POVM are diagonal in the Fock basis, its tomography is based on
coherent states rather than Fock states, because the latter are
extremely hard to produce. This causes large variances on the
reconstructed elements without a clear-cut distinction between
perfectly known and completely unknown components. When dealing with
such realistic experiments, the full-blown MaxEnt method is much more
preferable.

In its original formulation as a choice of prior, the
MaxEnt method has a number of shortcomings. One is that there appears
to be no satisfactory and rigorous way of choosing the parameter
$\alpha$. Secondly, the principle of maximum entropy does not
necessarily apply to the entropy of the states. In quantum tomography
we are dealing with a controlled system; the system is being prepared
in a predefined quantum state, to the best of the preparer's
abilities, and the tomography acts on a sequence of independent
identically prepared systems. In thermodynamical terms, this
corresponds to a system that could be as far from equilibrium as the
preparer wants it to be. This has to be contrasted with Jaynes' MaxEnt
principle, which has been inspired by the statistical mechanics of
systems in near-equilibrium, and which is based on the argument that
the probability of a macro-state should be proportional to the number
of microstates consistent with it, i.e.\ is proportional to its
thermodynamic entropy. For systems close to equilibrium, we agree that
it makes sense to choose a prior distribution that assigns more weight
to states with higher entropy. For controlled systems, and for those
systems lacking a fundamental notion of entropy, we are more tempted
to opt for a uniform distribution, as we have done in this work, and
incorporate the maximum entropy idea as a regularisation, as explained
below.

\subsubsection{Regularisation}\label{sec:regul}

Rather than apply the MaxEnt principle, which we deem not always
appropriate, one can adopt a more pragmatic approach in which the
entropy functional is no longer fundamental and can be replaced by
other functionals. And rather than replace the prior PDF with the
chosen functional, which implicitly changes the final posterior PDF,
and choose the maximum likelihood solution for that changed posterior,
the regularisation method does the following (Ref.~\cite{sivia},
Sec.~6.2): the prior PDF is unchanged, and within the confidence
region of the resulting posterior PDF (unchanged as well)
$$
(\bm{x}-\bm{\mu})^*\bm{\Sigma}^\dagger(\bm{x}-\bm{\mu}) \le \cM_{CR}^2,
$$
it finds the solution that maximises the chosen functional. When
expressing the functional as a cost, or penalty function, this would
be a minimisation.

Since the entropy is a concave functional, maximising it over a convex
set (such as the confidence region) is a convex problem and can be
efficiently solved numerically. Likewise, minimising a cost function
again gives a convex problem provided the cost
function is convex. Proper distance measures, for example, would
therefore be good cost functions.

Which cost function to use really depends on the problem setting. In
the example of the optical POVM mentioned above, theoretical
considerations suggested \cite{alvaro} that the \textit{smoothness} of the POVM
elements, defined as
\be
Q(\{\Pi^{(k)}\}_k) = \sum_{k=1}^K \sum_{i=1}^{d-1}(\Pi^{(k)}_{i+1} -
\Pi^{(k)}_{i})^2
\ee
could be appropriate. In fact, this smoothness is a
commonly used regularisation functional in image reconstruction
methods \cite{sivia}. It is immediately clear that this $Q$ is a
convex functional, as required. The appropriateness of this cost
function comes from the fact that it penalises the `wavyness' of the
centerline, as exemplified in Fig.~\ref{fig:loopy1}.

When the cost function is quadratic, like this smoothness term, the
minimisation problem is a quadratically constrained quadratic
programming (QCQP) problem. Such problems can be efficiently solved
using semi-definite programming (SDP) solvers \cite{boyd}. For the sake
of definiteness, let us consider the case where the states are quantum
states ($\rho\ge0$ and $\trace\rho=1$). The general form of a
quadratic cost function can then be written in terms of a matrix
$\bm{A}$ and a vector $\bm{b}$ as
$(\bm{A}\bm{\rho}-\bm{b})^*(\bm{A}\bm{\rho}-\bm{b})$. The SDP form of
the QCQP problem is then: minimise the (slack) variable $t$ over all
$t$ and $\rho$ under the combined quadratic and semi-definite constraints
\beas
\rho &\ge& 0 \\
\trace\rho&=&1 \\
(\bm{A}\bm{\rho}-\bm{b})^*(\bm{A}\bm{\rho}-\bm{b})&\le& t \\
(\bm{\rho}-\bm{\mu})^*\bm{\Sigma}^\dagger(\bm{\rho}-\bm{\mu}) &\le &
\cM_{CR}^2.
\eeas
This problem can be solved in a straightforward way by SDP solvers
like Sedumi \cite{sedumi}.


\section{Application 1: State reconstruction of an entangled 2-qubit state}
\label{sec:NathanState}

The methods introduced in this paper have all been tested on real sets
of tomographic data. In this Section and the next we report on two
such applications, one in state tomography and one in POVM tomography.

In the present Section, we consider the reconstruction of tomography
data of a source of polarisation-entangled photon pairs, obtained by
Langford \textit{et al} \cite{NathanPhD} and compare our results to their
reconstruction. The source is a BBO-crystal down-conversion source
operating in CW mode, pumped by an Argon laser. Two sets of tomography
data were taken, one directly on the crystal, and one on the single
mode fibres (SMF) attached to the crystal. In both cases, the sequence
of measurements is as given in Tab.~\ref{tab:NathanStateMeas}. This
measurement basis is over-complete because not all measurements are
needed to obtain a full state reconstruction. Nevertheless, it was
argued that by taking an over-complete basis a more accurate
reconstruction could be obtained.

\begin{table}[t]
\begin{tabular}{rl|rl|rl}
 1&$HH$ &13&$DH$&25&$RH$\\
 2&$HV$ &14&$DV$&26&$RV$\\
 3&$VH$ &15&$AH$&27&$LH$\\
 4&$VV$ &16&$AV$&28&$LV$\\
 5&$HD$ &17&$DD$&29&$RD$\\
 6&$HA$ &18&$DA$&30&$RA$\\
 7&$VD$ &19&$AD$&31&$LD$\\
 8&$VA$ &20&$AA$&32&$LA$\\
 9&$HR$ &21&$DR$&33&$RR$\\
10&$HL$ &22&$DL$&34&$RL$\\
11&$VR$ &23&$AR$&35&$LR$\\
12&$VL$ &24&$AL$&36&$LL$
\end{tabular}
\caption{
Sequence of measurements in the state tomography of the BBO
source of Application 1 (Section \ref{sec:NathanState}).
The labels $H$, $V$, $D$, $A$, $R$, $L$ refer to the
polarisation basis states: $H=(1,0)$, $V=(0,1)$, $D=(1,1)/\sqrt{2}$,
$A=(1,-1)/\sqrt{2}$, $R=(1,i)/\sqrt{2}$ and $L=(1,-i)/\sqrt{2}$.
\label{tab:NathanStateMeas}
}
\end{table}

A nice consequence of this choice for our reconstruction method is
that the projectors of the 36 basis states add up to a multiple of the
identity, $\sum_{k=1}^{36} |\psi^{(k)}\rangle\langle\psi^{(k)}|=9\id$.
As has been discussed in Sec.~\ref{sec:nopovm}, this allows us to
consider these projectors as if they were POVM elements of one big
over-complete POVM with normalisation factor $M=9$. We can thus take
all click frequencies and put them in one 36-dimensional vector
$\bm{f}$. Similarly, we have a 36-dimensional vector of probabilities
$\bm{p}$ such that $p/M$ is a genuine (normalised) probability
vector. As the measurements are obtained in CW mode, the frequencies
are Poissonian and after Bayesian inversion
($\bm{f}\longrightarrow\bm{p}$) we find that $\bm{p}/M$ is Dirichlet
distributed with parameters $\bm{f}$ and $N=\sum_{k=1}^{36} f_k$.

The upshot of all this is that the Kalman update equations have to be
executed exactly once, with $\bm{z}$ given by
$z=M\,\mu(\mbox{Dirichlet}(\bm{f}))$ and $\bm{\Theta}$ by $M^2$ times
the covariance matrix of Dirichlet$(\bm{f})$. This is particularly
convenient, because the issue of setting an initial prior and removing
it again after the Kalman updates (see Section \ref{sec:prior}) can be
resolved analytically, which allows us to choose an infinitely wide
initial prior $b=\infty$ without getting into numerical trouble. With
such a prior, the Kalman update yields the following posterior, as can
be checked with a modest amount of work:
\bea
\tilde{\bm{\mu}}' &=& (\tilde{\bm{H}}^* \tilde{\bm{\Theta}}^{-1}
\tilde{\bm{H}})^{-1}\,
\tilde{\bm{H}}^*\tilde{\bm{\Theta}}^{-1}\,\tilde{\bm{z}} \\
\tilde{\bm{\Sigma}}' &=& (\tilde{\bm{H}}^* \tilde{\bm{\Theta}}^{-1}
\tilde{\bm{H}})^{-1}.
\eea
Note that these formulas are stated in terms of the ``tilde
quantities'' [see Sec.~\ref{sec:exact}, Eq.~(\ref{eq:tildedefs})].
Both the state and the frequencies satisfy exact constraints,
$\trace\rho=1$, and $\sum_{k=1}^{36} f_k=N$, and we have chosen to
deal with these constraints in the numerically most stable way, by
``cutting off'' the kernels (zero eigenvalues) of the respective
operators. In the derivation of the above formulas care has to be
taken because the product $\tilde{\bm{H}}\tilde{\bm{H}}^*$ is not full
rank.

\begin{figure}[t]
\includegraphics[width=8cm]{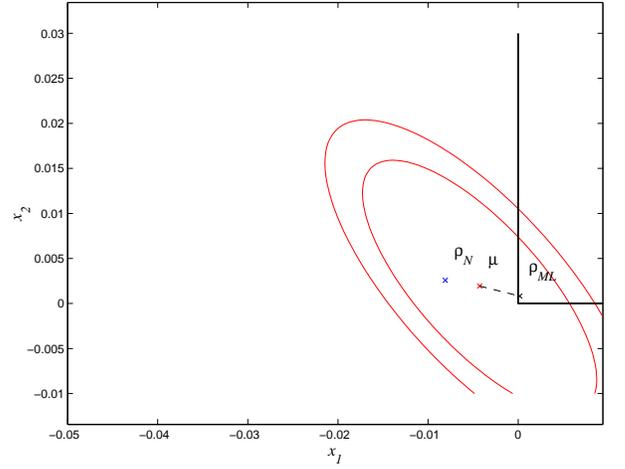}
\caption{
A display of the reconstruction results for Application 1, Section \ref{sec:NathanState},
showing a slice through state space illustrating the position of the
two-photon state, reconstructed from the data obtained by measuring directly at the BBO
crystal. What is shown is the projection of the state on the 2D subspace spanned by two
well-chosen pure state projectors. The two concentric ellipses
centered about the reconstruction mean $\bm{\mu}$ are the projections
of the 50\% (inner ellipse) and 95\% confidence regions (outer ellipse),
respectively; these ellipses are quite close together
due to the rather high dimension of the system ($d=15$).
The intersection of the physical set with the subspace is the triangle $x_1\ge0$,
$x_2\ge0$, $x_1+x_2\le 1$, of which the lower left corner is
shown. The projection of the MaxLik solution $\rho_{ML}$ is also
shown. This solution is well within the confidence region, as should be.
For comparison purposes we have also plotted the projection of the ``na{\"\i}ve'' least-squares reconstruction $\rho_N$.
\label{fig:NathanStateGood}
}
\end{figure}

\begin{figure}[t]
\includegraphics[width=8cm]{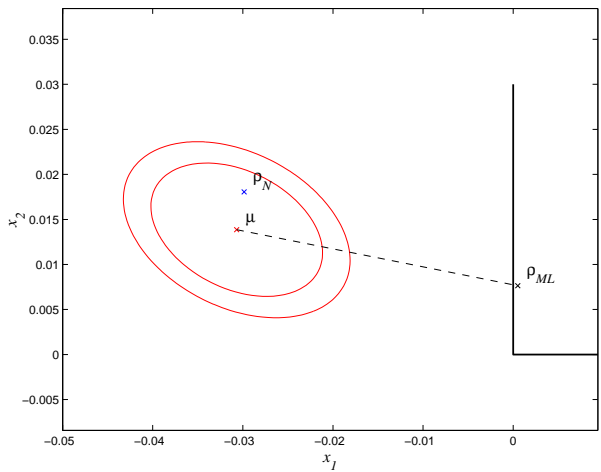}
\caption{
As Fig.~\ref{fig:NathanStateGood}, but with the state measured
at the SMF outputs. Because the duration of the tomography run was
twice as long as in the case of Fig.~\ref{fig:NathanStateGood}, the
confidence region is much smaller. The main difference, however, is
that the confidence region lies deep into non-physical space, meaning
that the MaxLik solution is far outside the confidence region.
This is not a deficiency of the KF reconstruction method, nor of its implementation,
but is actually a feature. It is a diagnostic feature that indicates that something is
wrong with the assumptions about the underlying noise model. A likely possibility is that
the measurements are subject to additional fluctuations. 
According to \cite{NathanPhD} the most likely source is
temperature-dependency of the spatial alignment of the SMFs, which caused the
measurements to drift.
To get a proper reconstruction this drift should be taken into account in
the noise model as an additional term.
\label{fig:NathanStateBad}
}
\end{figure}

We show the results of the tomographic reconstructions of the measurements at
the crystal and at the SMF in Figs.~\ref{fig:NathanStateGood} and
\ref{fig:NathanStateBad}, respectively. Obviously, we cannot show the
confidence regions in full 16-dimensional space, and we have chosen a
2D subspace spanned by two pure state projectors. We take the two
eigenvectors $\psi$ and $\phi$ of the reconstructed mean state
$\bm{\mu}$ that correspond to the 2 smallest eigenvalues (one of them
being negative). The parameters $x_1$ and $x_2$ are then given by the
mappings  $\rho\mapsto x_1=\langle\psi|\rho|\psi\rangle$ and
$\rho\mapsto x_2=\langle\phi|\rho|\phi\rangle$.

For both cases we calculate the least-squares solution and the MaxLik solution.
The least-squares solution $\rho_N$ is the state
$\rho_N = \sum_j c_j |\psi^{(j)}\rangle\langle\psi^{(j)}|/\sum_j c_j$,
where the coefficients $c_j$ are the least-squares solutions of the system
$f_i =\langle \psi^{(i)}|\,\,\sum_j c_j |\psi^{(j)}\rangle\langle\psi^{(j)}|\,\,|\psi^{(i)}\rangle
= \sum_j c_j |\langle \psi^{(i)} | \psi^{(j)}|^2$.
The MaxLik solution is the physical
state $\rho_{ML}$ for which the Mahalanobis distance from the
reconstructed mean state is minimal. We have implemented this in
Sedumi, as indicated in Sec.~\ref{sec:phys}.

In \cite{NathanPhD}, the MaxLik solution was calculated in a different
way, through the minimisation of a penalty function
$$
\Pi(\rho) := \sum_k \frac{[f_k-Ap_k(\rho)]^2}{[Ap_k(\rho)]^2}.
$$
Here $A$ is the unknown brightness factor of the experiment. This
MaxLik solution closely matches the MaxLik solution obtained through
our KF method. To obtain a quantification of the accuracy of the
MaxLik solution, Langford used a Monte Carlo calculation to estimate
the mean value of $\Pi(\rho_{ML})$ when $f_k$ is considered as a
Poissonian random variable with mean $Ap_k(\rho_{ML})$. From this mean
value, a fit quality parameter $Q$ is obtained by dividing the mean
value by the total number of measurements and taking the square
root. Ideally, the mean value of $Q$ should be 1.

Compared to the full error bars of the KF method, the $Q$ quantity
conveys little information about the statistical errors and it is not
clear what the acceptable values of $Q$ should be. Moreover, the Monte
Carlo calculation needed to find $Q$ is several orders of magnitude
slower than the KF algorithm. Langford reports MC running times of
about 150 seconds for 200 MC iterations. In contrast, our KF algorithm
runs in 0.12 sec (about 1000 times faster), while at the same time
offering much more error information, with a clear statistical
interpretation.

\section{Application 2: Reconstruction of an Optical POVM}
\label{sec:loopy}

Following a proposal of Ref.~\cite{banaszek}, in Ref.~\cite{achilles}
an experimental realisation was reported of an optical detector with
photon-number resolving capabilities. The basic idea is to carve up
an optical pulse into 8 portions and detect the presence of photons in
each of these portions. More precisely, this setup simulates a cascade
of beam splitters and eight avalanche photo-detectors (APDs), with the
probability of a photon arriving at a certain APD being roughly 1 in
8. The number of detectors clicking therefore gives an indication of
the photon numbers in the pulse. The detector is implemented using two
Franson interferometers, an additional balanced beam splitter, two
avalanche photo-detectors, and two identical circuits for performing
time binning.

The behaviour of this composite detector can be described by a
9-element POVM, where each of the outcomes corresponds to the number
of APD's clicking (from 0 to 8). We denote the POVM elements by
$\{\Pi^{(k)}\}_{k=0}^8$, where the elements $\Pi^{(k)}$ are positive
semi-definite and add up to the identity matrix. In principle, the
elements are infinite dimensional (corresponding to photon numbers
being unbounded), but we will truncate them at a certain dimension $d$
(in our calculations we have chosen values of $d$ of up to 170). Since
this detector has no phase reference, it is insensitive to phase,
which means that the POVM elements have to be diagonal in the Fock
basis.

To obtain a precise characterisation of the POVM elements, a
tomography experiment has been performed \cite{alvaro} by which a
large number of pulses consisting of coherent states $|\alpha\rangle$
of ever increasing power ($\propto\!|\alpha|^2$) were sent to
the composite detector and the resulting numbers of detectors clicking
were recorded. The parameter $\alpha$ was sweeped from 0.4 to 11, in steps
of about 0.01, and for each value of $\alpha$, $N=38084$ measurements
were taken. Per value of $\alpha$ the measurement record consisted of
the number of pulses $f_k$ that caused $k$ detectors to click,
for $k=0,\ldots,8$; obviously, $\sum_{k=0}^8 f_k=N$.

Using these data, a reconstruction of the POVM elements (without error
bars) was obtained and presented in \cite{alvaro}. Here we take the
same data and perform a reconstruction based on the
KF method, yielding a maximally likely solution with in addition a
definite confidence region. To avoid any confusion, we stress that the
object under scrutiny is a POVM and the measurement is made using
prepared quantum states. In other words: the POVMs are states and the
state is a POVM.

We have calculated the (unphysical) mean value $\mu$ and covariance
matrix $\Sigma$ using Kalman filtering, including $T$ projectors for
including the exact constraints that the POVM elements must add up to
the identity matrix. Then we applied the KF
method for restricting to the physical set, giving physical mean value
and covariance matrix. Finally, we calculated the maximally smooth
solution within the physical confidence region.

\begin{figure*}[ht]
\begin{tabular}{ccc}
\includegraphics[width=6.8cm,angle=90]{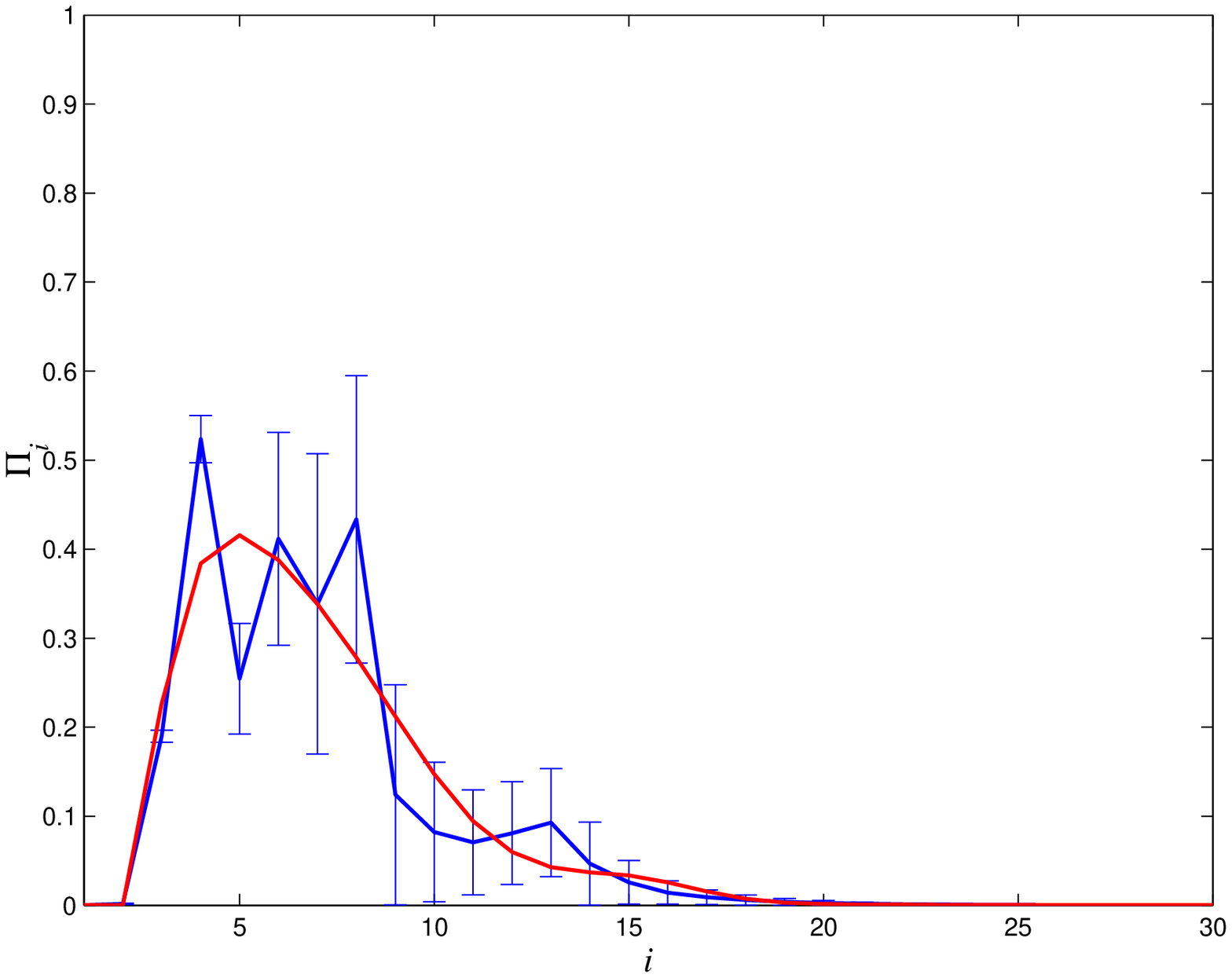} &
\includegraphics[width=6.8cm,angle=90]{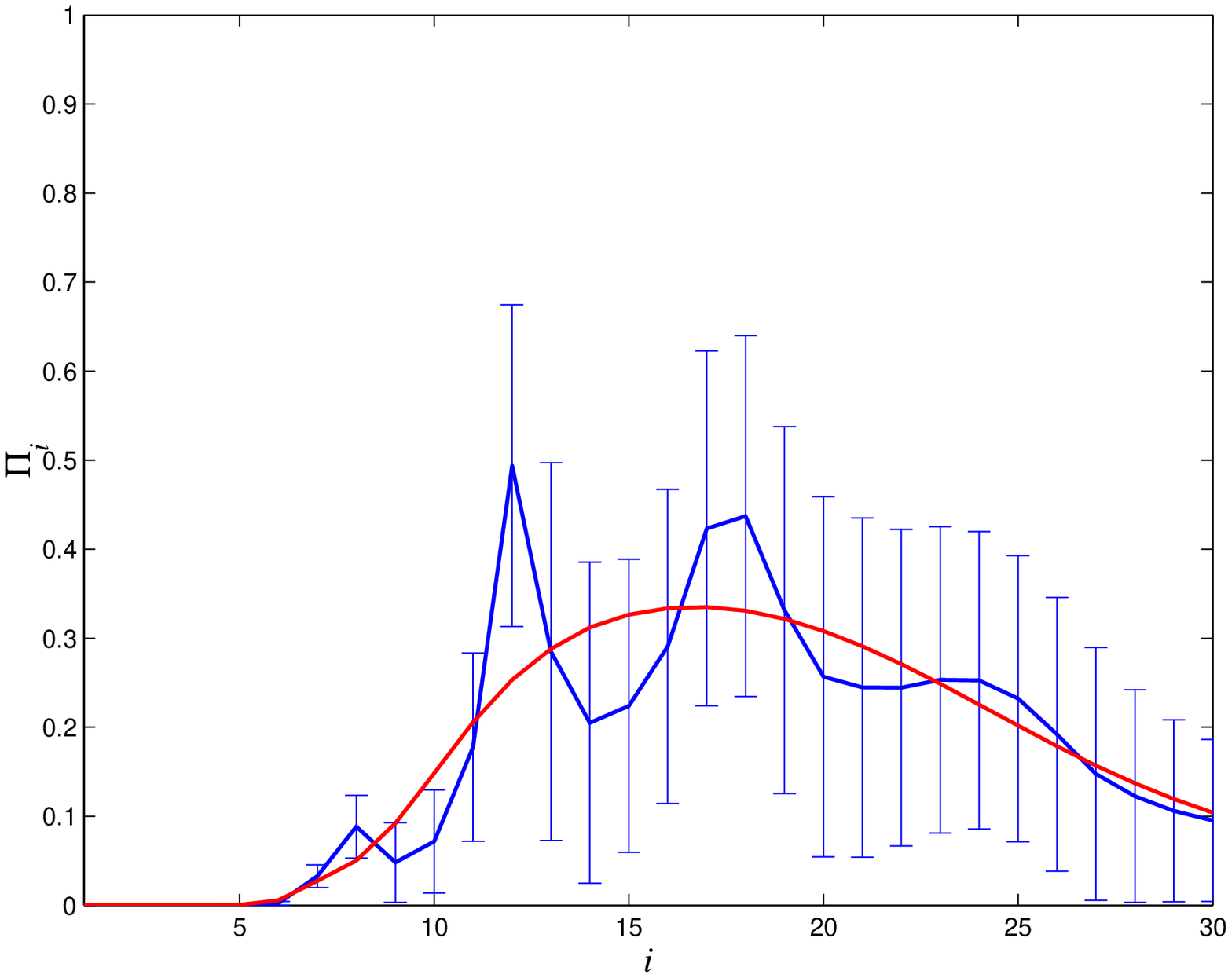} &
\includegraphics[width=6.8cm,angle=90]{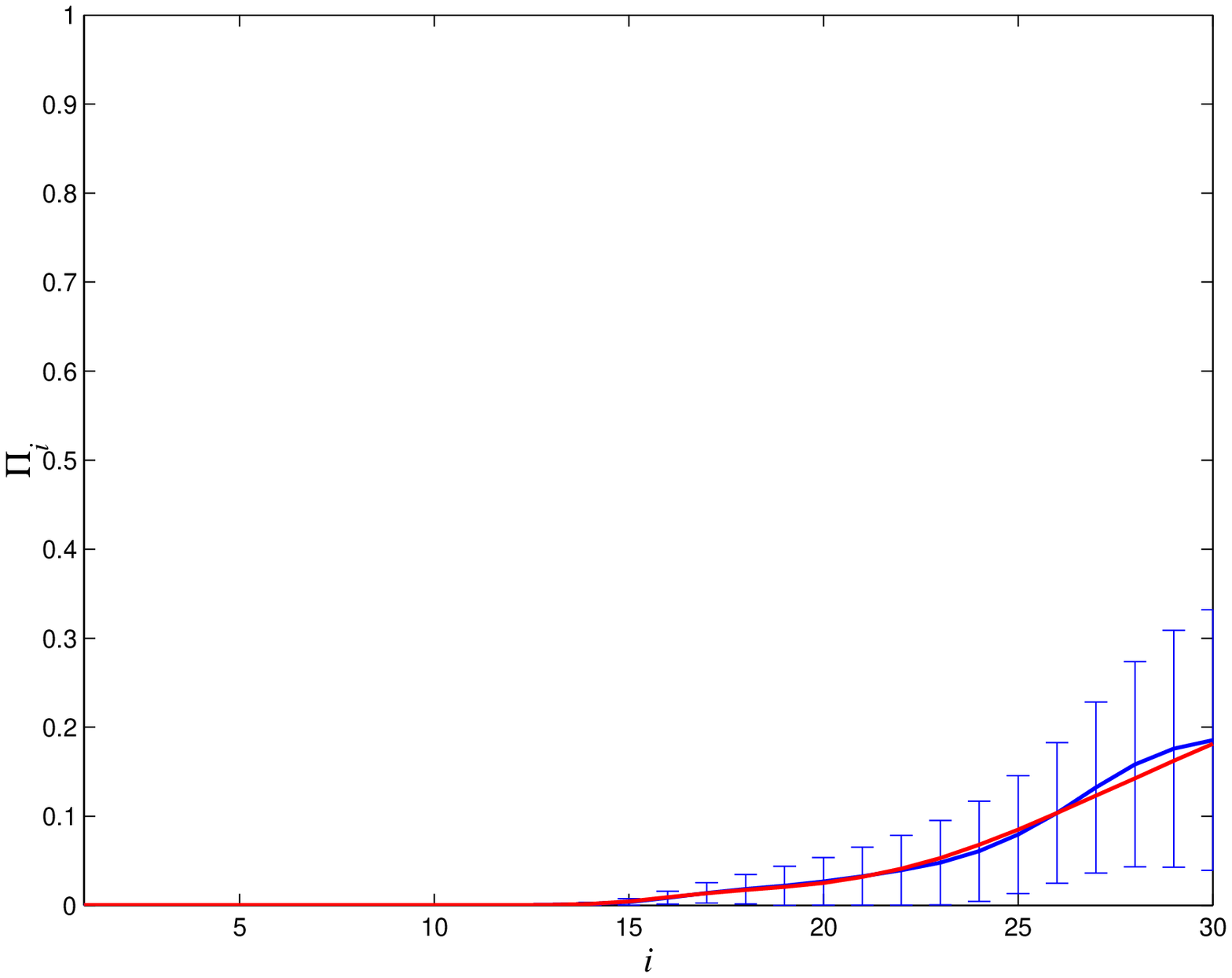} \\
\includegraphics[width=6.8cm,angle=90]{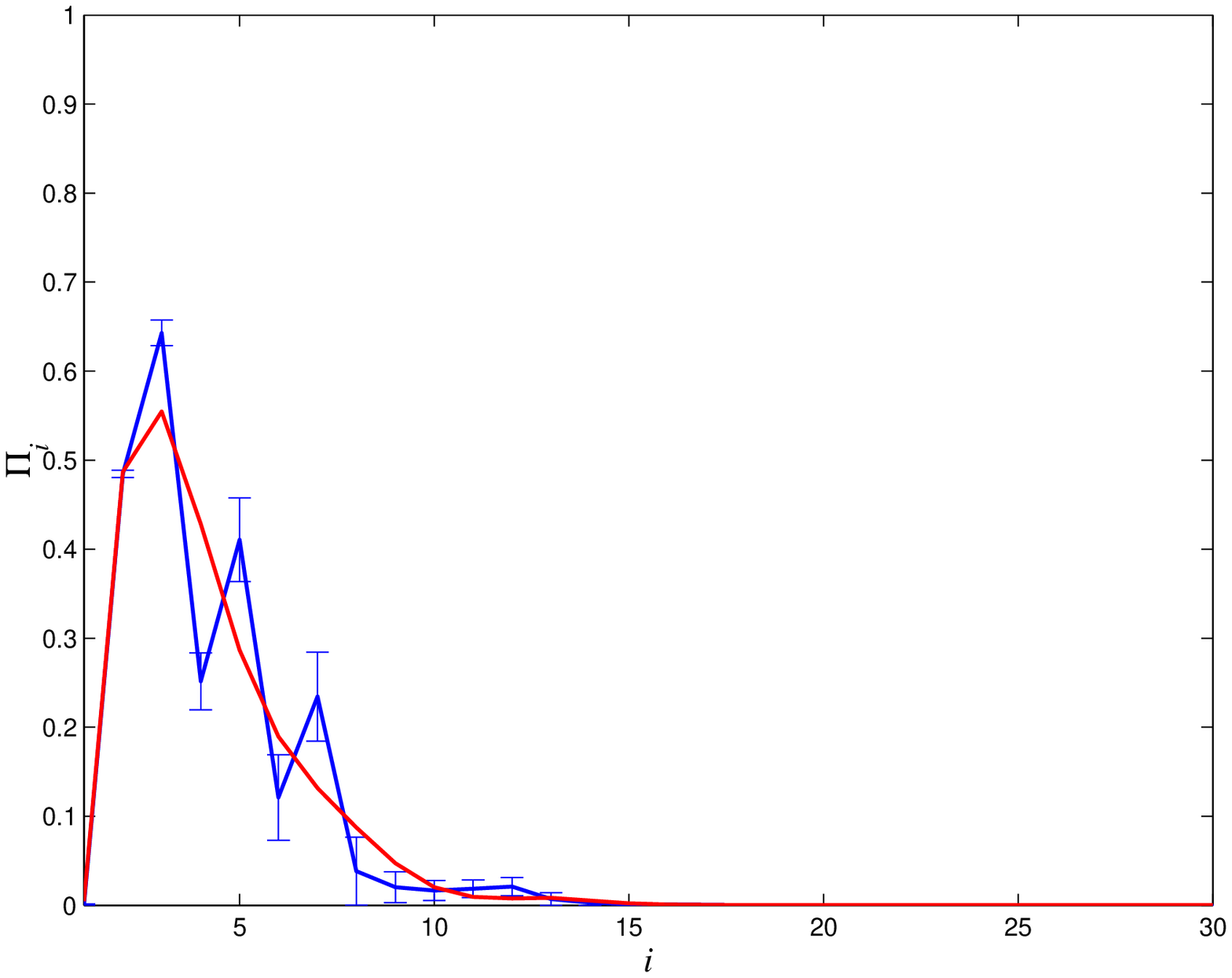} &
\includegraphics[width=6.8cm,angle=90]{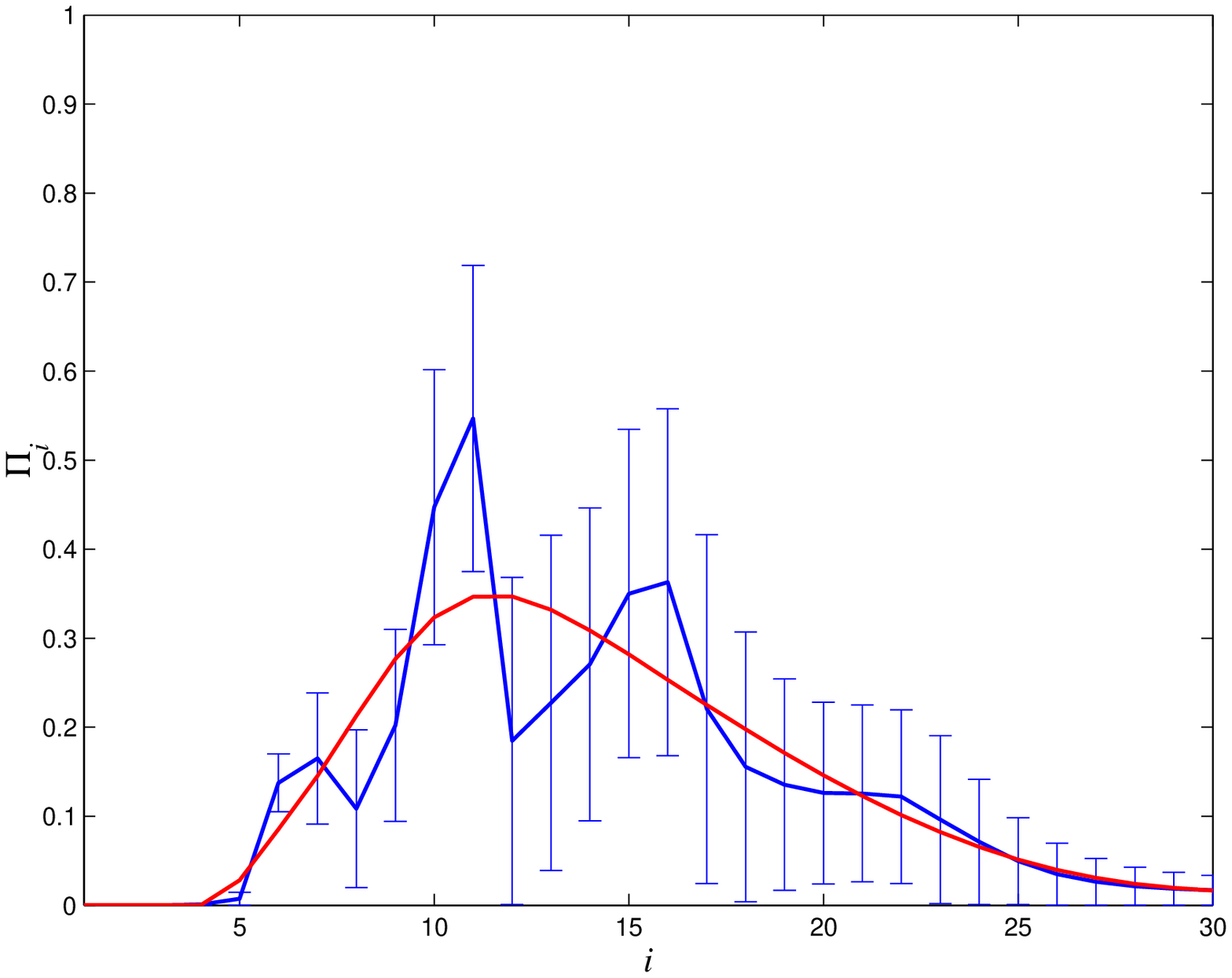} &
\includegraphics[width=6.8cm,angle=90]{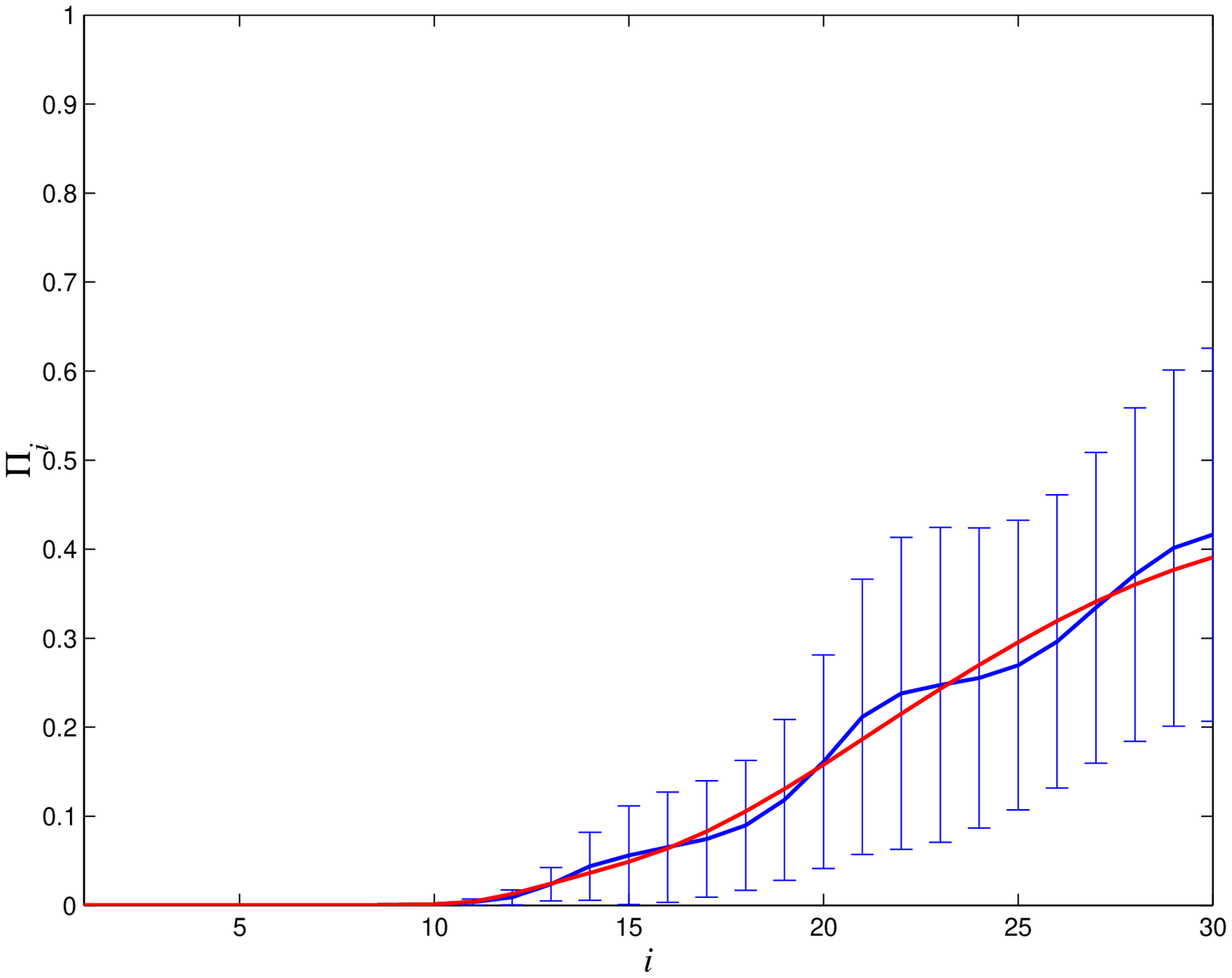} \\
\includegraphics[width=6.8cm,angle=90]{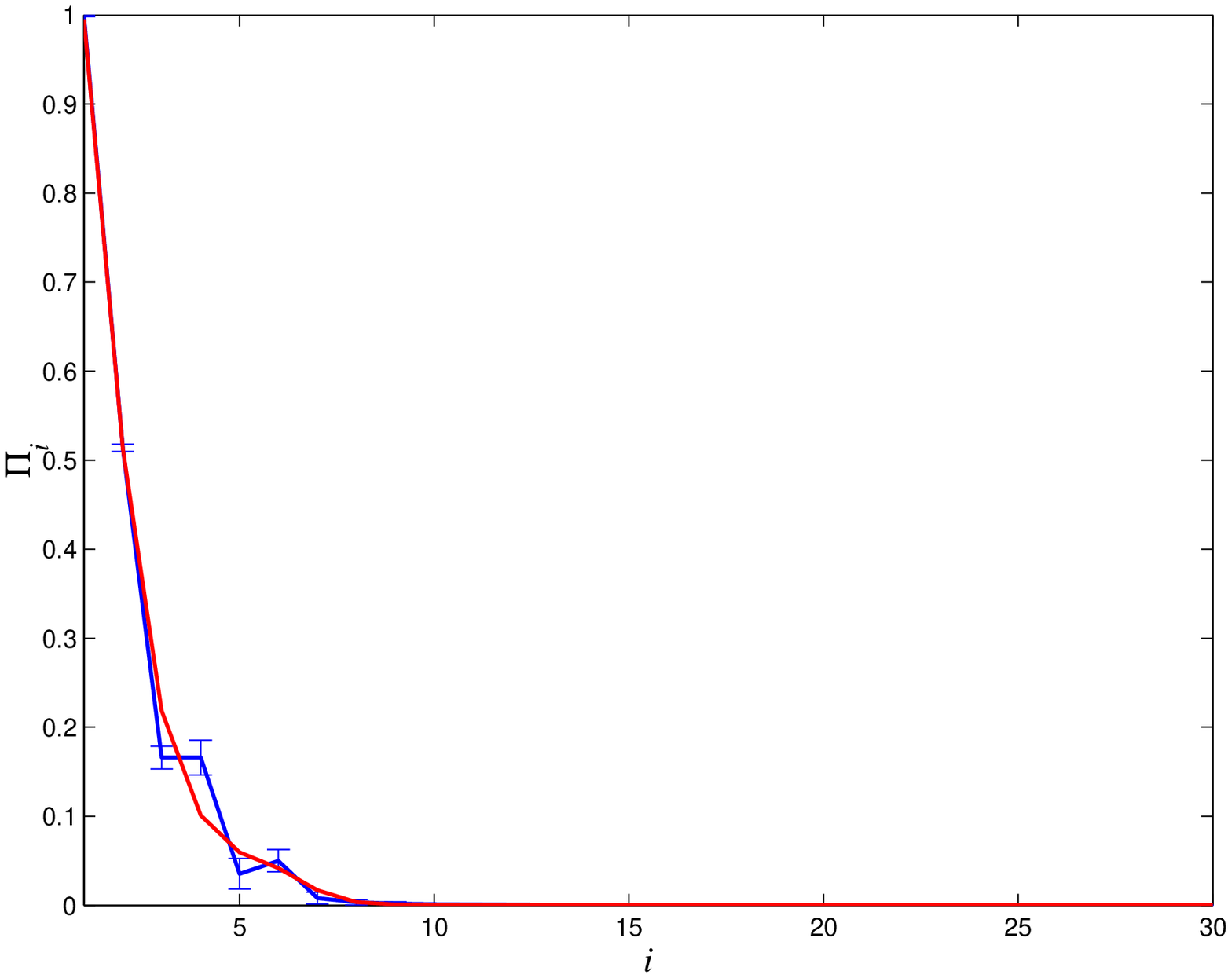} &
\includegraphics[width=6.8cm,angle=90]{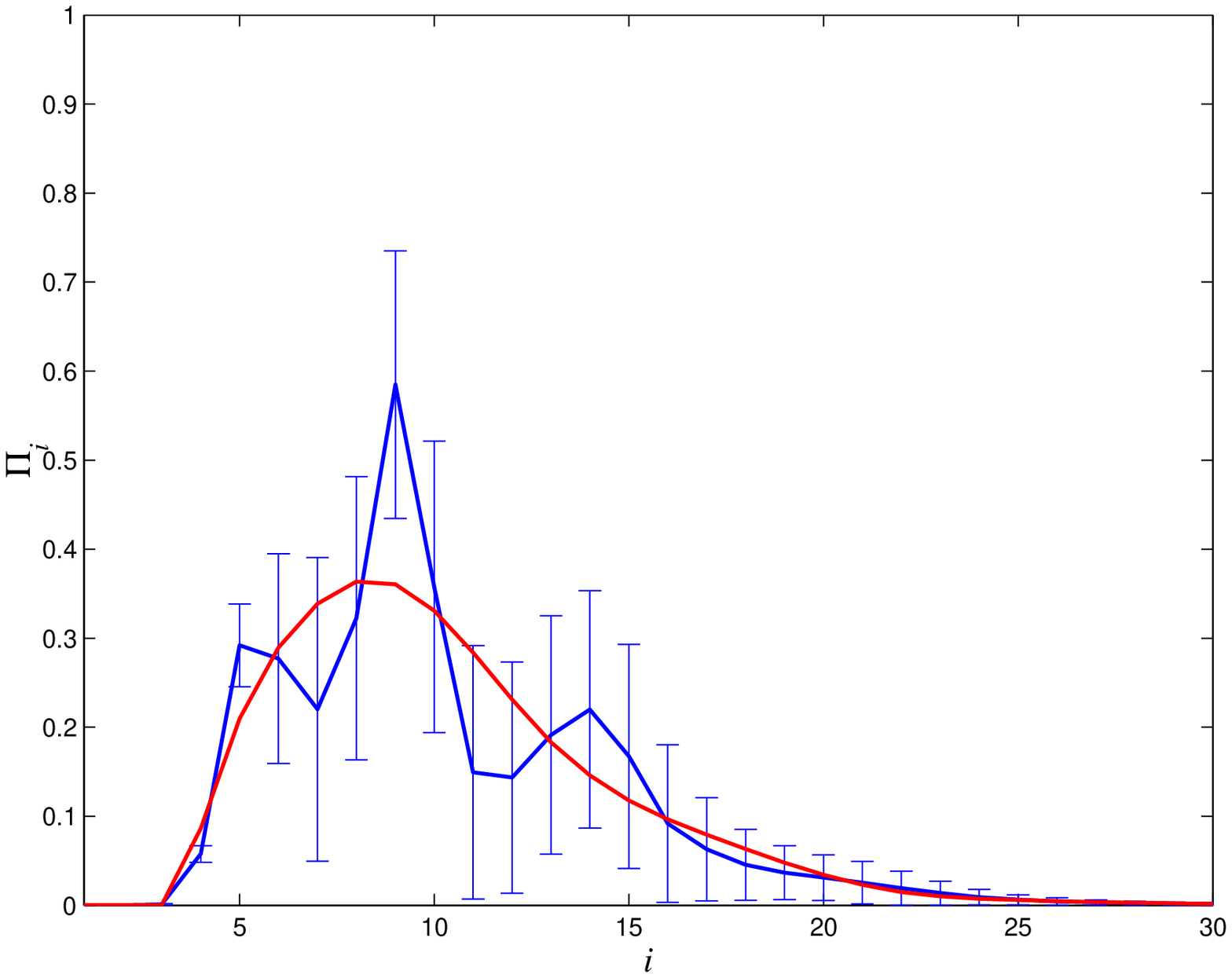} &
\includegraphics[width=6.8cm,angle=90]{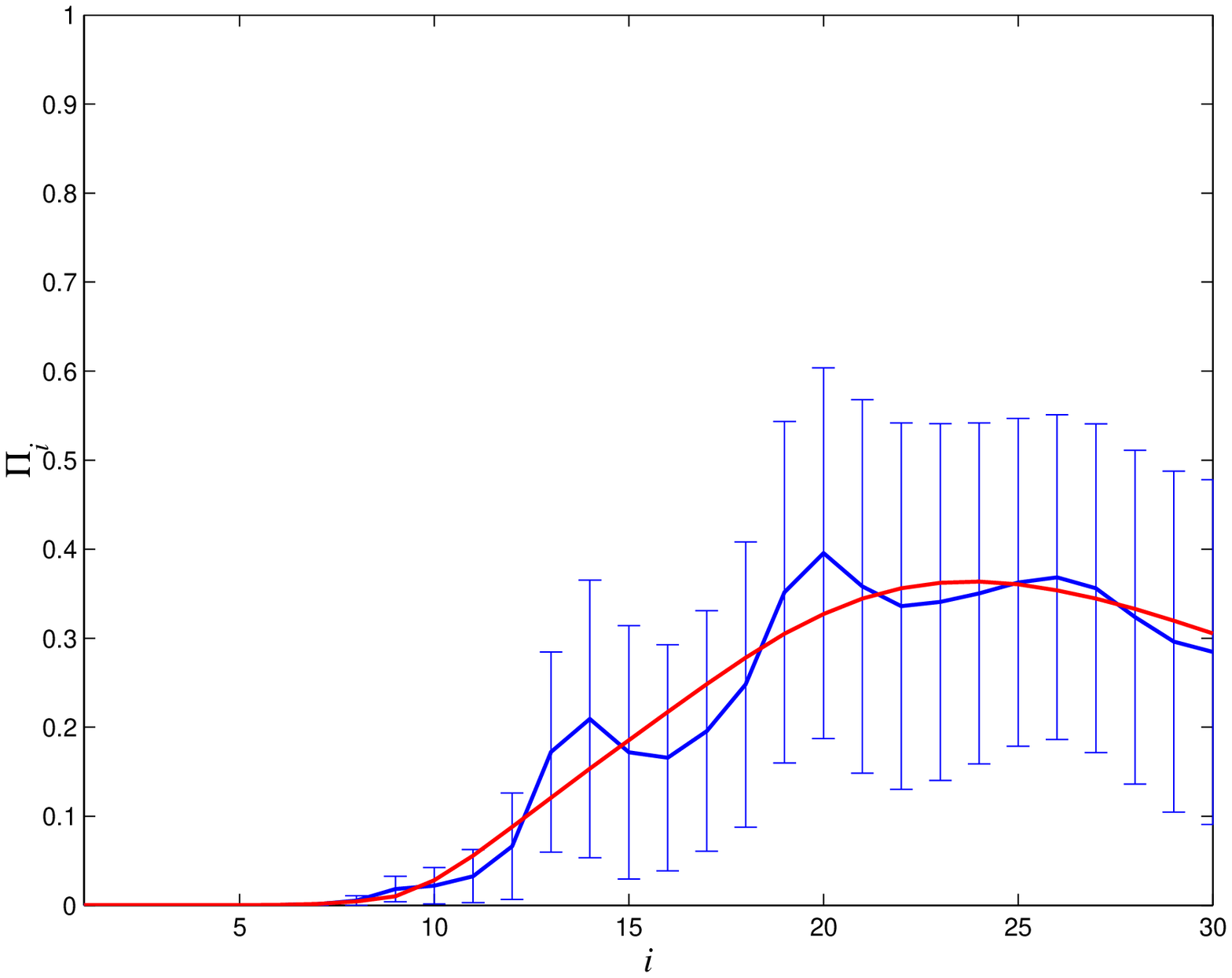}
\end{tabular}
\caption{
\label{fig:loopy1}
(Colour online) Reconstruction of the POVM elements of the Optical
POVM of Sec.~\ref{sec:loopy}. The graph depicts the mean value
solution $\mu_i$ of the diagonal entries (central wavy line (blue))
and their marginal standard deviations $\sqrt{\Sigma_{ii}}$
($\pm1\sigma$ error bars). As the actual variations on the diagonal
entries are correlated, this plot can only give an indication. Along
with the mean value solution, the regularised solution is plotted
(central smooth line (red)).
}
\end{figure*}
\begin{figure}[t]
\includegraphics[width=8cm]{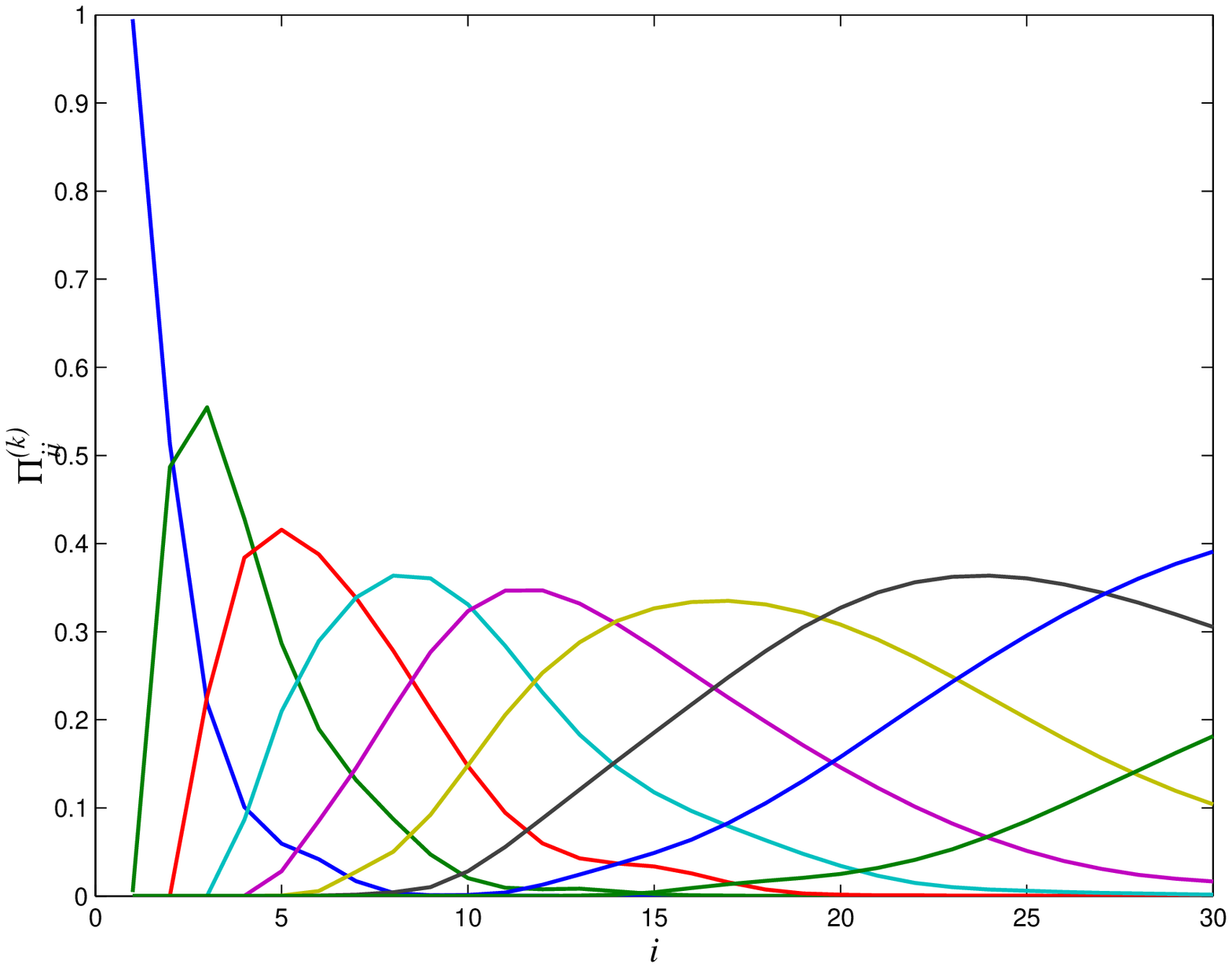}
\caption{
\label{fig:loopySm}
Reconstruction of the POVM elements $\Pi^{(k)}_{ii}$ of the Optical
POVM of Sec.~\ref{sec:loopy} (up to photon number 30 only). This is
the regularised solution, i.e.\ the solution with maximal smoothness
that is still within the confidence region obtained from the Kalman
Filter method. The solution obtained is in complete agreement with the
solution from \cite{alvaro}, which was obtained in a completely
different way. }
\end{figure}

In Fig.~\ref{fig:loopy1} we depict the final results for each of the
POVM elements, showing the physical mean value solution, the error
bars, and the smoothed solution. The smoothed solution of all POVM
elements together is depicted separately in
Fig.~\ref{fig:loopySm}. The results are in very good correspondence
with both the reconstruction of \cite{alvaro} and the theoretical
model of the POVM (based on independent measurements of the
reflectivities of the beam splitters and the overall photon loss).

To illustrate how the mean values and error bars change after each
KF
iteration, we have created a movie, where each frame
consists of a plot similar to the one of Fig.~\ref{fig:loopy1},
generated after each iteration. We refer the reader to \cite{extra}
for this animation, the MatLab routines used, and other related
material.

\begin{figure}[t]
\includegraphics[width=8cm]{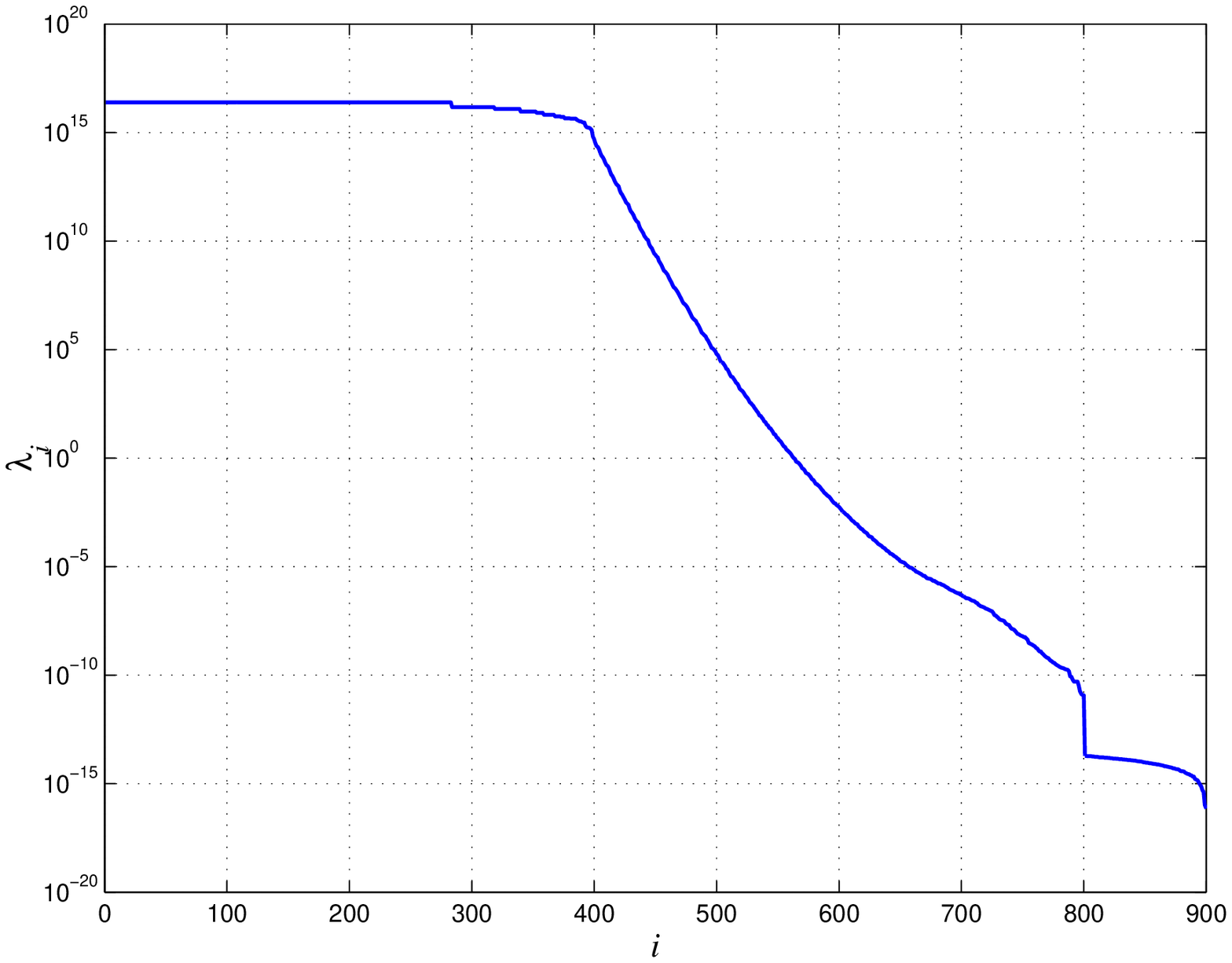}
\caption{
\label{fig:loopyspec}
Spectrum of the corrected covariance matrix $\Sigma_{\text{corr}}$ for
the Optical POVM, before restricting to the physical set; the
eigenvalues are plotted on a logarithmic scale. The values saturate at
about $10^{16}$, due to numerical imprecision in the calculation of
$\Sigma_{\text{corr}}$. Eigenvalues 800 and higher correspond to the
exact constraints imposed by the projector $T_X$ and are nominally 0.}
\end{figure}
In order to infer how many measurements are needed to reduce the
errors, one has to look at the unconstrained confidence region. We
have plotted the spectrum of the unphysical covariance matrix in
Fig.~\ref{fig:loopyspec}. This graph allows to estimate the number of
experimental runs $N$ necessary to achieve a certain final
precision. It is evident from the graph that only 110 of the 800 free
components have standard deviation less than $0.001$
($\lambda_i=\sigma^2_i\le 10^{-6}$). Since variances scale as $1/N$,
to double that number to 220, say, $N$ should be increased by a factor
of no less than about 100,000 (to get the $\sigma^2_{580}$ of
$10^{-1}$ below $10^{-6}$), i.e.\ from 38,000 to the rather
impractical 3,800,000,000. Hence, to really achieve higher precision
with this kind of experiment, another setup should be considered.

\section{Discussion}\label{sec:discuss}
\subsection{Comparison to other Methods}
The reconstruction method that matches ours most closely is the one reported in \cite{rehacek08}, 
which is also based on the likelihood function and also yields a covariance matrix. 
Hence, this method allows to calculate confidence regions in the same way as ours. 
The main differences are that in \cite{rehacek08}
the point of maximum likelihood is calculated first,
the covariance matrix is calculated as the inverse of the Hessian 
(the second derivative matrix $\partial^2/\partial x_i\partial x_j$)
of the logarithm of the likelihood function, taken in the mode of that function, and the restriction to
the physical set is imposed beforehand. 
In contrast, our method amounts to calculating the mean of the log-likelihood function, 
its Hessian in that mean, and the restriction to the physical set is made afterwards.

First of all, we believe that our approach yields results that better match the exact confidence region.
The likelihood function is highly skewed, whenever there are a lot of
low-probability measurement outcomes; this appears to be the rule rather than the exception. 
In those cases the mean is statistically more meaningful than the
mode, especially when one restricts to the mode over the physical set from the outset.
Second, restricting to the physical set only in a post-processing phase yields valuable diagnostic information
about correctness of the assumed noise model and also about the ultimate accuracy allowed by the particular
tomographic data.
As illustrated in Application 1,
the mode over the physical set can be really far from the mean, due to unforeseen 
noise/error contributions, and the mean has to be calculated in order to see that.
Also, to infer how many more measurements would be needed to improve the reconstruction accuracy, one
needs to look at the covariance matrix before restricting to the physical set,
as illustrated in Application 2.

Other reconstruction methods also calculate the MaxLik solution and derive error measures from Monte Carlo
simulations. As such, they suffer from the same drawbacks as the method of \cite{rehacek08} in that
the restriction to physical space is made from the outset. Moreover,
the time required for the Monte Carlo calculations rapidly becomes prohibitive with increasing system
dimensions. Even for two-qubit systems, our method is orders of magnitude faster than Monte Carlo methods. 
A further problem with the Monte Carlo method is the difficulty of obtaining a reliable
stopping criterion.
\subsection{Computational Resources}
The memory requirements of our method are easily calculated. They are essentially governed
by the dimension of the subspace $\cS_X$ on which the state $\rho$ is supported.
If this dimension is $D$ then
storage for $\tilde{\bm{\mu}}$ consists of $D^2$ complex numbers, while for the (tilde) covariance
matrix it is the square of that, $D^4$. This means that for the full reconstruction of $n$-qubit
states, $2^{2n}$ elements are needed for the state, and $2^{4n}$ for the covariance matrix.

The computation time for the Kalman filter update (executed once per measurement setting)
is dominated by a fixed number of matrix multiplications (of $D^2\times D^2$ matrices) and
one matrix inversion (of a $K\times K$ matrix, where $K$ is the number of outcomes per measurement
setting and therefore is typically much smaller than $D$).
As the computation complexity of a matrix multiplication for two $k\times k$ matrices
is $O(k^3)$ (or somewhat less), we get a computational complexity of $D^6$.

The optional post-processing steps of calculating the MaxLik and/or MaxEnt solution
require solving a semi-definite program. In all reported applications this turned out to be the most
time-consuming step.
\section{Conclusion}
In this work we have introduced a novel Bayesian tomographic
reconstruction method based on Kalman filtering that does not just
give a maximum likelihood solution but also produces error bars, in
the form of a confidence region around a mean value solution. It must
be stressed that the error bars are directly derived from the
measurement data, unlike in Monte-Carlo methods, where they are
produced from simulations.

We have shown that to properly deal with low-probability events (e.g.
measurement outcomes with very few clicks) one has to consider the
conjugate distribution of the noise model, in the spirit of Laplace's
rule of succession. That is, if click frequencies are distributed
multinomially or Poissonian, this yields a distribution of the
underlying click probabilities that is Dirichlet distributed. This
avoids the incorrect assignment of zero probability to an outcome that
has not been observed.
Furthermore, we have introduced a novel method of ensuring that the
reconstruction is physical. This method is again based on Kalman
filtering, and has the benefit that it is very fast and again produces
appropriate error bars.

Finally, we have applied the method to two real world applications. In
the first example, the state reconstruction of an entangled two-qubit
state, the reconstruction process reduces to a single application of
the Kalman update equations which, apart from its numerical stability,
reduces the computational effort. Compared to Monte Carlo methods for
calculating error bars the computational effort is reduced
by several orders of magnitude. The
Kalman filter method also revealed the necessity to adjust the underlying
noise model by taking into account additional error sources.
The second example concerned the reconstruction of an optical
POVM. There the advantages of Kalman filtering also became evident in one's
capability to estimate the number of experimental runs necessary to
achieve a certain final precision. Both examples indicate that our
KF method can be an invaluable diagnostic tool.

In future work we will consider how to deal with measurement
imperfections, including drift in the tomographic and system
components. We will investigate how the present method can be applied
to tomography with continuous variable outcomes. A further topic of
study will be the integration of the Kalman filter method within
adaptive tomographic setups, as the method is very much an online
method, updating the covariance matrix as it goes. Among the more
technical issues, we will study the convergence properties of our
proposed Kalman filter method for restricting the reconstruction to
physical space.

We are confident that our reconstruction method, due to its
statistically well-founded nature, can be the basis of a dependable, easily
adaptable, and universal reconstruction algorithm.

\acknowledgments
SS thanks the UK Engineering and Physical Sciences Research Council
(EPSRC) for support. We thank the following people who have kindly
provided us with ample of tomographic data for developing and testing
the Kalman Filter method: I.~Walmsley, A.G.~White, J.~O'Brien and
N.K.~Langford. It is fair to say that without their assistance this
paper would have been very theoretic and equally useless.
We also thank M.~Plenio, A.~Feito, R.~Schack, T.~Osborne and T.~Sharia
for illuminating discussions.
Last but not least, we thank Z.~Hradil for sharing his thoughts on the issues considered
in our work.
%

\appendix
\section{Proof of the bound (\ref{eq:crbound}) on the physical confidence value
}
\label{sec:boundproof}

For definitions we refer back to Sec.~\ref{sec:cr}. We start with a Lemma.
\begin{lemma}
Define the function
$$
g(r,a) := \int_0^r \md x\, x^{d-1}e^{-(x+a)^2/2}.
$$
Then for $a\ge 0$, and $r\le R$ the relation
$$
\frac{g(r,a)}{g(R,a)} \ge  \frac{g(r,0)}{g(R,0)}
$$
holds.
\end{lemma}
\textit{Proof.}
Consider three integrable functions on the interval $[0,R]$, $f$, $g$,
and $h$. Let $f$ be non-negative, and $g$ and $h$ non-increasing.
It is easily shown that these functions satisfy the inequality
\bea
\lefteqn{
\int_0^R \md x f(x)g(x)h(x)\,\,\int_0^R \md x f(x) } \nonumber\\
&\ge& \int_0^R \md x f(x)g(x)\,\,\int_0^R \md x f(x)h(x).
\label{eq:lemmaineq}
\eea
To see this, subtract the right-hand side from the left-hand side,
rewrite the integrals as double integrals over the square
$[0,R]\times[0,R]$, split up this square into two equal parts along
the diagonal $x=y$, and enjoy the benefits of the integrand's
symmetry, giving:
\beas
\lefteqn{\int_0^R \md x \int_0^R \md y f(x)f(y)g(x)(h(x)-h(y))} \\
&=& \int_0^R \md x \int_x^R \md y f(x)f(y)\Big(g(x)\left(h(x)-h(y)\right) \\
&& \qquad\qquad +g(y)\left(h(y)-h(x)\right)\Big) \\
&=& \int_0^R \md x \int_x^R \md y f(x)f(y)\left(g(y)-g(x)\right)\left(h(y)-h(x)\right) \\
&\ge& 0.
\eeas
Now specialising the inequality (\ref{eq:lemmaineq}) to the functions
\beas
f(x) &=& x^{d-1}e^{-x^2/2} \\
g(x) &=& e^{-ax} \\
h(x) &=& \Phi(0\le x\le r)
\eeas
gives the inequality of the lemma.
\qed


We start from the unphysical reconstruction, that is the mean $\mu$
and the covariance matrix $\Sigma$. Let $\cP$ be the physical set, and
let $\rho_0$ be the maximum likelihood solution, i.e. the state that
is closest to the mean $\mu$, in the Mahalanobis distance.
In what follows we will use the Hilbert space representation of
states, i.e. a representation as vectors. As before, we will denote
this by math boldface. The discussion becomes easier by going over to
a new, ``standardised'' coordinate system, in which the mean $\mu$ is
the origin and the covariance matrix is the identity matrix. The
Mahalanobis distance is then just the Euclidean distance, and the
confidence regions are spheres centered around the origin.

In quantum mechanics, the physical set $\cP$ is convex. By definition,
$\bm{\rho}_0$ is on the boundary of $\cP$. Therefore, $\cP$ can be
decomposed into infinitesimal cones with center $\bm{\rho}_0$,  each
pointing to a different direction $\bm{\Omega}$, having cross-section
$\md\bm{\Omega}$, and cut to certain length $R(\bm{\Omega})$, where
the latter function determines the overall shape of $\cP$.

In standardised coordinates, the unphysical posterior $f$ is given by
$f(\bm{x})=C\exp(-x^2/2)$, with $C$ the normalisation constant, and
$x=||\bm{x}||$. We now want to calculate the cumulative distribution
function (CDF) of the physical posterior, which is the normalised
integral of $f$ over the intersection of $\cP$ with the ball of radius
$x$, $g(x)/g(\infty)$, with
\beas
g(x) &:=& \int_\Omega \md\bm{\Omega}\int_0^{R(\bm{\Omega})} \md r\,
r^{d-1} \Phi(||r\bm{\Omega}+\bm{\rho}_0||\le x) \\
&&\times \exp(-||r\bm{\Omega}+\bm{\rho}_0||^2/2).
\eeas

Let us also define the non-negative function
\beas
g(x,\bm{\Omega}) &=& \int_0^{R(\bm{\Omega})} \md r\,
r^{d-1} \Phi(||r\bm{\Omega}+\bm{\rho}_0||\le x) \\
&& \times \exp(-||r\bm{\Omega}+\bm{\rho}_0||^2/2).
\eeas
Then we have
\beas
g(x)/g(\infty)
&=& \frac{\int_\Omega \md\bm{\Omega} g(x,\bm{\Omega})}%
{\int_\Omega \md\bm{\Omega} g(\infty,\bm{\Omega})} \\
&=& \int_\Omega \md\bm{\Omega} \frac{g(\infty,\bm{\Omega})}%
{\int_\Omega \md\bm{\Omega}' g(\infty,\bm{\Omega}')}
\,\,\frac{g(x,\bm{\Omega})}{g(\infty,\bm{\Omega})}.
\eeas
The first factor of the integrand, which we will denote by
$w(\bm{\Omega})$, is a PDF, in that it is a non-negative function
integrating to 1 over $\Omega$. We have thus shown the following
statement:


\noindent
\textbf{Statement C:} \textit{The function $g(x)/g(\infty)$ is a
weighted average of the functions
$g(x,\bm{\Omega})/g(\infty,\bm{\Omega})$ over $\Omega$.}


Let us now fix $\bm{\Omega}$. The value of $g(x,\bm{\Omega})$ no
longer changes for $x$ beyond $R(\bm{\Omega})$. We define $R_x$ as
that value of $r$ for which $||r\bm{\Omega}+\bm{\rho}_0|| = x$. Thus,
for $R_x\ge R(\bm{\Omega})$, we have
$g(x,\bm{\Omega}) = g(\infty,\bm{\Omega})$.

Consider now the case that $x$ is small enough so that $R_x\le
R(\bm{\Omega})$. Let $\rho=||\bm{\rho}_0||$ (the 2-norm of the vector
representation of $\rho_0$). In fact, $\rho=\cM_{ML}$ as used in the
bound (\ref{eq:crbound}). Let $\theta$ be the angle between a normal
to $\bm{\rho}_0 - \bm{\mu}$ and $\bm{\Omega}$.
Because $\bm{\rho}_0$ is the nearest point in $\cP$ to $\bm{\mu}$,
this angle is between 0 and $\pi/2$.

In this case we have
$$
g(x,\bm{\Omega}) = \int_0^{R_x}\md r\,r^{d-1}\exp[-\xi(r)^2/2],
$$
with
\beas
\xi(r)^2 &=& ||r\bm{\Omega}+\rho_0||^2 \\
&=& \rho^2+r^2+2\rho r\sin\theta \\
&=&(r+\rho\sin\theta)^2+\rho^2\cos^2\theta.
\eeas
This gives us
\beas
g(x,\bm{\Omega}) &=& \exp(-(\rho^2\cos^2\theta)/2) \\
&& \times \int_0^{R_x} \md r\, r^{d-1} \exp(-(r+\rho\sin\theta)^2/2).
\eeas
The factor in front of the integral is independent of $x$ and cancels
out in the quantity of interest $g(x,\bm{\Omega})/g(\infty,\bm{\Omega})$.
Applying the lemma we now get
\beas
\frac{g(x,\bm{\Omega})}{g(\infty,\bm{\Omega})}
&=& \frac{\int_0^{R_x} \md r\, r^{d-1} \exp(-(r+\rho\sin\theta)^2/2)}
{\int_0^{R(\bm{\Omega})} \md r\, r^{d-1}
\exp(-(r+\rho\sin\theta)^2/2)} \\
&\ge& \frac{\int_0^{R_x} \md r\, r^{d-1}
\exp(-r^2/2)}{\int_0^{R(\bm{\Omega})} \md r\, r^{d-1} \exp(-r^2/2)} \\
&\ge& \frac{\int_0^{R_x} \md r\, r^{d-1} \exp(-r^2/2)}{\int_0^{\infty}
\md r\, r^{d-1} \exp(-r^2/2)}.
\eeas


Now $R_x$ satisfies the triangle inequality:
$$
x\le \rho+R_x.
$$
Thus if we replace $R_x$ as upper integration limit by its lower bound
$x-\rho$, or 0 if the difference is negative, then we get a lower
bound on the integral too, giving
\beas
\frac{g(x,\bm{\Omega})}{g(\infty,\bm{\Omega})}
&\ge& \frac{\int_0^{x-\rho} \md r\, r^{d-1}
\exp(-r^2/2)}{\int_0^{\infty} \md r\, r^{d-1} \exp(-r^2/2)}.
\eeas
The upshot of this step is that the right-hand side is now completely
independent of $\bm{\Omega}$, which allows us to invoke
\textbf{Statement C} and get that $g(x)/g(\infty)$ satisfies the same
inequality:
\beas
\frac{g(x)}{g(\infty)}
&\ge& \frac{\int_0^{x-\rho} \md r\, r^{d-1}
\exp(-r^2/2)}{\int_0^{\infty} \md r\, r^{d-1} \exp(-r^2/2)}.
\eeas
The right-hand side is the CDF of the chi distribution (with $d-1$
degrees of freedom) evaluated in $x-\rho$, i.e.\ the CDF is shifted to
the right by an amount $\rho=\cM_{ML}$. Its confidence region is
therefore the interval $[0,\cM_{CR,unphys}+\cM_{ML}]$. The left-hand
side is the CDF of the restricted posterior, with confidence region
$[0,\cM_{CR,phys}]$. Because of the inequality, the latter confidence
region is contained in the former. That proves the bound
(\ref{eq:crbound}). \qed

\section{Properties of the Dirichlet estimator}
\subsection{Mode v Confidence Region}
\label{sec:modeCR}

Here we give the promised proof that the mode of the Dirichlet
distribution lies within the confidence region as defined in
(\ref{eq:MahBound}), with $\bm{\mu}$ and $\bm{\Sigma}$ given by
(\ref{eq:dirichletmean}) and (\ref{eq:dirichletcov}).


\textit{Proof.}
Let $\bm{x}$ be the mode of the Dirichlet distribution,
$\bm{x}=\bm{f}/N$, and $\bm{\mu}$ be its mean,
$\bm{\mu}=(\bm{f}+1)/(N+d)$. Then
$\bm{x}-\bm{\mu} = (d\bm{f}-N)/(N(N+d))$;  as the sum of the entries
of $\bm{x}-\bm{\mu}$ is 0, $\bm{x}-\bm{\mu}$ lies in the subspace on
which $\bm{G}$ of (\ref{eq:dirichletsigmp}) projects. Thus we have
\beas
\cM^2 &=& (\bm{x}-\bm{\mu})^*\bm{\Sigma}^\dagger(\bm{x}-\bm{\mu}) \\
&=& \frac{N+d+1}{N^2(N+d)} \sum_{i=1}^d \frac{(df_i-N)^2}{f_i+1} \\
&=& \frac{(N+d+1)}{N^2}\left((N+d)\sum_{i=1}^d (f_i+1)^{-1} - d^2\right).
\eeas
If $r$ is the number of non-zero components of $f$ (thus
$1\le r\le d$) and if we put $f_i=Np_i$, fixing $p_i$, then this
expression can be expanded as $d-r+O(1/N)$. The term $\sum_{i=1}^d
(f_i+1)^{-1}$ is maximal for $\bm{f}=(N,0,\ldots,0)$, giving the sum
$(N+1)^{-1}+d-1$. In this way we get the upper bound
$$
\cM^2 \le \frac{N+d+1}{N+1}(d-1) = d-1+O(1/N).
$$
For not too small values of $N$, this bound is approximately equal to
$d-1$, which is also the number of degrees of freedom $\nu$ in this
case. As $d-1$ is the mean value of the $\chi_{d-1}^2$ distribution,
the value $d-1$ lies within any reasonable confidence
interval. Therefore, the mode of the Dirichlet distribution lies
within the confidence region of its normal approximation.
\qed

\subsection{Wald statistic}
\label{sec:waldproof}

Suppose the actual state under consideration is $\rho$, and a
measurement is made using a $d$-outcome POVM, so that the
probabilities of the outcomes are given by the probability vector
$\bm{p}$. In an experiment this gives rise to certain outcome
frequencies $\bm{f}$, drawn from a multinomial distribution with
parameters $N$ and $\bm{p}$. From these frequencies $\bm{f}$ one can
derive an estimation $\hat{\bm{P}}$ of $\bm{p}$, Dirichlet distributed
with parameter $\bm{f}$ according to the prescription of
Sec.~\ref{sec:charlik}.  Let $\bm{\mu}$ and $\bm{\Sigma}$ be the
moments of this Dirichlet estimation.

We want to study how well the actual $\bm{p}$ fits within the
confidence region obtained from this estimation. To do so, we
construct the Wald statistic
\beas
z&:=&(\bm{p}-\bm{\mu})^*\bm{\Sigma}^\dagger(\bm{p}-\bm{\mu}) \\
&=& \frac{N+d+1}{N+d}\sum_{i=1}^d \frac{((N+d)p_i-(f_i+1))^2}{f_i+1} \\
&=& (N+d+1)\,\left((N+d)\sum_i \frac{p_i^2}{f_i+1}-1\right).
\eeas
If the distribution involved was Gaussian, this statistic would be
$\chi^2_{d-1}$ distributed. In reality, the distribution only tends to
a Gaussian and the Wald statistic is only asymptotically
$\chi^2_{d-1}$ \cite{wald}.

An exact calculation yields the first two moments of $z$, given that
$\bm{F}$ is distributed as $\bm{F}\sim\mbox{Mtn}(N,\bm{p})$, in terms
of $\bm{p}$:
\begin{widetext}
\bea
\mu(Z) &=& (N+d+1)\left(\frac{N+d}{N+1}\Big(1-\sum_i
p_i(1-p_i)^{N+1}\Big) -1\right) \label{eq:muz}\\
\sigma^2(Z) &=&
\frac{(N+d+1)^2(N+d)^2}{(N+1)^2}\,\,
\Bigg\{
    \frac{N+1}{N+2}\sum_{i\neq j} p_i p_j
    \left[1+(1-p_i-p_j)^{N+2}-(1-p_i)^{N+2}-(1-p_j)^{N+2}\right]
    \nonumber\\
&&  +(N+1)\sum_i g(p_i,N)
    -\Big[1-\sum_i p_i(1-p_i)^{N+1}\Big]^2
\Bigg\},\label{eq:sigz}
\eea
\end{widetext}
where the function $g(p,N)$ is defined as
\beas
g(p,N) &:=& p^4 \sum_{k=0}^N {N+1\choose k}\frac{1}{N+1-k}\,
p^{N-k}(1-p)^{k} \\
&=& p^3 \sum_{k=1}^{N+1} {N+1 \choose k}\,\frac{1}{k}\,
p^{k}(1-p)^{N+1-k}.
\eeas

\begin{figure}[t]
\includegraphics[width=8.7cm]{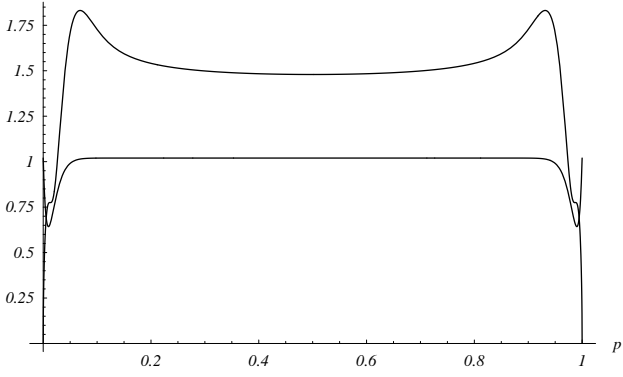}
\caption{
\label{fig:musig}
Graph of $\mu(Z)$ (lower curve) and $\sigma(Z)$ (upper curve),
from Eq.~(\ref{eq:muz}) and the square root of (\ref{eq:sigz}), as
function of $p$, for $d=2$ and $N=100$.
}
\end{figure}

The sum $g(p,N)$ is related to the first inverse moment of the
positive (i.e. non-zero) binomial distribution
$$
\mu_{-1}(p,N) = \sum_{k=1}^N \frac{1}{k} {N \choose k}
p^k(1-p)^{N-k}
$$
by
$$
g(p,N) = p^3 \mu_{-1}(p,N+1).
$$
No closed form for inverse moments exists, but several expansions are
known (see, e.g. Ref.~\cite{moments} and references therein). For
large $N$, one can approximate the binomial distribution by a Poisson
distribution with mean $\mu=Np$. For the first inverse moment, this
gives an approximation by the known first inverse moment of the
Poisson distribution, with relative error of the order $1/N$:
$$
\mu_{-1}(p,N) \approx f(Np), \quad f(\mu) =
e^{-\mu}(\mbox{Ei}(\mu)-\log\mu-\gamma);
$$
here, $\mbox{Ei}(x)$ is the exponential integral function and $\gamma$
is the Euler-Mascheroni constant. Thus, we get
$$
g(p,N)\approx (N+1)^{-3}\, \mu^{3}e^{-\mu}(\mbox{Ei}(\mu)-\log\mu-\gamma)
$$
with $\mu=(N+1)p$. To obtain $\sigma^2(Z)$, however, $g$ has to be
multiplied by a constant of order $O(N^3)$, and as $\sigma^2(Z)$
turns out to be of order $O(1)$, we need to know $g$ with a relative
precision of order $O(1/N^3)$. This requires correction terms of
$\mu_{-1}$ of up to second order. According to the recipe described in
\cite{moments}, the required approximate formula for $\mu_{-1}$ is
given by
$$
\mu_{-1}(p,N) \approx
\frac{A(\mu) f(\mu)
+B(\mu) e^{-\mu}
+C(\mu)}{24(N+1)^2}
$$
with $\mu=(N+1)p$ and
\beas
f(\mu) &=& e^{-\mu}(\mbox{Ei}(\mu)-\log\mu-\gamma) \\
A(\mu) &=& 3\mu^4-8\mu^3-12(N+1)\mu^2+24(N+1)^2 \\
B(\mu) &=& 12\mu^3-6\mu^2-24(N+1)\mu-(12N+10) \\
C(\mu) &=& -3\mu^3+5\mu^2+(12N+14)\mu+(12N+10).
\eeas
For the 2-dimensional case, with $N=100$, the resulting values for
$\mu(Z)$ and $\sigma(Z)$ are plotted as function of $p$ in
Fig.~\ref{fig:musig}.  When $p$ is sufficiently far removed from the
endpoints 0 or 1, one sees that $\mu$ and $\sigma$ converge to their
$\chi^2$ values $1$ ($\mu_{\chi^2_d}=d-1$) and $\sqrt{2}$
($\sigma_{\chi^2_d}=\sqrt{2(d-1)}$).

More generally, good convergence occurs when the smallest $p_i$ is
still larger than about $20/N$, i.e. when every outcome has at least
20 clicks. Numerical studies reveal that the highest value of $\sigma$
occurs roughly when the smallest $p_i$ is about $7.2/N$. In turn, this
highest value of $\sigma$ is maximal when all $p_i$ bar one are equal
to $7.2/N$. This worst case value is approximately given by the
empirical formula $1.285(1+2(d-3/2)/N)\sigma_{\chi^2}$.

This gives us the following conservative approach:
\textit{Take the chi-square value for $\sigma(Z)$ whenever the smallest
$p_i$ is larger than $20/N$, and $1.285(1+2(d-3/2)/N)$ times the
chi-square value otherwise.}



\begin{thebibliography}{99}
\bibitem{vogel89} K.~Vogel and H.~Risken, Phys. Rev. A \textbf{40},
2847 (1989).
\bibitem{smithey93} D.T.~Smithey, M.~Beck, M.G.~Raymer and
A.~Faridani, Phys. Rev. Lett. \textbf{70}, 1244 (1993).
\bibitem{leonhardtBook} U.~Leonhardt, \textit{Measuring the quantum state of light},
(Cambridge University Press, Cambridge, 1997).
\bibitem{welschRev} D.-G.~Welsch, W.~Vogel and T.~Opatrn\'y,
``Homodyne detection and Quantum state reconstruction'', in:
\textit{Progress in Optics XXXIX}, ch.~II (1999).
\bibitem{rehacek08} J.~\v{R}eh\'a\v{c}ek, D.~Mogilevtsev and Z.~Hradil,
New J. Phys. \textbf{10},
043022 (2008).
\bibitem{awg} J.L.~O'Brien, G.J.~Pryde, A.~Gilchrist, D.F.V.~James,
N.K.~Langford, T.C.~Ralph and A.G.~White,
Phys. Rev. Lett. \textbf{93}, 080502 (2004).
\bibitem{blatt} M.~Riebe, M.~Chwalla, J.~Benhelm, H.~Haeffner,
W.~Haensel, C.F.~Roos and R.~Blatt,
New J. Phys. \textbf{9}, 211 (2007).
\bibitem{martinis} M.~Steffen et al,
Science \textbf{313}, 1423 (2006).
\bibitem{leonhardt96} U.~Leonhardt, M.~Munroe, T.~Kiss, Th.~Richter
and M.G.~Raymer,
Opt. Commun. \textbf{127}, 144 (1996).
\bibitem{dariano} G.M.~D'Ariano, C.~Macchiavello and M.G.A.~Paris,
Phys. Rev. A \textbf{50}, 4298 (1994).
\bibitem{leonhardt} U.~Leonhardt, H.~Paul and G.M.~D'Ariano,
Phys. Rev. A \textbf{52}, 4899 (1995).
\bibitem{dwm95} T.~Dunn, I.A.~Walmsley and S.~Mukamel,
Phys. Rev. Lett. \textbf{74}, 884 (1995).
\bibitem{wwv97} L.J.~Waxer, I.A.~Walmsley and W.~Vogel,
Phys. Rev. A \textbf{56}, R2491 (1997).
\bibitem{zvww99} A.~Zucchetti, W.~Vogel, D.-G.~Welsch, and I.A.~Walmsley,
Phys. Rev. A \textbf{60}, 2716 (1999).
\bibitem{dsimon} D.J.~Simon, \textit{Optimal State Estimation}, First
Edition, (Wiley, New York, 2006).
\bibitem{belavkin} V.P.~Belavkin,
Rep. Math. Phys. \textbf{43}, 405 (1999).
\bibitem{geremia} J.M.~Geremia, J.K.~Stockton, A.C.~Doherty and
H.~Mabuchi,
Phys. Rev. Lett. \textbf{91}, 250801 (2003).
\bibitem{verstraete} F.~Verstraete, A.C.~Doherty and H.~Mabuchi,
Phys. Rev. A \textbf{64}, 032111 (2001).
\bibitem{alvaro} J.S.~Lundeen, A.~Feito, H.~Coldenstrodt-Ronge,
K.L.~Pregnell, Ch.~Silberhorn, T.C.~Ralph, J.~Eisert, M.B.~Plenio and
I.A.~Walmsley, ``Tomography of Quantum Detectors,'' Nature Physics,
published online: 16 Nov.\ 2008, DOI: 10.1038/NPHYS1133 (2008).
\bibitem{banaszek07}  M.~Karpinski, C.~Radzewicz and K.~Banaszek,
J. Opt. Soc. Am. B \textbf{24}, 668 (2008).
\bibitem{zeilinger} G.~Molina-Terriza, A.~Vaziri, J.~\v{R}eha\v{c}ek,
Z.~Hradil and A.~Zeilinger,
Phys. Rev. Lett. \textbf{92}, 167903 (2004).
\bibitem{polzik} J.~Sherson, H.~Krauter, R.K.~Olsson, B.~Julsgaard,
K.~Hammerer, I.~Cirac and E.S.~Polzik,
Nature \textbf{443}, 557 (2006).
\bibitem{lukin} L.~Childress et al, Science \textbf{314}, 281 (2006).
\bibitem{julsgaard} L.~Rippe, B.~Julsgaard, A.~Walther, Y.~Ying and
S.~Kr\"oll,
Phys. Rev. A \textbf{77}, 022307 (2008).
\bibitem{kalman2} K.~Audenaert and S.~Scheel, In preparation.
\bibitem{hager} W.W.~Hager,
SIAM Review \textbf{31}, 221 (1989).
\bibitem{kbj} S.~Kotz, N.~Balakrishnan, and N.L.~Johnson,
\textit{Continuous Multivariate Distributions, Volume 1: Models and
Applications}, Second Edition, (New York, Wiley, 2000).
\bibitem{sivia} D.S.~Sivia, with J.~Skilling,
\textit{Data Analysis, a Bayesian Tutorial}, second edn.
(Clarendon Press, Oxford, 2006).
\bibitem{laplace} P.S.~Laplace,
M\'emoires de l'Acad\'emie Royale des Sciences \textbf{6}, 621 (1774).
\bibitem{thatcher} A.R.~Thatcher,
J. Roy. Statist. Soc. Series B (Methodological) \textbf{26}, 176 (1964).
\bibitem{fienberg} S.E.~Fienberg and P.W.~Holland,
J. Am. Statist. Assoc. \textbf{68}, 683 (1973).
\bibitem{abramowitz} M.~Abramowitz and I.A.~Stegun (eds.),
\textit{Handbook of mathematical functions}, (Dover, New York, 1972).
\bibitem{mogilevtsev06} Z.~Hradil, D.~Mogilevtsev and J.~\v{R}eh\'a\v{c}ek,
Phys.\ Rev.\ Lett.\ \textbf{96}, 230401 (2006).
\bibitem{mogilevtsev07} D.~Mogilevtsev, J.~\v{R}eh\'a\v{c}ek and Z.~Hradil,
Phys.\ Rev.\ A \textbf{75}, 012112 (2007).
\bibitem{barlow} R.~Barlow, \textit{Statistics}, (Wiley, New York, 1989).
\bibitem{jkb94} N.L.~Johnson, S.~Kotz, and N.~Balakrishnan,
\textit{Continuous Univariate Distributions, Volume 1}, Second
Edition, (Wiley, New York, 1994).
\bibitem{EvansSwartz} M.~Evans and T.~Swartz,
Statistical Science \textbf{10}, 254 (1995).
\bibitem{mckaybook} D.J.C.~MacKay, \textit{Information Theory,
Inference, and Learning Algorithms}, (Cambridge University Press,
Cambridge, 2003).
\bibitem{sedumi}  J.F.~Sturm, Optimization Methods and Software
\textbf{11--12}, 625-653 (1999). Free software (running under Matlab)
available at {\tt http://sedumi.mcmaster.ca}.
\bibitem{jaynes} E.T.~Jaynes,
Phys. Rev. \textbf{106}, 620 (1957).
\bibitem{rehacek} J.~\v{R}eh\'a\v{c}ek and Z.~Hradil, ``MaxEnt assisted
MaxLik tomography'', Proc. Maxent 2003 (2003 August 3--8, Jackson
Hole, WY, USA), see also arXiv:physics/0404121 (2004).
\bibitem{boyd} L.~Vandenberghe and S.~Boyd,
SIAM Review \textbf{38}, 49 (1996).
\bibitem{NathanPhD} N.K.~Langford, PhD Thesis, University of
Queensland (2007).
\bibitem{banaszek} K.~Banaszek and I.A.~Walmsley,
Opt. Lett. \textbf{28}, 52 (2003).
\bibitem{achilles} D.~Achilles, Ch.~Silberhorn, C.~Sliwa, K.~Banaszek
and I.A.~Walmsley,
Opt. Lett. \textbf{28}, 2387 (2003).
\bibitem{wald} A.~Wald,
Trans.  AMS \textbf{54}, 426 (1943).
\bibitem{moments} K.~Audenaert, ``Inverse moments
of univariate discrete distributions via the Poisson expansion'', Submitted (2008).
%
%
%
\bibitem{extra} Additional material available at\\
http://personal.rhul.ac.uk/usah/080/Kalman.htm
\end{thebibliography}
\end{document}